\documentclass{aa}
\usepackage{txfonts,natbib,graphicx,longtable}
\bibpunct{(}{)}{;}{a}{}{,}

\begin{document}

\title{A Catalog of Galaxy Clusters Observed by XMM-Newton
\thanks{Based on observations obtained with {\it XMM-Newton}, 
an ESA science mission with instruments and contributions 
directly funded by ESA Member States and NASA}
\thanks{Figures~\ref{fig:prof-01}$-$\ref{fig:im-07} and 
Table~\ref{tbl:clusterdetails} are only available in electronic form 
via http://www.edpsciences.org }}

\author{S. L. Snowden\inst{1} 
\and R. M. Mushotzky\inst{2}
\and K. D. Kuntz\inst{3}
\and D. S. Davis\inst{4,}\inst{5}}

\institute{Code 662, NASA/Goddard Space Flight Center, Greenbelt, MD 20771, 
steven.l.snowden@nasa.gov
\and
Code 662, NASA/Goddard Space Flight Center, Greenbelt, MD 20771,
richard.f.mushotzky@nasa.gov
\and
Henry A. Rowland Department of Physics and Astronomy, The Johns 
Hopkins University, 3400 N. Charles Street,
Baltimore, MD 21218, kuntz@pha.jhu.edu
\and
Department of Physics, University of Maryland,
Baltimore County, 1000 Hilltop Circle, Baltimore, MD 21250
\and
CRESST and the Astroparticle Physics Laboratory,  NASA/GSFC, Greenbelt, MD 20771, USA,
david.s.davis@nasa.gov}
\authorrunning{Snowden et al.}

\date{Received 0 Month 2007 / Accepted 0 Month 2007}

\abstract
{}
{We present a uniform catalog of the images and radial profiles of the temperature,
abundance, and brightness for 70 clusters of galaxies observed by {\it XMM-Newton}.}
{We use a new ``first principles'' approach to the modeling and removal
of the background components; the quiescent particle background,
the cosmic diffuse emission, the soft proton contamination, 
and the solar wind charge exchange emission.
Each of the background components demonstrate significant spectral variability,
several have spatial distributions that are not described 
by the photon vignetting function,
and all except for the cosmic diffuse emission are temporally variable.
Because these backgrounds strongly affect the analysis of low surface brightness objects,
we provide a detailed description our methods of identification, characterization,
and removal.}
{We have applied these methods to a large collection of XMM-Newton observations
of clusters of galaxies and present the resulting catalog.
We find significant systematic differences 
between the {\it Chandra} {\rm and} {\it XMM-Newton} {\rm temperatures.}}
{}

\keywords{x-rays: observations, clusters of galaxies, analysis: methods}

\maketitle

\section{Introduction}

Clusters of galaxies are the largest and most massive collapsed objects in the 
universe, and as such they are sensitive probes of the history of structure 
formation. While first discovered in the optical band in the 1930s 
\citep[for a review see][]{b97}, 
in some ways the name is a misrepresentation since 
most of the baryons and metals are in the diffuse hot X-ray emitting intercluster 
medium and not in the galaxies.  Clusters are fundamentally ``X-ray objects'' 
as it is this energy range where this preponderance of the baryons is visible.   
Studies of cluster evolution can place strong constraints 
on all theories of large scale structure and determine 
precise values for many of the cosmological parameters. 
As opposed to galaxies, 
clusters probably retain all the enriched material created in them, 
and being essentially closed boxes 
they provide a record of nucleosynthesis in the universe. 
Thus measurement of the elemental abundances 
and their evolution with redshift provides 
fundamental data for the origin of the elements. 
The distribution of the elements in clusters 
reveals how the metals moved from stellar systems into the IGM. 
Clusters should be fair samples of the universe and 
studies of their mass and their baryon fraction 
should reveal the gross properties of the universe as a whole. 
Since most of the baryons are in the gaseous phase 
and clusters are dark-matter dominated, 
the detailed physics of cooling and star formation 
are much less important than in galaxies.  
For this reason, clusters are much more amenable to detailed simulation 
than galaxies or other systems 
in which star formation has been a dominant process. 

Clusters are luminous, extended X-ray sources and are visible out to 
high redshifts with current technology.  The virial temperature of most 
groups and clusters corresponds to $T\sim2-100\times10^6$~K 
($kT\sim0.2-10$~keV, velocity dispersions of $180-1200$~km~s$^{-1}$),
and while lower mass systems certainly exist we usually call them 
galaxies.  Most of the baryons in groups and clusters of galaxies
lie in the hot X-ray emitting gas that is in  rough virial equilibrium
with the dark matter potential well \citep[the ratio of gas to stellar 
mass is $\sim2-10:1$,][]{asf01}. This gas is enriched in heavy elements 
\citep{mea78} and it thus preserves a record of the entire history of 
stellar evolution in these systems.  The presence of heavy elements is 
revealed by line emission from H and He-like transitions as well as 
L-shell transitions of the abundant elements.  Most clusters are too 
hot to have significant ($>2$~eV equivalent width) line emission from 
C or N, although cooler groups may have detectable emission from these 
elements. However, all abundant elements with $z>8$ (oxygen) have strong 
lines from H and He-like states in the X-ray band and their abundances 
can be well determined.

Clusters of galaxies were discovered as X-ray sources in the late 1960's 
(see \citep[for a historical review see][]{m02} and large samples were first 
obtained with the {\it Uhuru} satellite in the early 1970's \citep{jf78}. 
Large samples of X-ray spectra and images were first obtained in the late 
1970's with the {\it HEAO} satellites \citep[for an early review
see][]{jf84}.  The early 1990's brought large samples of high quality 
images with the {\it ROSAT} satellite 
and good quality spectra with {\it ASCA} and {\it Beppo-SAX}. 
In the last few years there has been an enormous increase 
in the capabilities of X-ray instrumentation due to the launch 
and operation of {\it Chandra} and {\it XMM-Newton}. 
Both {\it Chandra} and {\it XMM-Newton} can find and identify clusters 
out to $z>1.2$ and their morphologies can be clearly discerned to $z>0.8$. 
Their temperatures can be measured to $z\sim1.2$ and {\it XMM-Newton} can 
determine their overall chemical abundances to $z\sim1$ with a sufficiently 
long exposure.  For example, a cluster at $z=1.15$ has recently had its 
temperature and abundance well constrained by a 1~Ms {\it XMM-Newton} 
exposure \citep{hea04}.

The temperature and abundance profiles of clusters out to redshifts of 
$z\sim0.8$ can be measured and large samples of X-ray selected clusters 
can be derived.  {\it Chandra} can observe correlated radio/X-ray structure 
out to $z>0.1$ and has discovered internal structure in clusters.  
The {\it XMM-Newton} grating spectra can 
determine accurate abundances for the central regions of clusters in a 
model independent fashion for oxygen, neon, magnesium, iron, and silicon.
Despite the stunning successes of the {\it Chandra/XMM-Newton} era,
clusters have not yet fulfilled their promise as a cosmological Rosetta 
stone; the most important tests of cluster theory require measurements 
of cluster properties to large radii ($R\sim R_{virial}$) 
which is observationally difficult.
The lack of consensus among the recent X-ray missions about,
for example, temperature profiles, is a large stumbling block
in the use of clusters for cosmological purposes.        

\subsection{Temperature Structure of Clusters}

As discussed in detail by \citet{e03}, we now have a detailed understanding
of the formation of the dark matter structure for clusters of galaxies.
If gravity has been the only important physical effect since the formation,
then the gas should be in rough hydrostatic equilibrium and its density and 
temperature structure should provide a detailed measurement
of the dark matter distribution in the cluster.
Recent theoretical work has also taken into account other processes,
such as cooling, which can be important.
The fundamental form of the \citet{nfw97} dark matter potential
and the assumption that the fraction of the total mass that is in gas
is constant with radius results in a prediction that the cluster gas
should have a declining temperature profile
at a sufficiently large distance from the center (in $R/R_{viral}$ units),
both from analytic \citep{ks01} and numerical modeling \citep{lea02}.
The size of the temperature drop in the outer regions
is predicted to be roughly a factor of 2 by $R/R_{viral}\sim0.5$.

Although some observational results appear consistent with the 
theoretical predictions
\citep[in particular,the {\it ASCA} results of ][]{mea98}, 
many others do not, and considerable controversy exists.
Much of the uncertainty of the pre-{\it Chandra}/pre-{\it XMM-Newton} 
data arises from insufficient spectral and spatial resolution
and the resultant difficulties in removing backgrounds,
modeling the spectra, and interpreting the results.
For example, the {\it ASCA} results of \citet{mea98}
were consistent with a decline in temperature with radius,
while the analysis of a similar sample of clusters by \citet{kea99},
\citet{wb00}, and \citet{w00} revealed a large number of isothermal 
clusters.  Similar results were obtained from {\it Beppo-SAX},
with \citet{dgea99} finding temperature gradients
and \citet{ib00} finding isothermality.
Simultaneous analysis of the higher angular resolution {\it ROSAT} 
data with the {\it ASCA} data did not resolve the issue;
\citet{fad01} finding gradients and \citet{ibe99} isothermal profiles.
The bulk of the problem with interpreting {\it ASCA} results
is the analysis of impact of the PSF on the profile \citep{ibe99}.

{\it XMM-Newton} and {\it Chandra} have significantly better spectral 
and angular resolution than the previous generation of missions
and might be expected to resolve the previous controversies.
The recent {\it Chandra} results of \citet{vea06}
show a temperature profile in good agreement
with the gradients seen by \citet{mea98} results
and predicted by the standard theory.
Analysis of samples of cooling flow clusters with {\it XMM-Newton}
\citep{pea05,app05,pea07} are also mostly consistent with the 
\citet{mea98} results.  However flatter, more isothermal profiles 
have also been found in both {\it Chandra} 
and {\it XMM-Newton} observations \citep{asf01,kea04,app05}.
Despite some early difficulties \citep[e.g.,][]{dea06},
the {\it Chandra} and {\it XMM-Newton} calibrations have stabilized
but agreement between the two great observatories is not assured 
\citep[e.g,][]{vea06}.  The difference in the PSF between the two 
instruments as well as different methods of background subtraction
often make direct comparison difficult.
Further, an agreement between {\it Chandra} and {\it XMM-Newton}
would not entirely resolve the problem;
the smaller FOV of current instruments
have led to observation of a somewhat higher redshift sample
than observed by the previous generation of instruments,
suggesting that part of the difference between 
the {\it XMM-Newton}/{\it Chandra} results
and the {\it ASCA}/{\it ROSAT}/{\it Beppo-SAX} results
may be due to a real difference between clusters at lower and higher redshifts.
 
The measurement of the cluster mass function can provide a sensitive
cosmological test but is sensitive, in turn,
to the parameters that are directly measurable,
and especially to the observed quantities at large radius.
Recent simulations show that cluster temperature profiles decline
with radius but less rapidly than is
shown by previous X-ray analysis \citep[e.g.,][]{ktea04}.
Since the total mass of the cluster is quite sensitive to the
measured temperature profile \citep{rea06},
particularly at large radii, these systematic differences
lead to significant uncertainties in the cosmological constraints.
Thus, there is an urgent need to understand
the temperature profiles of clusters at large radii
and to understand the source of the systematic differences
observed in the literature.
 
In this paper we consider a large sample of clusters observed with the 
{\it XMM-Newton} observatory and derive temperature, density and 
abundance profiles for many of these systems out to near the virial radius.
We present a new technique that should provide
more accurate background subtraction at large radii,
and are careful to correct for the effect of the finite {\it XMM-Newton} 
PSF.  A comparison of our measurements 
with {\it Chandra} measurements of the same clusters
shows a simple systematic difference between the two observatories.
Although we have not yet determined the source of that difference,
resolution of this relatively well defined issue
should significantly reduce the uncertainties in cluster cosmology.

\subsection{Analysis of Extended Sources}

The analysis of extended sources in X-ray astronomy is typically problematic 
and quite often very complex.  This is particularly true for objects which
subtend the entire field of view (FOV) of the observing instrument such as 
nearby galaxies, relatively nearby clusters of galaxies, many regions of 
galactic emission, and of course the cosmic diffuse background.  Even 
with objects smaller than the FOV, quite often the simple subtraction of a
nearby ``background'' region from the same data set is inappropriate due to 
spectral and spatial variations in the internal background and angular 
variations in the cosmic background.  The use of deep ``blank sky'' 
observations can also be inappropriate due to the same considerations, 
as well as the probability that many background components are temporally 
varying.  Because of the temporal variation of the background
and the angular variation of the cosmic background,
the average of the blank-sky data, even after normalization,
may not match the conditions of a specific observation of interest,
and so may yield an incorrect result.

While the cores of many clusters are relatively bright in X-rays so the 
treatment of the background is not such a significant consideration, at the 
edges of clusters it is absolutely critical.  Clusters fade gently into the
backgrounds at large radii, therefore improving the modeling of the 
backgrounds extends the reliable radial range for the determination of 
cluster parameters.

Critical to compensating for the various background components by filtering, 
subtraction, or modeling is a basic understanding of their origin and effects
on the detectors.  Unfortunately this usually takes a considerable amount of
time to develop, which is why useful methods for a specific observatory become 
available to the general community only years into the mission.  Even then, 
the methods will continue to evolve with greater understanding of the various 
background components and their temporal evolution, and the operation 
of the instruments.  In addition, the efforts are quite often undertaken by 
individuals who are not project personnel, but whose scientific interests 
require the improved analysis methods.

This is certainly true of the {\it XMM-Newton} mission and observations 
using the EPIC instruments.  Several groups have presented methods and 
published scientific results based upon them \citep{aea01,rp03,nml05,dlm04}.  
As opposed to these methods which derive 
background spectra from normalized blank-sky observations, this paper 
presents the details of a method based as much as possible on the 
specific understanding of the individual background components.  This 
method was used successfully in the paper identifying the solar wind 
charge exchange emission in the {\it XMM-Newton} observation of the 
Hubble Deep Field North \citep{sck04}.  

Section~\ref{sec:instrumentation} of this paper gives a short 
description of the {\it XMM-Newton} observatory, Sect.~\ref{sec:components} 
discusses the various background components and the suggested methods used 
to compensate for them, Sect.~\ref{sec:example} demonstrates the data
reduction method using the observation of \object{Abell 1795}. 
Sect.~\ref{sec:catalog} applies the methods to the 
determination of the temperature, abundance, and flux radial profiles of 
a catalog of 70 clusters of galaxies and presents the results, and 
Sect.~\ref{sec:conclusions} discusses the conclusions.  Note that the detailed
discussion of the science derived from these observations is deferred to
Paper~II.  

Currently the specific method and software package discussed here
are only applicable to EPIC MOS data. Although the MOS and pn experience 
the same backgrounds, the physical difference between the two detectors
(readout rates, fraction of unexposed pixels, etc.) make analysis of the 
pn background somewhat more difficult than that of the MOS. However, the 
analysis methods described here are being extended to the pn.

\section{{\it XMM-Newton} and the EPIC MOS Detectors}
\label{sec:instrumentation}

The {\it XMM-Newton} observatory \citep{eea05} orbits the Earth in a long 
period ($\sim48$~hours), highly elliptical path (the original perigee and 
apogee were $\sim6000$~km and $\sim115,000$~km but they have since evolved 
over the mission to $\sim19,000$~km and $\sim103,000$~km as of 2006 June). 
The scientific package of {\it XMM-Newton} is comprised of six 
independent but co-aligned instruments which operate simultaneously.  The 
European Photon Imaging Camera (EPIC) is comprised of three CCD imagers of 
two distinct technologies (MOS and pn), and are coupled to the three X-ray 
mirror assemblies. The EPIC instruments provide imaging over a $\sim30'$ 
FOV with moderate energy resolution.  Half of the light 
from two of the X-ray mirrors (those with the MOS detectors) is diverted 
by reflection gratings to the Reflection Grating Spectrometer (RGS), two 
instruments which provide high spectral resolution for point sources and 
small-scale extended objects ($<2'$).  The final scientific instrument is 
the Optical Monitor (OM) which is an optical/UV telescope with a FOV 
($17\arcmin\times17\arcmin$) somewhat smaller than that of the EPIC.

The EPIC MOS detectors are each comprised of seven individual CCDs where 
one is roughly centered on the optical axis and the others form a hexagonal 
pattern surrounding it.  The central CCD can be operated independently in 
several different observation modes (imaging, windowed imaging, and timing) 
while the outer CCDs always operate in the standard imaging mode.  There 
are three optical blocking filters (thin, medium, and thick) which can be 
set by the observer.  The filter wheel has a circular aperture with a $30'$ 
diameter which leaves portions of the outer CCDs shielded from exposure 
to the sky. 
These unexposed corners of the detectors play a vital role in the modeling 
of the quiescent particle background (QPB) as described below.  The filter 
wheel also has settings which expose the CCDs to an on-board calibration 
source (cal-closed position) and which block the sky (filter wheel closed 
position, FWC), data from the latter position 
are also used in modeling the QPB.  13 of the 14 MOS CCDs are still 
functioning as of 2007 September, one of the MOS1 outer CCDs (CCD~\#6) 
was lost to a micrometeorite hit on 2005 March 9.

\section{EPIC MOS Background Components}
\label{sec:components}

There are five major contributors to the background of EPIC MOS (and pn) 
observations that we consider here.  However, the characterization 
of some components as background is occasionally debatable as they may 
actually be the main scientific interest of an observation.  The 
first is the quiescent particle background, a continuum component 
produced by the interaction of high energy penetrating particles with 
the detectors.  Generally included with, but distinct from, the QPB are 
fluorescent X-rays
(FX) which are produced by the particle flux interacting with various 
components of the satellite and then are detected by the instruments.  
For the MOS the fluorescent X-rays are dominated by the lines Al~K$\alpha$ 
($E\sim1.49$~keV) and Si~K$\alpha$ ($E\sim1.75$~keV), but there are also 
lines from other elements at higher energies (Au, Cr, Mn, Fe, Ni, Zn).  
The continuum QPB dominates at high (above $\sim2$~keV) and low 
(below $\sim1.2$~keV) energies while the Al and Si lines dominate the 
$1.3-1.9$~keV band.  

The third background component is also produced locally at the detectors 
and is caused by soft protons  (SP, with energies less than a few 
100~keV\footnote{{\it XMM-Newton} Technical Note XMM-SOC-USR-TN-0014, 
P. M. Rodriguez-Pascual \& R. Gonz\'alez-Riestra, 
http://xmm.esac.esa.int/docs/documents/USR-TN-0014-1-0.pdf.}) 
which travel down the telescope light path and deposit their energy directly 
in the detectors.  The SP spectrum, as recorded by the EPIC detectors,
can be described by a power-law continuum 
and varies both in magnitude and slope.
The soft proton background is highly variable
and enhancements in the soft proton background
are often referred to as ``flares''.

For many observations the fourth component, the cosmic X-ray background 
(CXB), is a source of contamination although it can also be the scientific 
goal of the observation.  The diffuse CXB dominates below $\sim1$~keV and 
has a thermal spectrum dominated by emission lines.  It is the superposition 
of Galactic emission from multiple sources as well as the Galactic halo and 
perhaps even more distant emission, and is strongly variable over the sky.  
Included in the CXB is the unresolved emission from the superposition 
of cosmological objects \citep[e.g., AGN,][]{hm07} which dominates at 
higher energies and Galactic stars with a relatively minor contribution 
at lower energies \citep[e.g., ][]{kks01}.  The average spectrum of the 
cosmological emission is for the most part a power law 
continuum with a possible change in slope at lower energies.  There is 
thought to be a true cosmic variation of magnitude on the sky but there 
is also the obvious variation caused by the excision of point sources to 
various levels.    

The fifth background component, solar wind charge exchange emission 
\citep[SWCX, e.g.,][]{wea04,sck04}, like the 
CXB, can either be background or a source of scientific interest, although 
admittedly to a rather limited community.  SWCX in the MOS energy band is 
essentially all line emission at energies less than $\sim1.3$~keV 
and is strongly variable in both total magnitude and relative line 
strengths.  For the MOS detectors of {\it XMM-Newton} the strongest SWCX 
emission is from \ion{C}{vi}, \ion{O}{vii}, \ion{O}{viii}, \ion{Ne}{ix}, 
and \ion{Mg}{xi}, 
although this ignores the $\frac{1}{4}$ keV band where {\it ROSAT} 
observations were occasionally affected by very strong SWCX emission. 

\begin{figure}
\centering\includegraphics[angle=90,width=8.5cm]{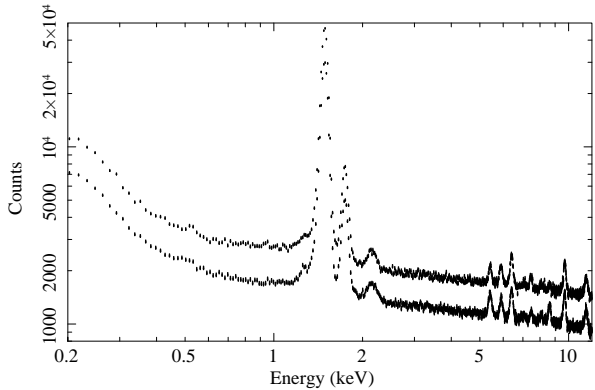}
\caption{Filter wheel closed spectra for the MOS1 (upper) and MOS2 
(lower) detectors.  The MOS2 data have been scaled by a factor of 1.5 
in order to separate the spectra for clarity.  The spectra are 
comprised of a general continuum from the
QPB and the FX lines of Al, Si, Au, and other elements.  The 
energy binning for the data is a constant 15~eV.
\label{fig:qpb-sp}}
\end{figure}

\subsection{Quiescent Particle Background}
\label{sec:qpb}

\begin{figure}
\centering{\includegraphics[width=8.5cm]{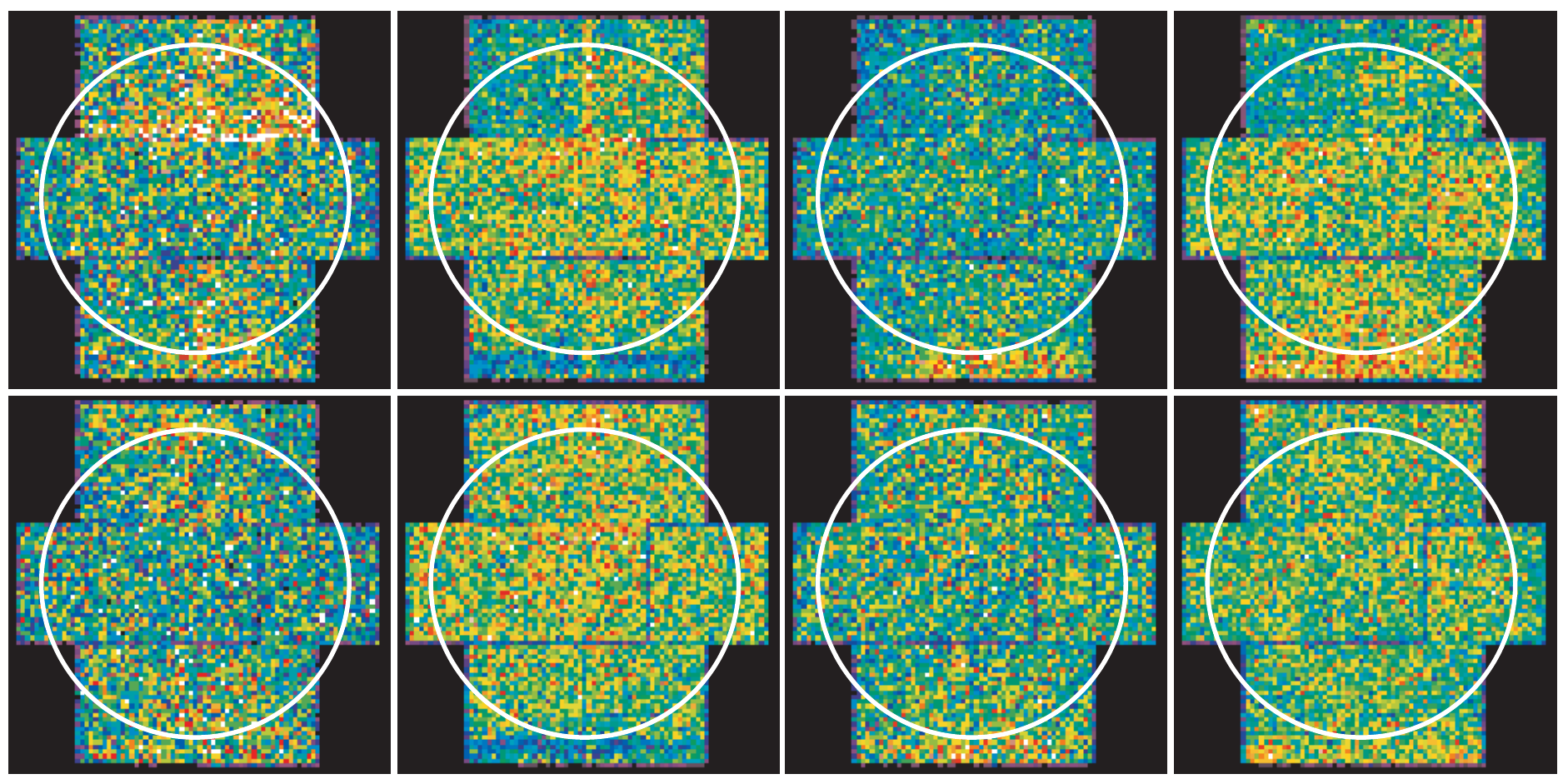}}
\caption{Images in detector coordinates of the FWC data 
for the MOS1 (upper row) and MOS2 (lower row) detectors.  The data are from 
(left to right) the $0.35-1.25$, $1.25-2.0$, $2.0-4.0$, 
and $4.0-8.0$~keV bands, and have been binned into $25''\times25''$ pixels.  
The $1.25-2.0$ band is most affected by the FX contamination, and it is likely
that there is some difference between the spatial distributions between FWC 
and open data due to the different source geometry.  Note that all of the 
bands show at least somewhat different structure.  The circles indicate the 
FOV regions of the instruments outside
of which the detectors are always shielded from cosmic X-rays.
\label{fig:qpb-im}}
\end{figure}

The QPB and FX for the EPIC MOS detectors has been well studied by 
\citet{ks07} (hereafter KS07) and is the dominant background above 
$\sim2.0$~keV.  In general it is 
relatively featureless and resembles a power law which is not folded 
through the instrumental effective area.  
Fig.~\ref{fig:qpb-sp} shows MOS1 and MOS2 spectra compiled from 
observations where the filter wheels were in their closed 
positions (FWC)
while Fig.~\ref{fig:qpb-im} shows FWC images in several bands.  
In this configuration no particles or X-rays passing through the optical 
system can penetrate to the detectors, nor are the on-board calibration 
sources visible to the detectors.  The FOV 
of the detectors for cosmic X-rays and soft protons is constrained by a 
circular aperture indicated by the circles in the figure.  The permanently 
shielded regions of the CCDs, i.e., the corner regions outside of the 
circles in Fig.~\ref{fig:qpb-im}, are read out the same as those within 
the FOV and experience roughly the same QPB flux.

The QPB spectra for the two detectors (Fig.~\ref{fig:qpb-sp}) are very 
similar and show a strong continuum with the Al~K$\alpha$ and Si~K$\alpha$ 
lines, as well as a few lines from other elements.  
Fig.~\ref{fig:qpb-im} shows that the distribution of counts over the 
detectors is clearly not uniform, and that 
the contributions from the Al~K$\alpha$ and Si~K$\alpha$ 
fluorescent lines are distributed somewhat differently from the QPB 
as well.

\begin{figure*}
\centering{\includegraphics[width=12.0cm]{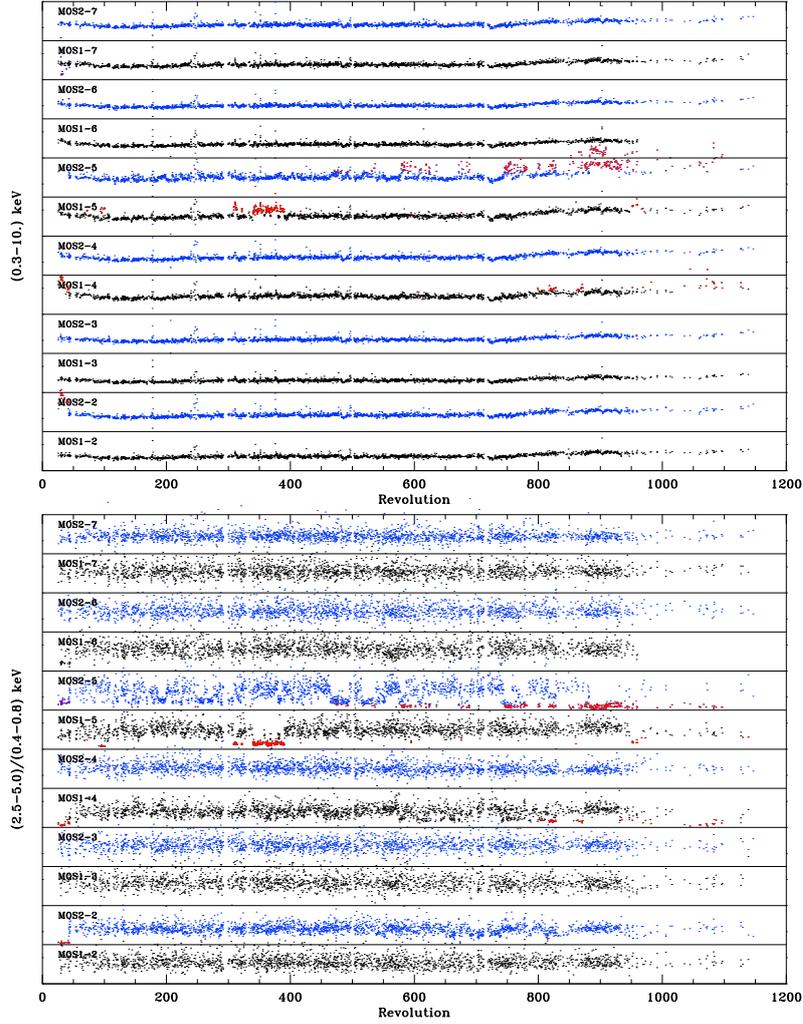}}
\caption{Top Panel: QPB count rate in the $0.3-10.0$~keV band from the 
out-of-FOV corners of the detectors from KS07 (their Fig.~6) 
for the individual CCDs from both MOS instruments.  The MOS1 data are shown 
in black, the MOS2 data are shown in blue, and time periods of anomalous CCD 
background behavior are shown in red. Bottom Panel: 
QPB ($2.5-5.0$~keV)/($0.4-0.8$~keV) hardness from the out-of-FOV 
corners of the detectors from KS07 (their Fig.~7) 
for the individual CCDs from both MOS instruments.
The MOS1 data are shown in black, the MOS2 data are shown in blue, and
time periods of anomalous CCD background behavior are shown in red.
The plot limits are $0-0.075$~counts~s$^{-1}$~chip$^{-1}$ for the 
count-rate plots and $0-7.5$ for the ratio plots. The data are linearly 
scaled in both cases.}
\label{fig:qpb-specvar}
\end{figure*}

In addition to the spatial variation of the QPB over the detectors, 
there is also 
a temporal variation in the spectra both in magnitude and in hardness.  
Fig.~\ref{fig:qpb-specvar} (top panel) shows the QPB count rates from the 
CCD corners outside of the FOV in the $0.3-10.0$~keV band.  The temporal 
variation is due both to changes
in the CCDs and their operating conditions as well as variations of the 
particle flux over the course of the solar cycle.  Some of the short-term
scatter is due to the varying conditions during the orbit ($\sim2$~days).  
Observations can take place both inside and outside of the 
magnetosheath and at various distances from the particle belts.  
Fig.~\ref{fig:qpb-specvar} (bottom panel) shows the QPB hardness ratio 
(the ratio of the $2.5-5.0$~keV band to the $0.4-0.8$~keV band) over the 
course of the mission for the individual 
CCDs.  Of note are the occasional deviations of CCD \#5 of both instruments 
as well as MOS1 CCD~\#4 from relatively nominal levels and the loss of 
MOS1 CCD~\#6 near revolution (orbit) 950.  The deviations are due to a
strong enhancement in the background below $E\sim1$~keV and are 
extensively discussed in KS07.

\subsubsection{Treatment of the QPB}
\label{sec:qpb-treat}

In all of the discussion above only the quiescent part of the particle 
background is considered, these are the time periods not affected by 
flares.  Frequently times of particle background flaring are so intense 
that the instruments must be turned off for their health.  Periods of 
less intense flaring are easily filtered out by light-curve screening, 
which is discussed in \S~\ref{sec:sp-treat}.

The QPB for an individual observation (primary observation, PO) can be 
modeled and subtracted with, in general, quite high reliability using the 
FWC data in conjunction with data from the unexposed corners of the CCDs
(KS07).  The modeling is a multi-step process, and is done for each 
detector and CCD individually.  The process creates a background 
spectrum tailored for the specific region of interest where the spectrum 
of an astrophysical object is extracted.  To summarize the process 
outlined in KS07: 1) After the PO has been screened for flares, the 
spectra from the unexposed corners of the outer CCDs are extracted.  
2) The magnitudes ($0.3-10.0$~keV band) and hardness ratios 
($2.0-5.0$~keV band to the $0.5-1.2$~keV band) for the spectra are 
determined. 3) A data base of all archived observations is searched for
observations (secondary observations, SO) whose unexposed corner spectra 
have similar magnitudes and hardness.  4) The PO corner spectra are then 
augmented by the SO corner spectra increasing the statistical significance of
individual spectral bins to a useful level.  This step makes the assumption 
that data collected from time periods of similar spectral magnitude and
hardness have in aggregate the same spectrum.  This appears to generally 
be the case, although CCDs \#4 and \#5 in their anomalous states can be 
problematic.  5) Spectra from the FWC data are extracted from CCD corners 
as well as from the region of interest.  If the region of interest is 
comprised of more than one CCD, the individual CCD spectra are kept separate.  
6) For the outside CCDs the FWC spectra from the region of interest are 
scaled, spectral bin by spectral bin, by the ratio of the augmented 
observation spectra from the CCD corners to the FWC spectra from the 
corners.  The central CCD has to be handled in a more complicated way 
(KS07).  7) For reasons discussed in \S~\ref{sec:fx-treat} below, the 
spectral region affected by the Al~K$\alpha$ and Si~K$\alpha$ lines 
($1.2-2.0$~keV) is cut out and replaced by an interpolated power law.
The EXPOSURE and BACKSCAL keywords in the background spectrum are
set to be consistent with the PO.  The spectrum is then included as 
the background in spectral fitting. 

\begin{figure*}
\centering{\includegraphics[width=17.0cm]{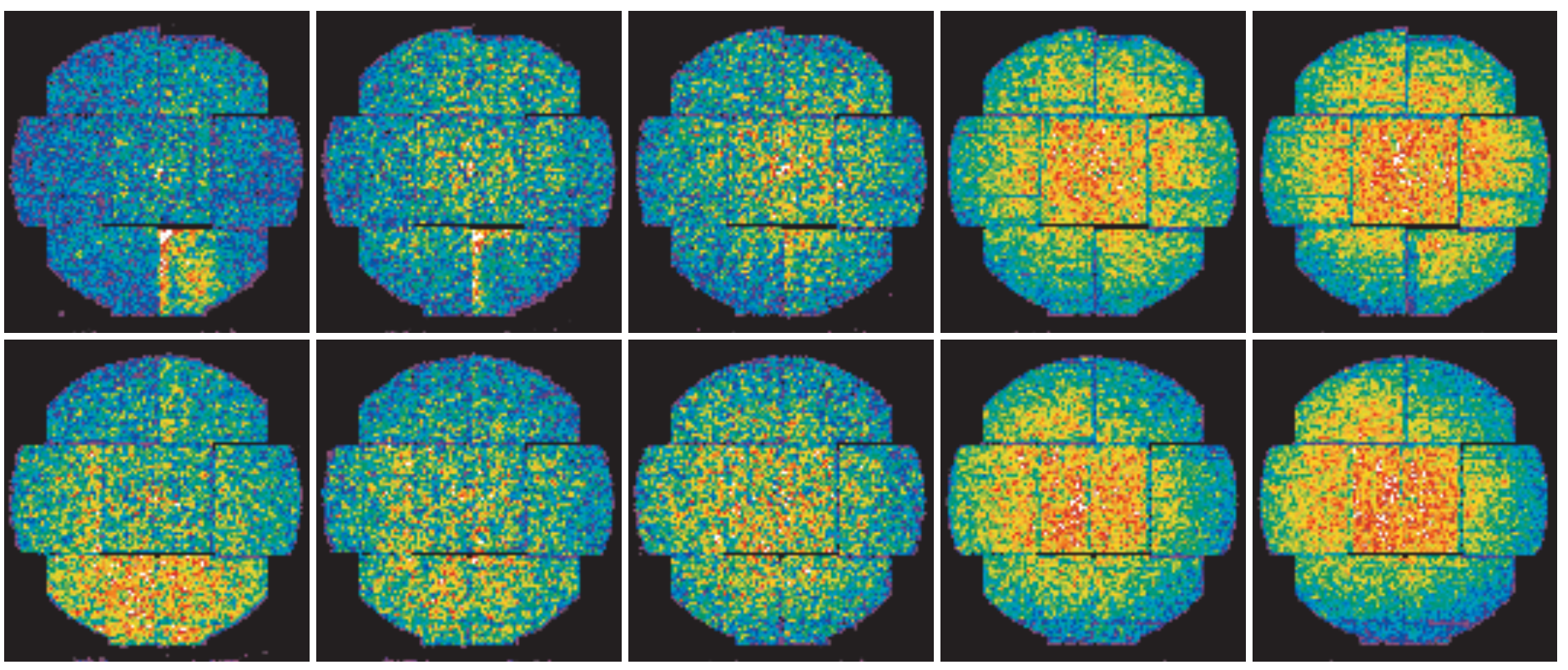}}
\caption{Image in detector coordinates of the SP data for the MOS1 (upper 
row) and MOS2 (lower row) detectors.  From left to right the data are from 
the $0.35-0.8$~keV, $0.8-1.25$~keV, $1.25-2.0$~keV, $2.0-4.0$~keV, and 
$4.0-8.0$~keV bands.  In the plots blue and green indicate lower 
intensities while red and white indicate higher intensities.  The data
are linearly scaled.  For better statistics, data are from the observations
using all filters have been combined as there is little difference between 
the distributions for the thin, medium, and thick filter observations 
separately.  Note that the distributions are not flat across 
the detectors nor are they symmetrically vignetted like cosmic X-rays.
As well, the distributions are not the same for different energies.
\label{fig:sp-im}}
\end{figure*}

\subsubsection{Treatment of the FX Background}
\label{sec:fx-treat}

There are two reasons why the Al-K$\alpha$ and Si-K$\alpha$ FX background 
can not be treated in the 
same manner as the QPB.  First, the environment with the filter wheel 
open (with the thin, medium, and thick filters) is different from that
with the filter wheel closed, and therefore the distribution and magnitude
of the FX background are unlikely to be the same.  Second, there are quite 
large numbers of counts in the lines providing high statistics.  Because 
of this, even the slight residual variations in the instrumental gains 
(within the gain uncertainty) when compared to the FWC data 
can produce large residuals.  The most 
straight-forward method for treating the lines is to fit them as separate 
Gaussian model components where the line energy is allowed to vary to 
achieve an acceptable fit.

\subsection{Soft Proton Background}
\label{sec:sp}

The SP background is produced by relatively low energy protons 
($<$~a few 100~keV) passing down the telescope tube, penetrating the 
filters, and depositing their energy directly in the CCDs.  This is 
a very problematic component which can vary from undetectable levels 
(by examination of the count rate) to strong flaring of over one 
hundred counts per second rendering the observation 
useless for the study of all but the brightest point sources.  The SP 
spectrum is a continuum with variable hardness.  The distribution of 
SP events across the FOV is different from both cosmic X-rays and the
QPB, and varies as well with energy.  Fig.~\ref{fig:sp-im} shows 
SP background images collected from time intervals of slightly 
enhanced background ($\sim1$~counts~s$^{-1}$) for several 
energy bands.  While there is a 
significant variation in the distributions from low to high energies, 
and between the two detectors, they are relatively similar at energies 
$>2.0$~keV for the individual detectors where the SP contribution is 
relatively stronger.

\subsubsection{Treatment of the SP Background}
\label{sec:sp-treat}

The primary treatment of the SP background is to filter the data by
creating a light curve and excluding all time intervals with a count 
rate greater than some selected threshold.  There are a number of 
different filtering methods in the literature but they all 
give basically the same results.  Since most, if not all observations 
contain residual SP contamination at some level, the setting of the 
threshold becomes dependent on a trade-off between the level of that 
contamination and the amount of the exposure left over after the 
screening process.  Our goal is to retrieve as much useful data as possible 
so rather than setting a strict limit to exclude all possible time
periods of SP contamination \citep[e.g.,][]{dlm04}, we follow the working
assumption that there will always be residual contamination which will be
modeled during the spectral fitting process.

The filtering light curve is usually extracted in a high-energy band 
(e.g., $2.5-12.0$~keV) and may or may not have had point sources 
excluded.  Only infrequently is there a source in the field 
which is bright, sufficiently hard, and variable enough to 
significantly affect the filtering process.  The light 
curve can be filtered either by setting a fixed absolute
threshold or, more creatively by using the light curve of the specific 
observation to set the threshold.  We use this method in our analysis 
of the clusters presented here (see \S~\ref{sec:example}).  In this method 
a histogram is made of the light curve count rate which typically has 
a roughly Gaussian peak with a high count-rate tail.  A Gaussian is 
then fit to the peak of the distribution and the threshold set at the 
mean value of the Gaussian plus some number of sigma (typically about 
1.5~$\sigma$).  A second threshold is 
set at the mean value minus the same number of sigma to avoid 
biasing the data to lower count rates.  The fitted width of the 
Gaussian can give an indication of residual low level contamination,
although examination of the light curve can often do the same.  The 
benefit of this more complicated screening method is that it works 
well for observations of bright, hard extended objects (e.g., clusters 
of galaxies).

As noted above, even after screening there may well be residual SP 
contamination in the data.  This can be accommodated in the spectral
fitting process by the inclusion in the model of a power law component 
which is not folded through the instrument effective area.  Care needs to 
be taken, however, as power from the source signal can be transferred
to the SP component.

\begin{figure}
\centering{\includegraphics[width=8.5cm]{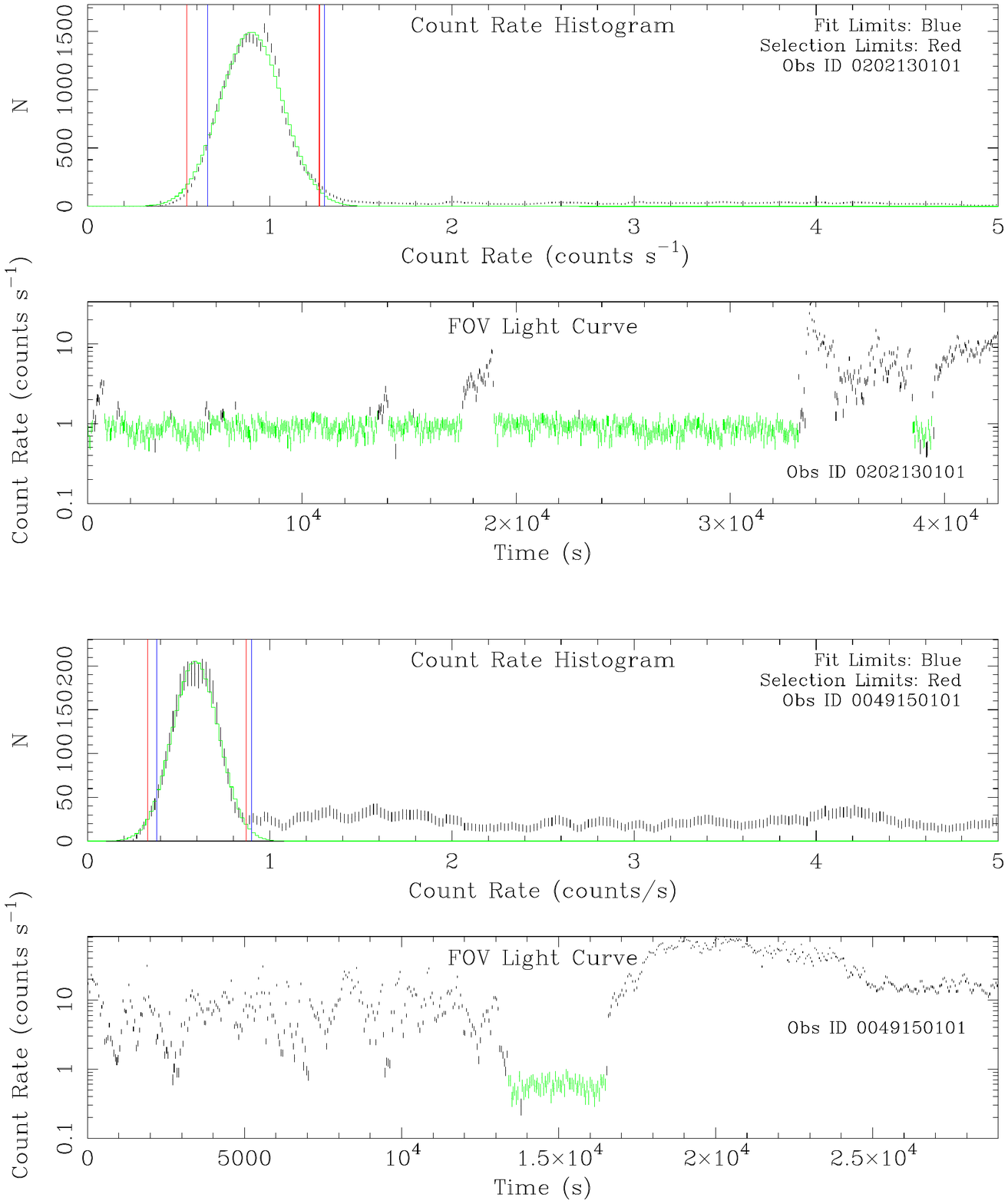}}
\caption{Sample light curves and light-curve histograms from two 
observations with different amounts of SP contamination.  The top 
two panels show the light-curve histogram and light curve for the
data from ObsID 0202130101 while the bottom two panels 
show the same for ObsID 0049150101.}
\label{fig:gen-lc}
\end{figure}

Also note, again, that the screening process is inherently a trade-off 
between the amount of data available for analysis and how clean those 
data are.  Fig.~\ref{fig:gen-lc} shows examples of two observation
light curves along with their light-curve histograms.  As can be seen, 
the extent of the contamination in a given observation is extremely 
variable, as well as the magnitude of that contamination.  Also be 
aware that even though a light curve may look relatively flat, there
is no guarantee that there is no contamination.  Although the longer 
that the observation count rate looks constant, the more likely it is 
that the level of contamination is minimal.  However, the data in 
Fig.~\ref{fig:gen-lc} present a clear example of why caution is 
necessary in considering the possibility of residual SP contamination.  
The two observations are of the same direction on the sky (a density 
enhancement in the Magellanic Bridge with no bright point sources or 
extended emission) and the greater ``nominal'' count rate in the upper 
panel (ObsID 0202130101) is due entirely to a strong residual SP flux.  
In this case a relatively flat light curve is extremely misleading.

\subsection{Solar Wind Charge Exchange Background}

This is an insidious contributer to the 
backgrounds of extended objects, and particularly of observations of 
the diffuse background.  SWCX emission is produced as the solar wind 
flows out from the Sun and interacts with material in the solar system.  
This includes both interstellar neutral material from the Local Cloud 
\citep{lal04} flowing through the solar system and exospheric material at 
Earth's magnetosheath \citep{rc03}.  The highly ionized atoms in 
the solar wind collide with the neutral material and pick up 
electrons in excited states from which they radiatively decay.  In 
the MOS energy band this includes emission from 
\ion{C}{vi}, \ion{O}{vii}, \ion{O}{viii}, \ion{Ne}{ix}, and \ion{Mg}{xi}
some of which are commonly used for plasma temperature, density, and 
ionization equilibrium diagnostics.  

\begin{figure}
\centering{\includegraphics[width=7.5cm,angle=-90]{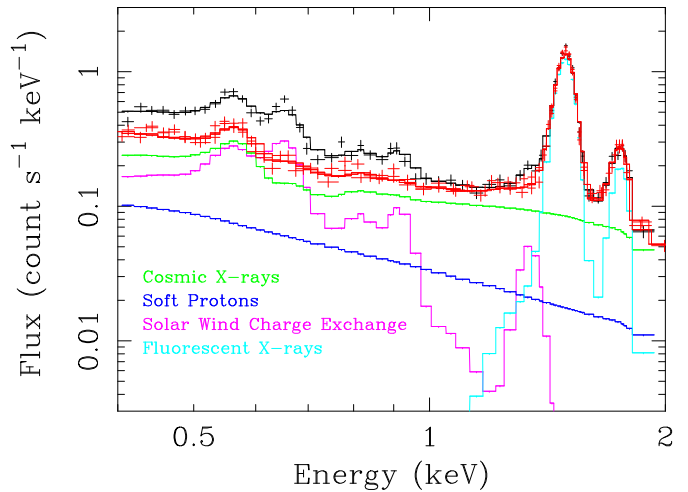}}
\caption{Spectra from two of the four {\it XMM-Newton} EPIC MOS 
observations of the Hubble Deep Field North 
(ObsID 0202130101 in black and  
ObsID 0049150101 in red).  The black data points
and curve show the spectrum from the contaminated observation while
the red data points and curve show an uncontaminated spectrum.  The
uncontaminated spectrum agreed well with the two other observations 
of this direction.  The additional curves show the fitted model 
contributions to the fits where all components besides the SWCX 
emission were fit simultaneously for the two spectra.}
\label{fig:swcx-sp}
\end{figure}

Fig.~\ref{fig:swcx-sp} shows the example 
of SWCX emission from \citet{sck04}, an analysis of four observations 
of the Hubble Deep Field North (HDF-N).  Displayed are two spectra from 
the same direction collected at different times (separated by two weeks).  
Since the cosmic background does not vary with time, the spectra should
be the same except for the possibility of SP contamination which would
be a continuum enhancement rather than the clear emission lines.
The \ion{O}{vii} (0.56 keV) and \ion{O}{viii} (0.65 keV) lines 
are particularly clearly seen as excesses.  For about 40~ks of the 
contaminated observation there 
was no significant indication in the $0.5-0.75$~keV light curve that
there was anything unusual happening.  If there were no other 
observations of the HDF-N and if the contaminated observation lasted 
only for that 40~ks period, there would have been no reason {\it a priori} 
to suspect the data.

Since some fraction of the SWCX emission is due to the interaction 
of the solar wind with the ISM flowing through the solar system,
SWCX emission must, at some level, contaminate all observations.
The contamination depends upon the look direction and the strength
of the solar wind. Usually, the temporal variation in the SWCX
is smaller than the uncertainty in the data, but is occasionally
significantly stronger. In a study of ``empty field'' lines of sight 
having multiple observations, KS07 found significant SWCX 
contamination in 12 of 46 observations. Of the large survey region 
near $\alpha,\delta\sim02^{hr}~25^{min},-03^\circ$, 5 of 26 observations
show significant SWCX contamination. This suggests that 10\% to 25\% of
observations may have significant SWCX contamination.

\subsubsection{Treatment of the SWCX Background}
\label{sec:swcx-treat}

Because the SWCX emission originates externally to the satellite and 
is unlikely to show any angular structure in the {\it XMM-Newton} FOV, 
it is inseparable from the cosmic background.  Depending on the length 
of the observation and the specific SWCX occurrence, the contamination 
may or may not be detectable in the observation light curve.  The 
emission is at energies less than 1.5~keV, primarily in the 
$0.5-1.0$~keV band, so a light curve of that band may show variation
in the diffuse count rate while the light curve in the hard band
($2.0-8.0$~keV) would not.  SWCX contamination may also be detectable 
in the spectrum.  There can be very strong 
\ion{O}{viii} and \ion{Mg}{xi} emission 
unfittable by any normal equilibrium or normal abundance plasma models.  
There are also certain observation geometries which may be more 
susceptible to SWCX contamination than others, specifically any
line of sight which passes near the subsolar point of Earth's 
magnetosheath \citep{rc03}.

\subsection{Cosmic X-ray Background}
\label{sec:cxb}

The CXB is comprised of many components which vary considerably over
the sky.  At high energies ($E>1$~keV) and away from the Galactic
plane the dominant component is the extragalactic power law.  Most 
of this power law represents the superposition of the unresolved 
emission from discrete cosmological objects (i.e., AGN).  There is 
considerable discussion concerning the uniformity of this emission 
over the sky and what the true form of the spectrum is 
\citep[e.g., whether the slope changes for energies less than 
1~keV,][]{tea06}.  The contribution 
of this component to the observed spectrum is clearly going to be 
dependent on the extent to which point sources
have been excluded from the analysis. The emission is also absorbed 
by the column of Galactic material along the line of sight.  

At lower energies there is a greater variety of components, most of
which have thermal emission spectra.  In the solar neighborhood the 
Local Hot Bubble \citep[LHB, ][ and references therein]{sea98} provides 
the dominant contribution near $\frac{1}{4}$ keV.  The LHB is a 
region of hot plasma ($T\sim10^6$~K) at least partially filling an 
irregularly shaped cavity in the neutral material of the Galactic 
disk surrounding the Sun with a radial extent of $\sim30$~pc to 
over 100~pc (preferentially extended out of the plane of the Galaxy).  
In the halo of the Galaxy there is 
additional plasma with $T\sim10^6$~K.  The distribution of this 
plasma is quite patchy and probably has a relatively low scale
height.  There is additional general diffuse emission at $\frac{3}{4}$ keV 
which may be associated with the Galactic halo or perhaps the local 
group \citep{mea02,ksm01}.  Except for the emission from the LHB, 
these components are all absorbed by the column density of the 
Galactic ISM.

\begin{figure}
\centering{\includegraphics[width=8.5cm]{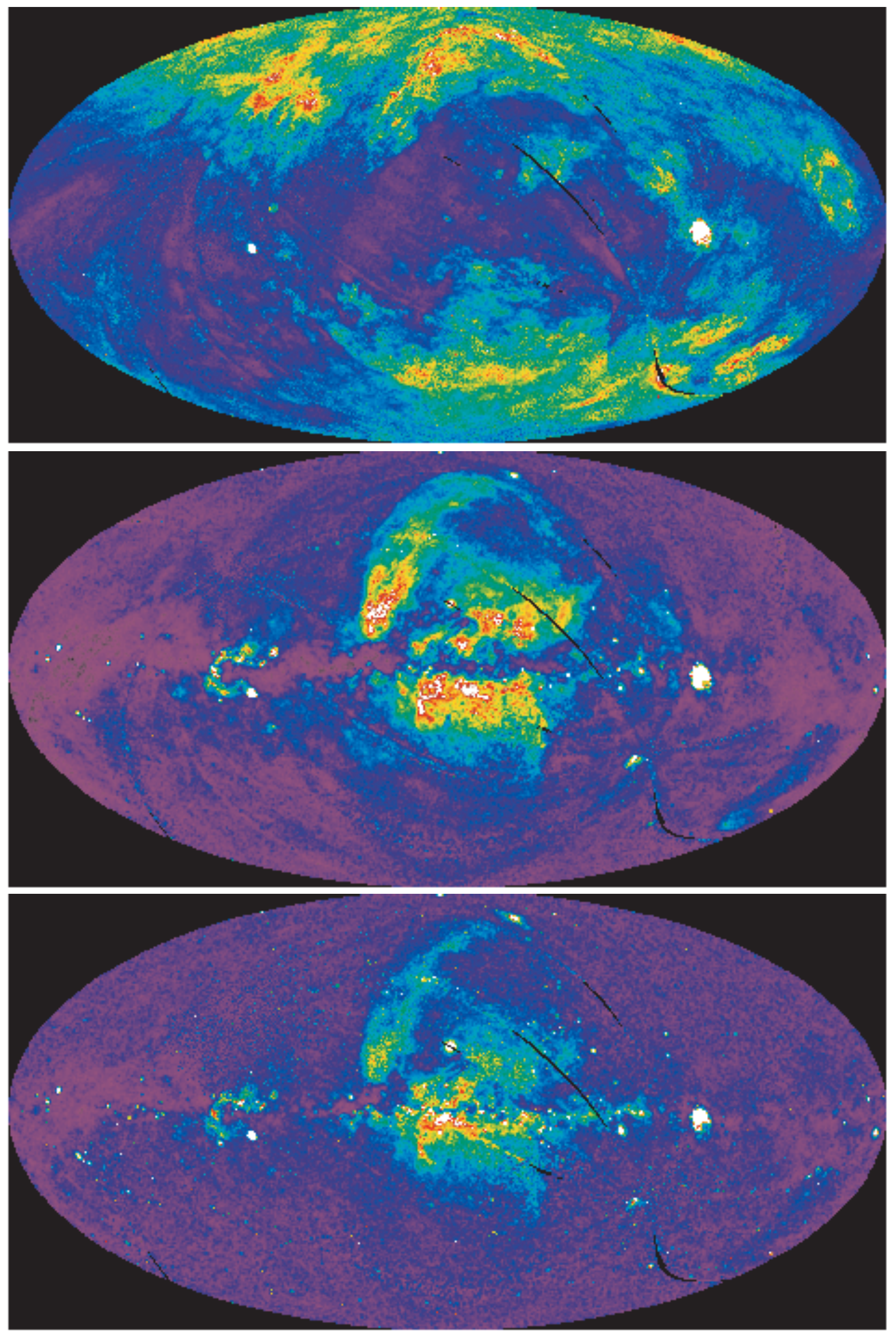}}
\caption{All-sky maps in the $\frac{1}{4}$ keV (upper), 
$\frac{3}{4}$ keV (middle), 
and 1.5~keV (lower) bands from \citet{sea97} in an Aitoff-Hammer 
projection with the Galactic center at the center with longitude
increasing to the left.  Red and white indicate higher intensities 
while purple and blue indicate lower intensities.
\label{fig:rass}}
\end{figure}

Also contributing to the cosmic X-ray background are a wide variety 
of distinct Galactic objects, some of which subtend large angles
on the sky.  Loop~I is a nearby superbubble which has a diameter 
of $\sim100^\circ$, and its emission is combined with the Galactic 
X-ray bulge which extends to $|b|>45^\circ$.  There are supernova 
remnants, the Galactic ridge, and the unresolved emission from stars 
all contributing to the CXB with varying spectra affected by varying
amounts of absorption.  The CXB at $\frac{1}{4}$ keV, 
$\frac{3}{4}$ keV, and 1.5 keV can 
vary by an order of magnitude over the sky, and it can vary 
independently between those bands (although to a lesser extent for 
the $\frac{3}{4}$ keV and 1.5 keV bands).  Fig.~\ref{fig:rass} displays the 
{\it ROSAT} All-Sky Survey (RASS) sky maps in the $\frac{1}{4}$ keV,
$\frac{3}{4}$ keV, and 1.5~keV band from \citet{sea97}.
Comparison of the $\frac{1}{4}$ keV and $\frac{3}{4}$ keV maps 
demonstrates the likely unsuitability of average blank sky data to 
sufficiently characterize the sky in any particular direction.

\subsubsection{Treatment of the Cosmic X-ray Background}
\label{sec:cxb-treat}

The CXB is the dominant background component at energies less than 
1.35~keV, i.e., below the Al~K$\alpha$ and Si~K$\alpha$ FX 
lines.  It is significant in all directions and it can not be modeled 
as a single spectrum independent of position on the sky.  The variation 
in both spectral shape and magnitude makes it very problematic to 
separate from the source of interest when the source covers a large
fraction or all of the instrument FOV.  This is particularly 
troublesome for the study of objects like clusters of galaxies 
where the source emission fades into the background at an uncertain 
rate and radius.  As noted in the introduction, several 
unanswered scientific questions are 
dependent on the true temperature radial profile and mapping that 
profile to the greatest possible radius is critical.

In the absence of an otherwise source-free region within the field 
of view there is no way to directly subtract the CXB from the source 
spectrum.  And, as noted above, the use of blank-field data as a 
spectral template may be inappropriate.  
For this reason, the CXB should be modeled as part of the fitting 
process.  Unfortunately, it is easy to transfer significant power 
between the various background components of a source with low surface 
brightness.  It is therefore desirable to constrain the fits to the 
greatest extent possible.  One method for doing so for the CXB is to 
use spectra from the {\it ROSAT} All-Sky Survey.  A publicly-available 
tool\footnote{http://heasarc.gsfc.nasa.gov/cgi-bin/Tools/xraybg/xraybg.pl}
at the High Energy Science Archive Research Center (HEASARC) 
extracts seven-channel spectra from the data of \citet{sea97} 
for user-defined regions (circular or annuli).  These data can be
simultaneously fit, after proper correction for the observed solid 
angle, with the {\it XMM-Newton} MOS data by a standard model for
the CXB.  For example (and this will be demonstrated in 
\S~\ref{sec:example} below for \object{Abell 1795}) a CXB RASS spectrum can 
be extracted for an annulus surrounding the cluster, but not 
including it.  With the assumption that the annulus spectrum is a
good representation of the CXB in the direction of the cluster, 
a model including 1) an unabsorbed $\sim0.1$~keV thermal spectrum 
representing the LHB, 2) an absorbed $\sim0.1$~keV thermal spectrum 
representing the cooler Galactic halo emission, 3) an absorbed 
$\sim0.25$~keV thermal spectrum representing the hotter halo emission 
(and/or emission from the local group), and 4) an $E^{-1.46}$ power law 
representing the unresolved cosmological emission \citep[e.g.,][]{ks00} 
can be fit to the RASS and MOS data, 
with additional components representing the 
cluster, SP, and FX components fit only to the MOS data.

\section{Abell 1795 -- A Case Study}
\label{sec:example}

\object{Abell 1795} is a well-studied nearby cluster of galaxies.  It was 
chosen for the example presented here as it was used by \citet{nml05}
for their discussion of the analysis of {\it XMM-Newton} observations 
of extended objects.  The observation (ObsID 0202130101) 
was taken on 2000 June 26 with an exposure of $\sim49.6$~ks.  The 
pointing direction was $\alpha,\delta=207.2208^\circ,26.5922^\circ$.

The preparation of the data for analysis presented below uses the
XMM-ESAS\footnote{http://heasarc.gsfc.nasa.gov/docs/xmm/xmmhp\_xmmesas.html} 
package of perl scripts and FORTRAN programs, which also require The
XMM-Newton Standard Analysis Software 
(SAS\footnote{http://xmm.esac.esa.int/external/xmm\_sw\_cal/sas\_frame.shtml})
package.  XMM-ESAS was prepared by the NASA/Goddard Space Flight Center 
XMM-Newton Guest Observer Facility (GOF) in conjunction with the ESA 
Science Operations Center (SOC) and the Background Working Group.
The software is publicly available through both the GOF and SOC and
is provided with documentation.

\subsection{Temporal Filtering}
\label{sec:ex-temporal}

\begin{figure}
\centering{\includegraphics[width=7.5cm,angle=-90]{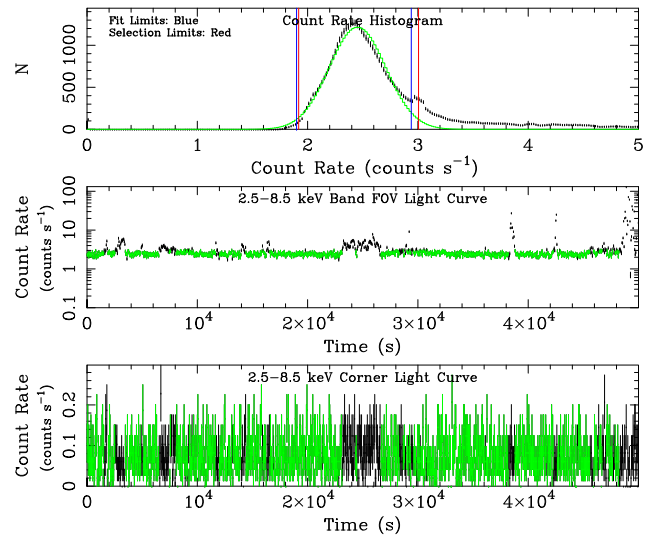}}
\caption{Temporal filtering results for the MOS1 \object{Abell 1795} cluster 
observation with ObsID 0097820101.  The upper panel plots the light 
curve histogram for the $2.5-12.0$~keV band from the FOV, the middle 
panel displays the $2.5-12.0$~keV band FOV light curve, and the lower 
panel displays $2.5-12.0$~keV band light curve from the unexposed corners 
of the instrument.  The histogram is derived from the smoothed light
curve.  In the upper panel, the blue vertical lines show the range for
the Gaussian fit, the green curve shows the Gaussian fit, while the 
red vertical lines show the upper and lower bounds for filtering 
the data.  In the bottom two panels green points indicate accepted data
while black points indicate data excluded by the filtering algorithm.
The high count rate excursions are produced by soft protons rather than 
a particle background flare as the latter case would produce a similar 
increase in the corner data.
\label{fig:A1795-lc}}
\end{figure}

The \object{Abell 1795} observation was relatively clean by visual observation
of its light curve with just a few excursions to high count rates 
from SP contamination.  Fig.~\ref{fig:A1795-lc} shows the results from 
the temporal filtering algorithm.  Filtering the data reduced the 
exposure to 36.5~ks, roughly 75\% of the original observation.  However, 
the slight ripple in the light curve indicates that there is likely to
be some residual SP contamination.  

In the screening process a light curve with a 1~s binning in the 
$2.5-8.5$~keV band was first created from the photon event file (PEF). 
This light curve, binned by 50~s, is shown in the the middle panel 
of Fig.~\ref{fig:A1795-lc}.  The light curve is smoothed with a 
50~s running average and a histogram created from the smoothed data 
(upper panel).  The presence of the SP contamination is shown by the 
high count-rate tail of the of the otherwise relatively Gaussian 
distribution.  That the flaring in the light curve is not caused by 
an increase in the high-energy particle background is shown by the 
corner count rate (lower panel) not having similar enhancements.
The histogram is searched for the maximum and a Gaussian is fit to 
the data surrounding the peak.  A count-rate cut of the light curve
is made by setting thresholds at $\pm1.5~\sigma$ on either side of 
the fitted peak channel.  Note that the setting of these thresholds
is somewhat arbitrary, and that there is no absolute answer.  With 
cleaner data wider limits can be set, but there is always a trade-off 
between the amount of accepted data and how clean those data are.

\subsection{Extraction of Spectra}
\label{sec:ex-spectra}

After the data were screened spectra were extracted and model 
background spectra created.  For this analysis of \object{Abell 1795}
the goal is the determination of the temperature radial profile, 
thus the extracted spectra were from concentric annuli.  

Extraction selection expressions consistent with the requirements
for the SAS task {\it evselect} were required for the annuli.  These
were most easily created using SAS and the {\it xmmselect} task 
and its interface with the 
{\it ds9}\footnote{http://hea-www.harvard.edu/RD/ds9/} 
\citep{jm03} image 
display software.  
From {\it xmmselect} an image was created in detector coordinates 
({\it DETX} and {\it DETY}).  The detector coordinates of the center 
of \object{Abell 1795} were determined from the image, and then the desired 
region descriptions defined.  As an example of the region selection
descriptors, \\
((DETX,DETY) IN circle(201,-219,2400)) \\ 
\ \hskip 0.1cm \&\&!((DETX,DETY) IN circle(201,-219,1200)) \\
selects data from the MOS1 detector from the $1'-2'$ annulus.
The numbers 201 and -219 are detector coordinates ({\it DETX} 
and {\it DETY}) of the cluster center while the numbers 1200 and 
2400 are the inner and outer radii of the annulus, all in units 
of 0.05 arc seconds.  The annulus is created by selecting all 
data within the first circle but not within the second circle 
(the ``\&\&'' symbol is used for the Boolean ``and'' and the ``!''
symbol is used for the Boolean ``not'').  Note that the DETX and 
DETY positions for a given sky position in the MOS1 and MOS2 
detectors will be different. 

\begin{figure}
\centering{\includegraphics[width=8.5cm]{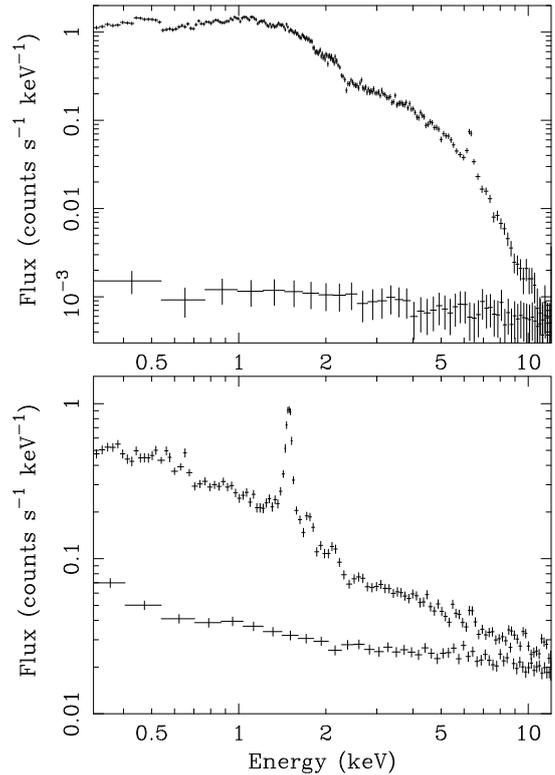}}
\caption{Spectra from two annuli from the \object{Abell 1795} analysis, 
$1'-2'$ (upper panel) and $10'-15'$ (lower panel).  In each panel 
the upper spectrum is the total spectrum while the lower spectrum 
is the modeled QPB spectrum.  The data have not been normalized
for solid angle, otherwise the $1'-2'$ spectrum would be relatively
brighter by about two orders of magnitude.
\label{fig:A1795-spec}}
\end{figure}

\subsection{Modeling the Particle Background}
\label{sec:ex-QPB}

The model particle spectra were created using the XMM-ESAS package 
which follows the process as outlined in \S~\ref{sec:qpb} above.
Fig.~\ref{fig:A1795-spec} displays total and model QPB spectra from 
an inner and an outer annulus of the \object{Abell 1795} analysis.  
As expected the fainter outer annulus is much more strongly affected 
by the various background components, in particular the FX 
contamination is clearly represented by the Al-K$\alpha$ line and the 
residual SP contamination which is responsible for the difference 
between the spectra above $E\sim8$~keV.

\subsection{Modeling the Cosmic Background}
\label{sec:ex-CXB}

Modeling and constraining the CXB was a two-part process.  First, 
the RASS spectrum of the CXB in the direction of interest was 
obtained from the HEASARC ``X-ray Background Tool'' 
(see \S~\ref{sec:cxb-treat} above).  Since the object of interest in
this analysis is a discrete object and not the CXB itself, an 
annular extraction region was used where the inner annulus 
radius was large enough to exclude cluster emission.  The outer 
annulus radius was limited so that the spectrum could be as  
appropriate as possible for the cluster region (and in addition 
so that the {\it ROSAT}-spectrum statistics would not 
dominate the spectral fitting 
process).  For this analysis of \object{Abell 1795}, inner and 
outer radii of $1^\circ$ and $2^\circ$, respectively, were used.

\subsection{The Fitted Spectral Model}
\label{sec:ex-model}

The model for this example (below and Table~\ref{tbl:modeldefs}) 
is rather extensive as it represents most of the emission components 
along the line of sight to and including the \object{Abell 1795} 
cluster as well some local background components.  To complicate 
the process even further, the fitted 
parameters for some of the components will differ between the
different annuli.  

$S = P_1 + G_1 + G_2 + {C_1}\times{C_2}\times(A_1 + (A_2 + A_3 + \\
P_2)\times{e^{-\sigma{N_{Hg}}}} + AC\times{e^{-\sigma{N_{Hc}}}})$

\begin{table}
\begin{minipage}[t]{\columnwidth}
\caption{Spectral Model Definitions}
\label{tbl:modeldefs}
\centering
\renewcommand{\footnoterule}{}
\begin{tabular}{cl}
\hline\hline
Parameter & Definition \\
\hline
$P_1$ & Power law representing the residual \\
      & SP contamination. \\
$G_1$, $G_2$ & Gaussians representing the Al~K$\alpha$ \\
      & and Si~K$\alpha$ FX lines. \\
$C_1$, $C_2$ & Constant representing the different \\
      & solid angles of the extraction \\
      & annuli and calibration offsets between \\
      & the two detectors. \\
$A_1$ & CXB LHB thermal component. \\
$A_2$ & CXB cooler halo thermal component. \\ 
$A_3$ & CXB hotter halo thermal component. \\ 
$P_2$ & CXB extragalactic power law  \\
      & component. \\
$N_{Hg}$ & Column density of Galactic hydrogen. \\
$AC$  & Cluster thermal component. \\
$N_{Hc}$ & Column density foreground to the \\ 
         & cluster, includes both Galactic \\ 
         & hydrogen and material associated \\
         & with the cluster. \\
\hline
\end{tabular}
\end{minipage}
\end{table}

\begin{figure*}[t!]
\centering{\includegraphics[angle=0,width=17.0cm]{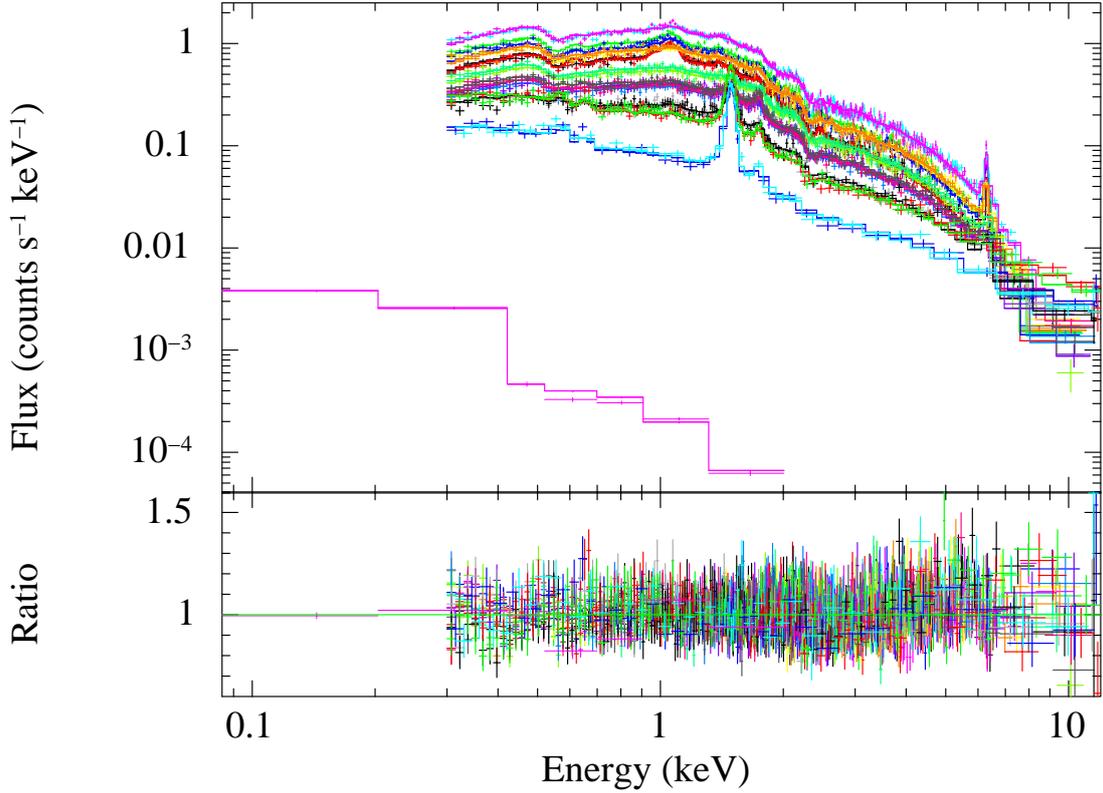}}
\caption{Spectral fit to the data from \object{Abell 1795}.
MOS1 and MOS2 spectra are shown for all ten annuli,
as well as the {\it ROSAT} spectral energy distribution.  The lower
panel shows the ratio of the data to the model and demonstrates that 
the fit is reasonably good over the full dynamic range.
\label{fig:A1759-spec}}
\end{figure*}

The equation above for the fitted spectrum includes a fairly  
complete model for the non-cluster component of the observed 
spectrum in the cluster analysis.  $P_1$ is a power law representing 
the residual SP contamination.  This is not folded through the 
instrumental effective areas.  $G_1$ and $G_2$ are Gaussians 
which represent the Al~K$\alpha$ and Si~K$\alpha$ FX lines.
$C_1$ and $C_2$ are constant scale factors which represent the 
different solid angles of the extraction annuli and any relative
calibration offsets between the two detectors.  For consistency 
with the RASS data, the $C_1$ parameter is set 
to the solid angle in units of square arc minutes (in practice,
this is the spectrum BACKSCAL keyword value produced by SAS divided 
by 1440000).  The cosmic background is represented by the three 
thermal components $A_1$ for the LHB, $A_2$ for the cooler halo 
component, $A_3$ for the hotter halo component, and the extragalactic 
power law, $P_2$.  $N_{Hg}$ is the column density of Galactic neutral 
hydrogen.  To model the cluster emission we use a simple absorbed 
thermal model where the abundance (a single overall scaling) and  
absorption are allowed to vary.  The spectral fitting is done using 
the Xspec package with Astrophysical Plasma Emission Code 
(APEC\footnote{http://cxc.harvard.edu/atomdb/sources\_apec.html}) 
thermal models and the \citet{mm83} absorption model (Wisconsin 
Absorption, WABS).

\begin{table*}
\begin{minipage}[t]{\textwidth}
\caption{Spectral fit parameters}
\label{tbl:params}
\centering
\renewcommand{\footnoterule}{}
\begin{tabular} {cccccc}
\hline\hline
Spectral & Model & Parameter & Initial & Initial & Final \\
Component & Component & & Guess & Constraint & Constraint \\
\hline
SP    & $P_1$    & $\gamma_1$      & 0.9                 & Fix   & Free \\
$-$   & $-$      & Normalization   & $10^{-5}$           & Free  & Free \\
FX    & $G_1$    & Energy          & 1.49 keV            & Fix   & Free \\
$-$   & $-$      & Width           & 0.0 keV             & Fix   & Free \\
$-$   & $-$      & Normalization   & $10^{-5}$           & Free  & Free \\
FX    & $G_2$    & Energy          & 1.75 keV            & Fix   & Free \\
$-$   & $-$      & Width           & 0.0 keV             & Fix   & Free \\
$-$   & $-$      & Normalization   & $10^{-5}$           & Free  & Free \\
Scale & $C_1$    & Solid Angle     & Set                 & Fix   & Fix \\
Scale & $C_2$    & Scale Factor    & 1.0                 & Fix   & Fix/Free\footnote{The MOS1 and RASS
scale factors were fixed at 1.0 and the MOS2 scale factor was allowed to vary.} \\
CXB\footnote{The abundances and redshifts of the cosmic
thermal components are fixed at 1.0 and 0.0, respectively.}   
& $A_1$    & $kT$            & 0.1                 & Fix   & Free \\
$-$   & $-$      & Normalization   & $5.0\times10^{-6}$  & Free  & Free \\
$-$   & $A_2$    & $kT$            & 0.1                 & Fix   & Free \\
$-$   & $-$      & Normalization   & $5.0\times10^{-6}$  & Free  & Free \\
$-$   & $A_3$    & $kT$            & 0.25                & Fix   & Free \\
$-$   & $-$      & Normalization   & $10^{-6}$           & Free  & Free \\
$-$   & $P$      & $\gamma$        & 1.46                & Fix   & Fix \\
$-$   & $-$      & Normalization   & $8.88\times10^{-7}$ & Fix   & Fix/Free\footnote{Whether 
the extragalactic power law normalization is fixed or allowed to vary must
be carefully examined.} \\
$-$   & $N_{Hg}$ & Galactic Column & $1.2\times10^{20}$  & Fix   & Free \\
A1795 & $A_4$    & $kT$            & 5.0 keV             & Free  & Free \\
$-$   & $-$      & Abundance       & 0.5                 & Free  & Free \\
$-$   & $-$      & Redshift        & 0.06                & Free  & Free \\
$-$   & $-$      & Normalization   & $5.0\times10^{-4}$  & Free  & Free \\
$-$   & $N_{Hc}$ & Cluster Column  & $1.2\times10^{20}$  & Fix   & Free \\
\hline
\end{tabular}
\end{minipage}
\end{table*}

\subsection{The Data}

For this analysis we extracted data from 10 annuli for the 
cluster.  These are the same annuli which are used for the rest 
of the clusters in this catalog.  The size of the annuli were 
chosen to be 
reasonable, where reasonableness in this, and most cases, is not 
unique.  The dominant constraint is that the number of events in 
a specified annulus must be sufficient for a significant spectral 
fit. 

\subsection{The Fit}

The setting up of the spectral fit was a time-consuming process.
For the number of spectra (20 MOS1 and MOS2 spectra and 1 RASS
spectrum) and the complex model used for the fit, there are 546
parameters.  Clearly if all 546 parameters are fit independently 
convergence of the fit would take place only on geologic time 
scales.  However, many of the parameters can be either linked or
frozen to known values, some of which may be later allowed to 
vary once the fit is relatively accurate.  (It is occasionally 
easy for the fitting engine to get ``stuck'' 
in a local minima.) The cosmic background
is the same for all spectra and so the parameters can be 
linked (the redshifts and abundances of the thermal components 
were frozen to 0.0 and 1.0, respectively).  The solid angle 
scale factors were frozen to their appropriate values and 
the instrument scale factors were linked.  The normalizations 
for the SP contributions were linked using the model 
distribution available in the XMM-ESAS package and the power
law index was also linked.  For the cluster contribution to the 
spectra, the redshift can be linked.  Table~\ref{tbl:params} 
lists suggested initial parameters and whether they should be frozen 
(fix) or allowed to vary (free).  In practice, the abundances 
for many of the outer annuli were effectively unconstrained.  In
such cases the abundance of the outer most annulus with a free 
abundance was linked to that of the next inner annulus and the 
data refit.  This process was repeated until a S/N of $\sim3$ 
was achieved. In addition, abundances which went to unphysical 
values, e.g., zero, were also linked to that of the next inner 
annulus. 

There are further complications to the fitting process.  First, 
because of the finite PSF of the EPIC instruments, some X-rays 
which originate in one annulus on the sky are detected in a 
different annulus.  In cases where there are strong spectral
gradients, e.g., for clusters with a strong cooling flow, this
can significantly affect the results with the inner annulus 
having a higher fitted temperature and the neighboring annuli 
having cooler fitted temperatures than their true values.  
The fitted value for the flux is also likely to be 
different than the true value.  The arfgen task of SAS now has the 
capability (using the crossregionarf parameter) of calculating the 
``cross-talk'' effective area file (ancillary region file, 
ARF) for X-rays originating in one region but which are detected 
in another.  The cross-talk contribution to the spectrum of a 
given annulus from a second annulus is treated in Xspec V12 as an 
additional model component.  The spectrum from the second annulus 
is folded through the cross-talk ARF linking the two annuli and 
then the redistribution matrix (RMF) of the first annulus.  
Note that the ARF for the contributions of X-rays originating in 
one region of the sky to a second region on the detector is 
typically not the same as the ARF for the contribution of X-rays 
originating in the second region on the sky to the first region 
on the detector. Second, the use of Xspec V12 requires that the 
SP power law be included as a separate model with a separate response 
matrix.  This response matrix is diagonal with unity elements.
For the cluster analysis presented here we fitted the {\it XMM-Newton} 
spectra over the $0.3-12.0$~keV energy range where statistics 
permitted.  Quite often the range was limited to $0.3-10.0$~keV.

\subsection{Abell 1795 Results}

The final fit for the \object{Abell 1795} data is relatively good with a
$\chi^2$ value of 1.25 for 3958 degrees of freedom.  The data,
model fits, and residuals are shown in Fig.~\ref{fig:A1759-spec}.
However, the distribution of the residuals does show some systematic
variation with energy, most noticeably at energies above 2~keV. 
The variation is rather limited in extent and could be due to the 
simplicity of the model for the cluster emission,  
residual calibration errors, or errors in the model background 
(both QPB and SP).  The latter is less likely as all annuli show 
the systematic, including the inner ones which are not 
significantly affected by backgrounds.

\begin{figure}
\centering{\includegraphics[angle=0,width=8.5cm]{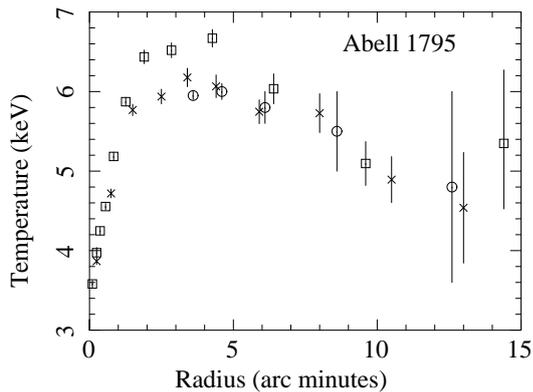}}
\caption{Comparison of results for the A1795 temperature radial 
profile from {\it Chandra} \citep[square,][]{vea05}, and 
{\it XMM-Newton} analysis from \citet{nml05} (circle) and this 
analysis (cross).  The radii for the {\it XMM-Newton} points have been
slightly offset in the plot for clarity.
\label{fig:A1759-compare}}
\end{figure}

\begin{figure}
\centering{\includegraphics[angle=0,width=8.5cm]{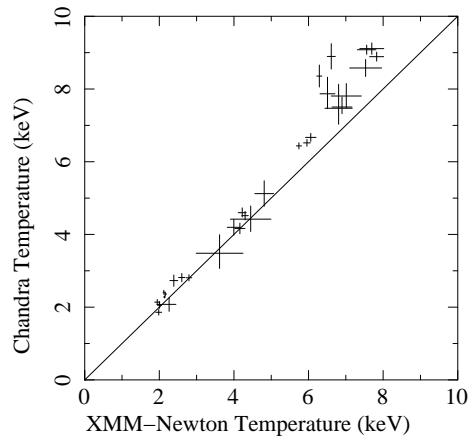}}
\caption{Comparison of results for the temperature radial 
profiles for various clusters in their $\sim1'-5'$ annuli from 
{\it Chandra} \citep{vea05} and {\it XMM-Newton} (this analysis).  
\label{fig:grand-compare}}
\end{figure}

Fig.~\ref{fig:A1759-compare} shows the comparison between the 
{\it Chandra} \citep{vea05}, {\it XMM-Newton} \citep{nml05}, and 
current analysis of \object{Abell 1795}.  As expected, there is
reasonable agreement between the {\it XMM-Newton} results.  
However, the {\it Chandra} results are very significantly 
different from those of {\it XMM-Newton} at intermediate radii.  
This discrepancy is
consistent for the higher temperature clusters which have been 
compared.  The sense of the difference is that the higher the 
fitted temperature the more likely it is that {\it Chandra} will 
find a higher temperature than {\it XMM-Newton} with the effect 
typically becoming significant above $kT\sim5-6$~keV.  
Fig.~\ref{fig:grand-compare} displays this difference in the fitted
temperatures for clusters in their $\sim1'-5'$ annuli \citep[{\it Chandra} 
data from][]{vea05}.  These annuli are used for comparison 
purposes since their signal to noise ratio are high, the effects of 
background subtraction is minimal, and the PSF issues are minor.
This discrepancy between {\it Chandra} and {\it XMM-Newton} 
can lead to significant differences in the fitted temperature 
profiles causing the {\it Chandra} observations to have greater 
fall-offs in temperature at higher radii.

\begin{figure}
\centering{\includegraphics[width=8.5cm]{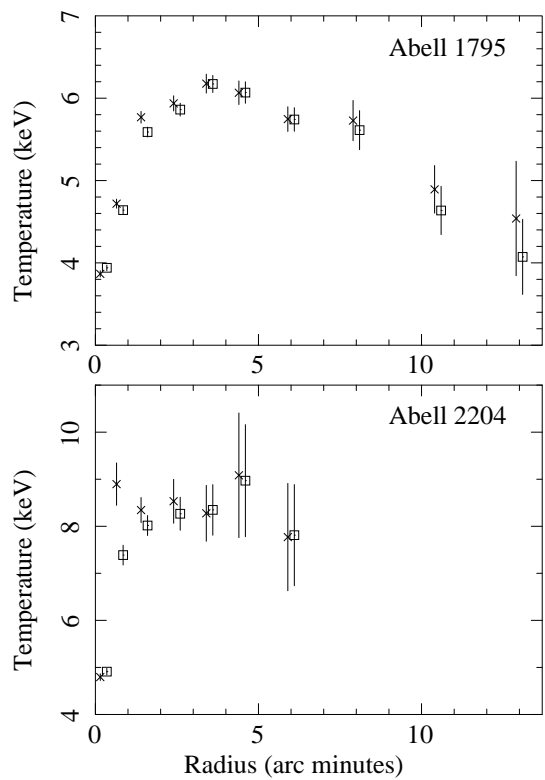}}
\caption{Comparison of results for the Abell 1795 (top panel) and 
Abell~2204 (bottom panel) temperature radial profiles from analysis 
including (cross) and not including (square) the effect of PSF 
smearing (crosstalk between adjacent annuli).  The radii have been
slightly offset in the plot for clarity.
\label{fig:cross}}
\end{figure}

One suggested explanation for the discrepancy was the effect of the 
finite PSF of {\it XMM-Newton} and the spreading of the cooler
X-rays from the cluster core to the inner annuli.  Indeed, this is 
what led to the development of the arfgen modification to account
for the cross-talk.  While the correction effect does go in the 
right direction (Fig.~\ref{fig:cross} top panel), for 
\object{Abell 1795} it is barely significant and not nearly sufficient to 
account for the difference.  Also, use of the {\it Chandra} image 
with its finer PSF for the calculation of the cross-talk contribution 
has no significant effect.  However, the effect can be significant 
in cases where the flux and temperature gradients are steeper (on an 
angular scale) and greater in magnitude.  Fig.~\ref{fig:cross} 
(bottom panel) shows a similar comparison for the cluster \object{Abell 2204}.
In this case the fitted temperature of the second annulus increases
by $\sim1.5$~keV when the correction for PSF smearing is applied.

\begin{figure}
\centering{\includegraphics[angle=0,width=8.5cm]{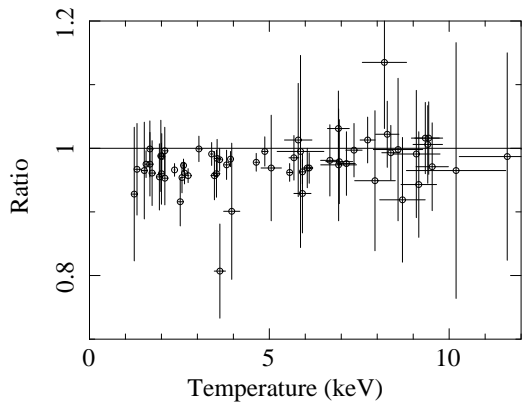}}
\caption{Ratio of the fitted temperatures for a selection of 
clusters analyzed using SAS~V7.1 and the calibration files of 2007 
September versus the calibration used for the cluster catalog.  
The horizontal line is set at a ratio of 1.0.
\label{fig:temp-area-rat}}
\end{figure}

In an effort to improve the cross-calibration between the MOS,
pn, and RGS detectors, new quantum efficiencies were released in 2007 
August\footnote{http://xmm.esac.esa.int/docs/documents/CAL-SRN-0235-1-0.ps.gz}.
The revisions decrease the effective area of the response at lower 
energies by increasing the absorption depth at the C, N, and O edges.
In order to gauge the significance of the change on the results reported 
in the cluster catalog, we reprocessed seven clusters with a range
of temperatures with SAS~V7.1 and the calibration files of 2007 September 14.  
Fig.~\ref{fig:temp-area-rat} shows the ratio 
of the reprocessed versus the cluster catalog temperatures.  There 
is a tendency for the reprocessed temperatures to be slightly lower 
although only at the $\sim1~\sigma$ level.  The average ratio is 
$\sim0.97$, or $\sim0.2$~keV at 6~keV.

\section{The Cluster Catalog}
\label{sec:catalog}

We applied the method described above for the \object{Abell 1795} 
data to 70 clusters of galaxies from the {\it XMM-Newton} archive
in a consistent manner.
The selection of the clusters was empirical; postage-stamp count 
images from the {\it ROSAT} All-Sky Survey were examined 
for each of the {\it XMM-Newton} cluster observations in the archive.  
Those which appeared to have (subjectively) reasonable extent and 
reasonable brightness were chosen for processing. A total of just
over 100 clusters were selected.

The initial step of the processing was to filter the data to 
exclude periods of SP flaring and to create count images.  Clusters 
where the accepted exposure time was less than $\sim8$~ks as well as 
clusters with a surface brightness insufficient to produce reasonable 
statistics for the cluster emission were excluded 
from further analysis.  The selection against overly contaminated 
observations excluded $\sim30$ clusters.  For those observations 
with filtered times acceptable for processing, roughly 25\% of
the original processing time was lost to flaring.  (This loss
does not include the useless exposures of observations with 
multiple exposures.)  A few other clusters were excluded from
the processing because of their extreme asymmetry or the presence
of strong substructures obviating the circular assumption.

\begin{figure*}
\centering\includegraphics[angle=-0,width=6.5in]{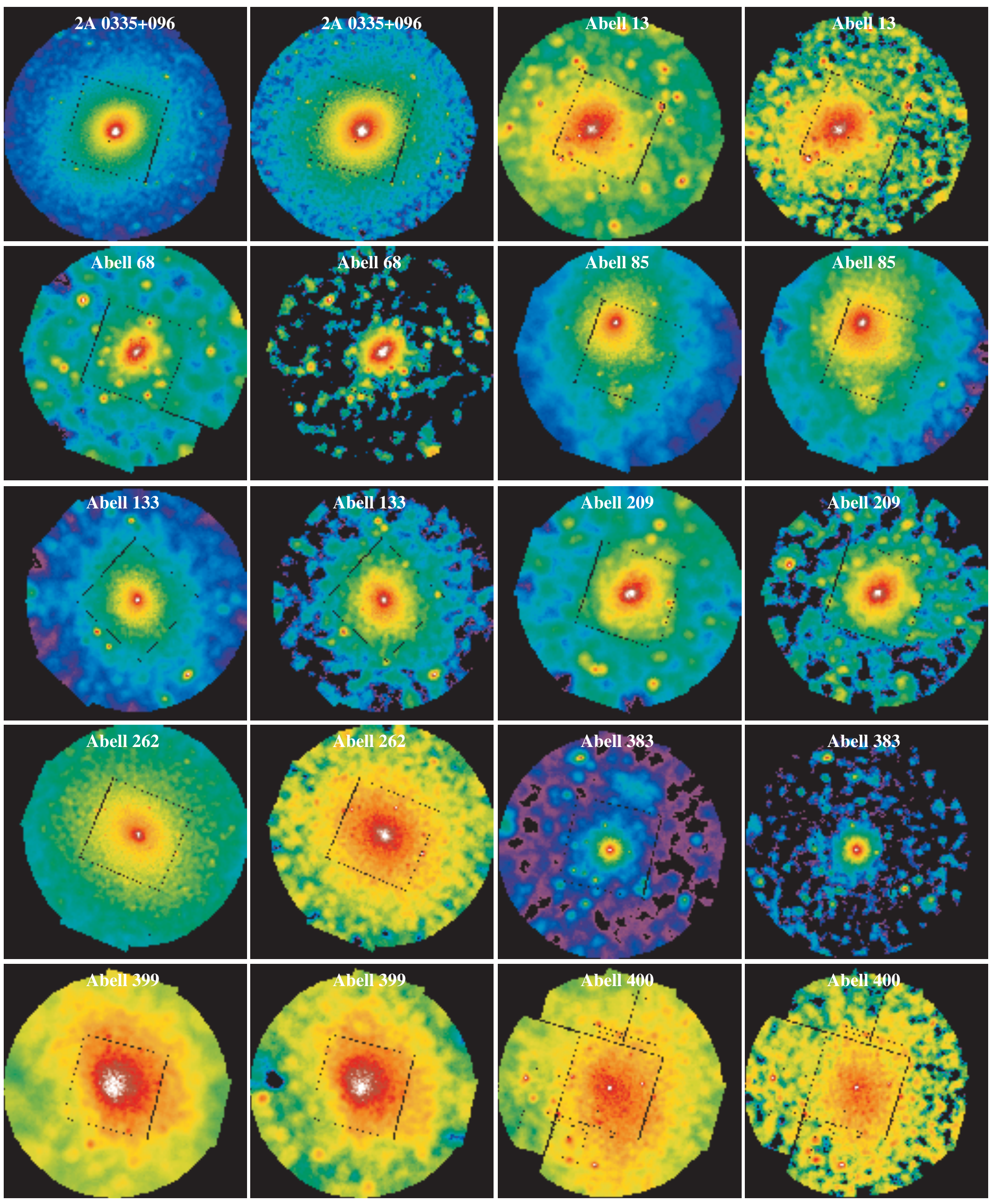}
\caption{Soft (left) and hard (right) band images of the clusters. 
\label{fig:im-01}}
\end{figure*}

\begin{figure*}
\centering\includegraphics[angle=-0,width=6.5in]{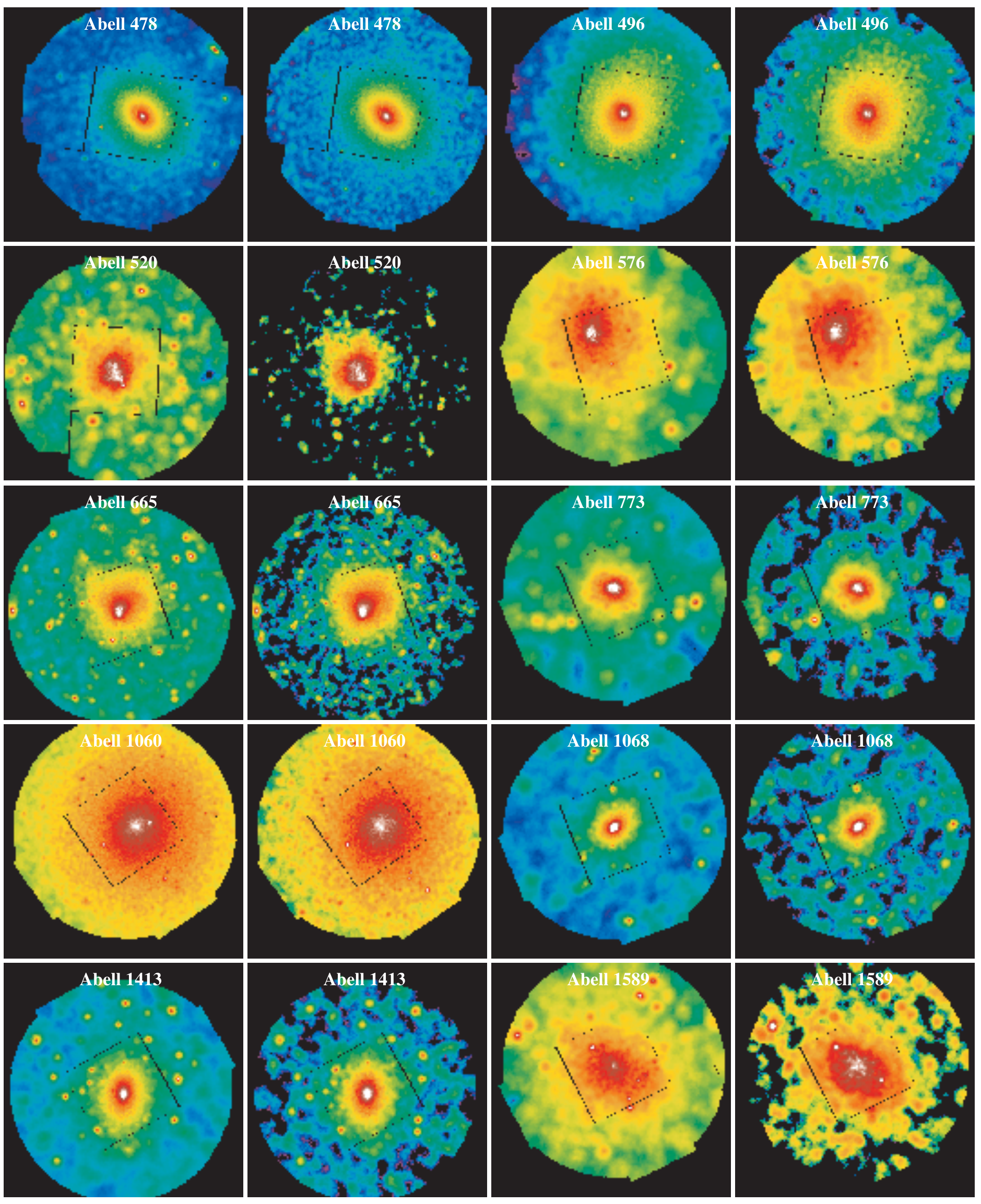}
\caption{Soft (left) and hard (right) band images of the clusters.
\label{fig:im-02}}
\end{figure*}

\begin{figure*}
\centering\includegraphics[angle=-0,width=6.5in]{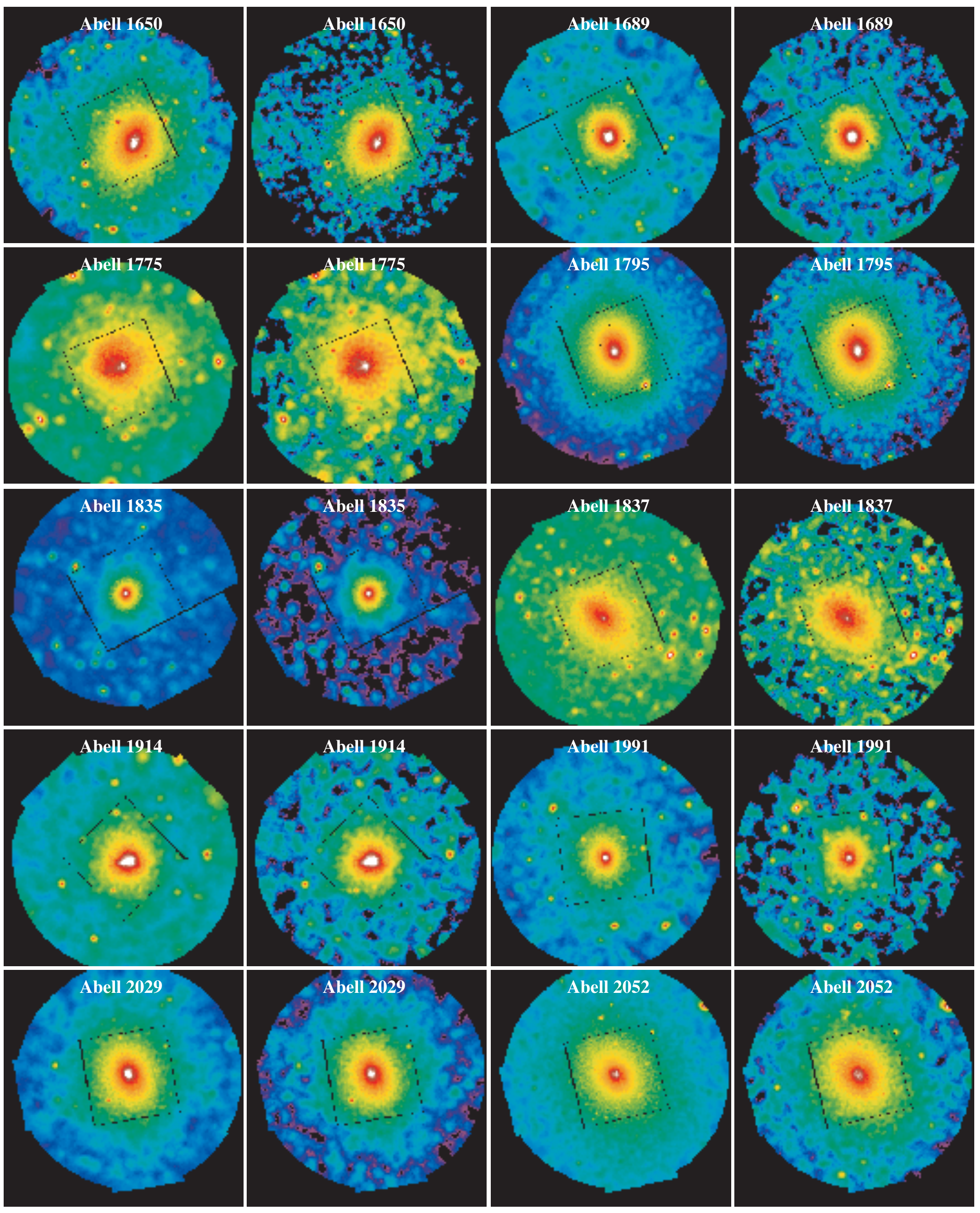}
\caption{Soft (left) and hard (right) band images of the clusters.
\label{fig:im-03}}
\end{figure*}

\begin{figure*}
\centering\includegraphics[angle=-0,width=6.5in]{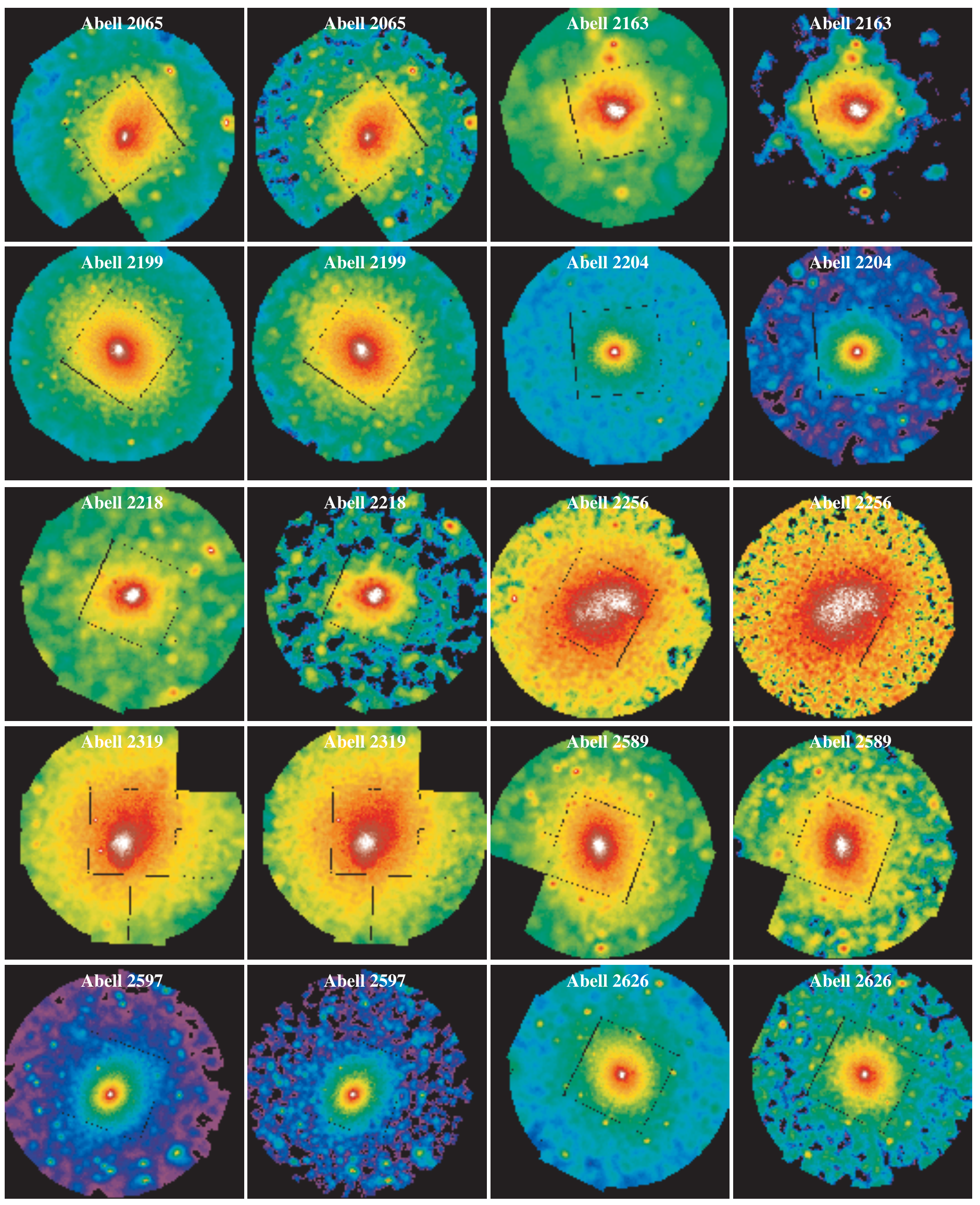}
\caption{Soft (left) and hard (right) band images of the clusters.
\label{fig:im-04}}
\end{figure*}

\begin{figure*}
\centering\includegraphics[angle=-0,width=6.5in]{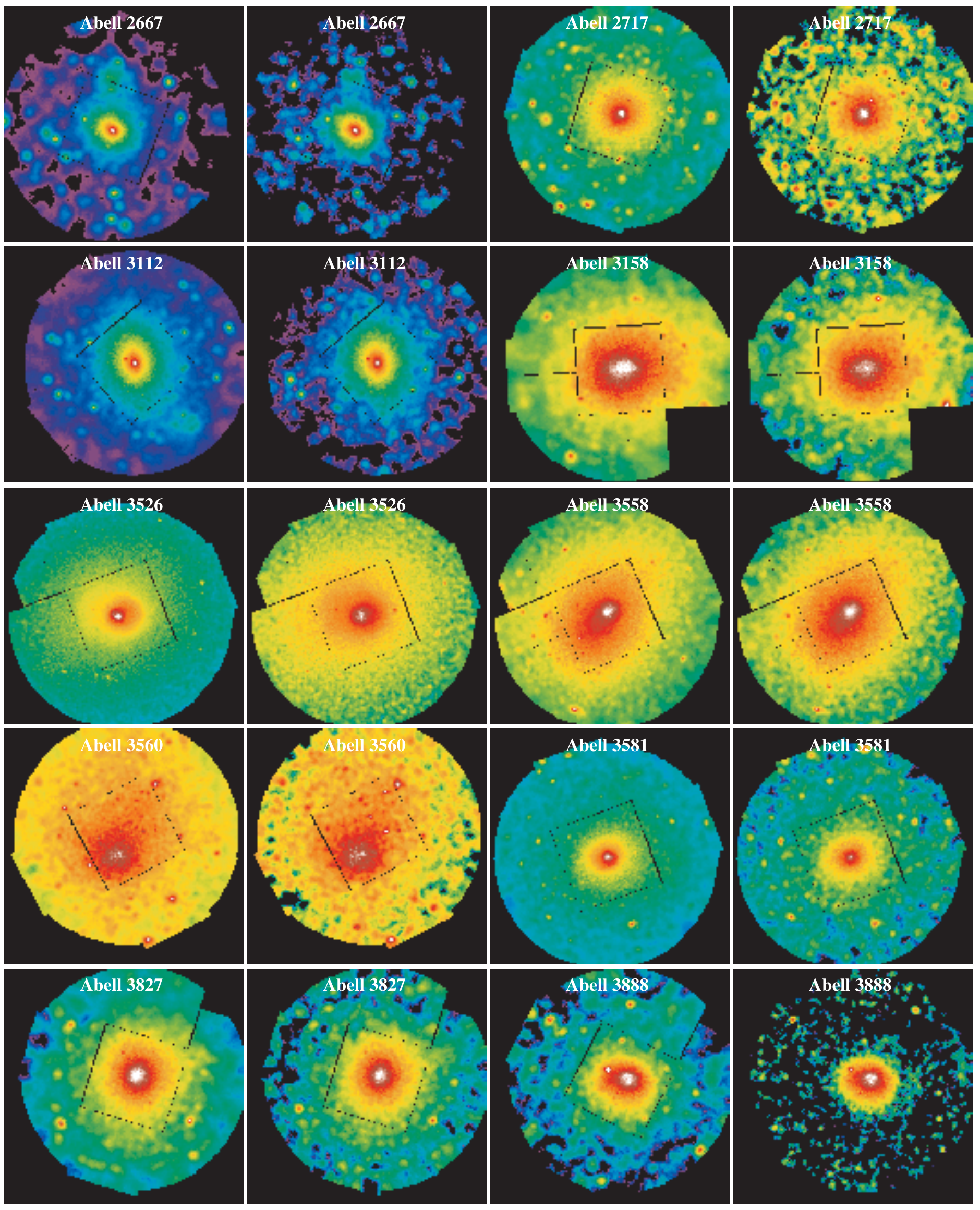}
\caption{Soft (left) and hard (right) band images of the clusters.
\label{fig:im-05}}
\end{figure*}

\begin{figure*}
\centering\includegraphics[angle=-0,width=6.5in]{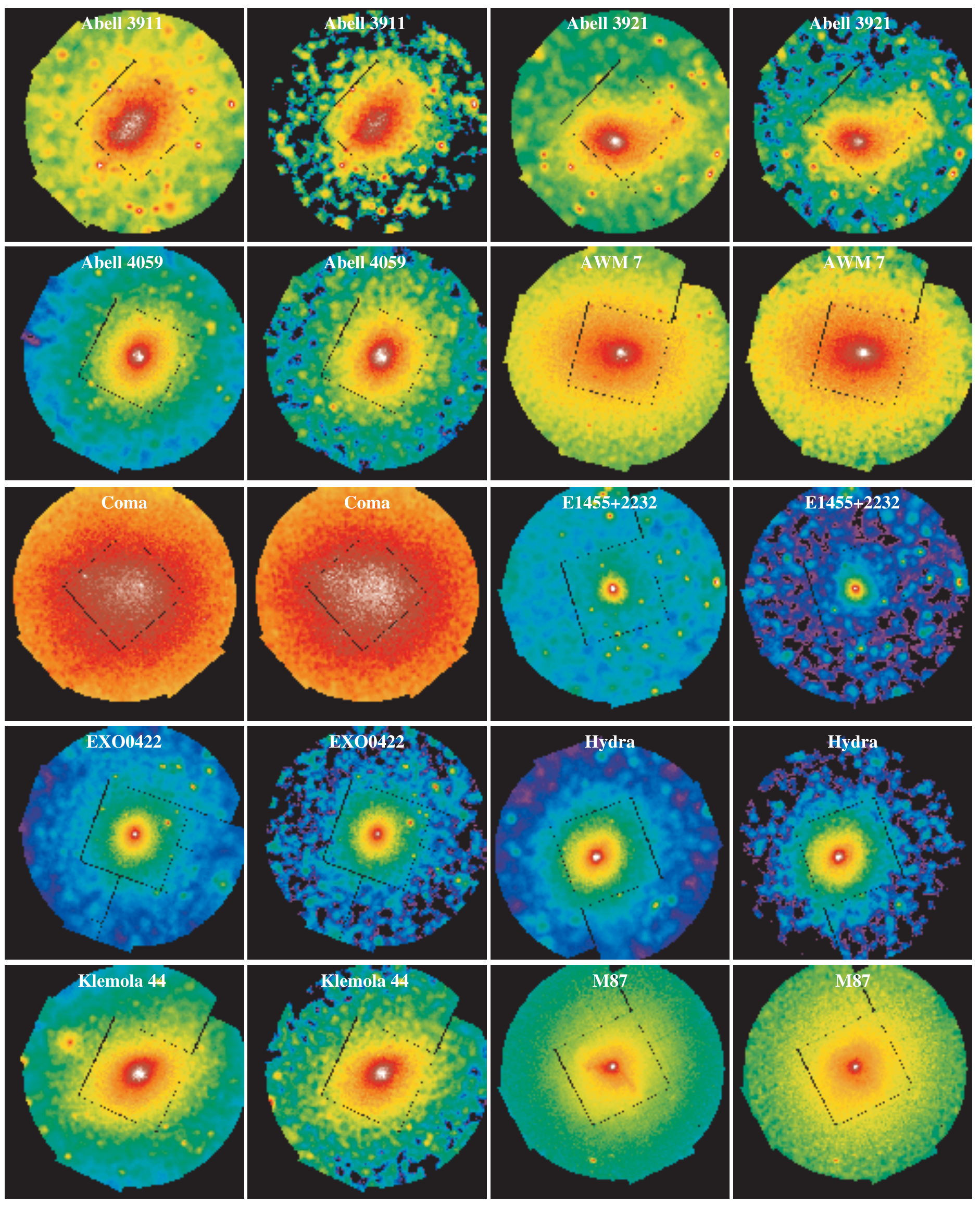}
\caption{Soft (left) and hard (right) band images of the clusters.
\label{fig:im-06}}
\end{figure*}

\begin{figure*}
\centering\includegraphics[angle=-0,width=6.5in]{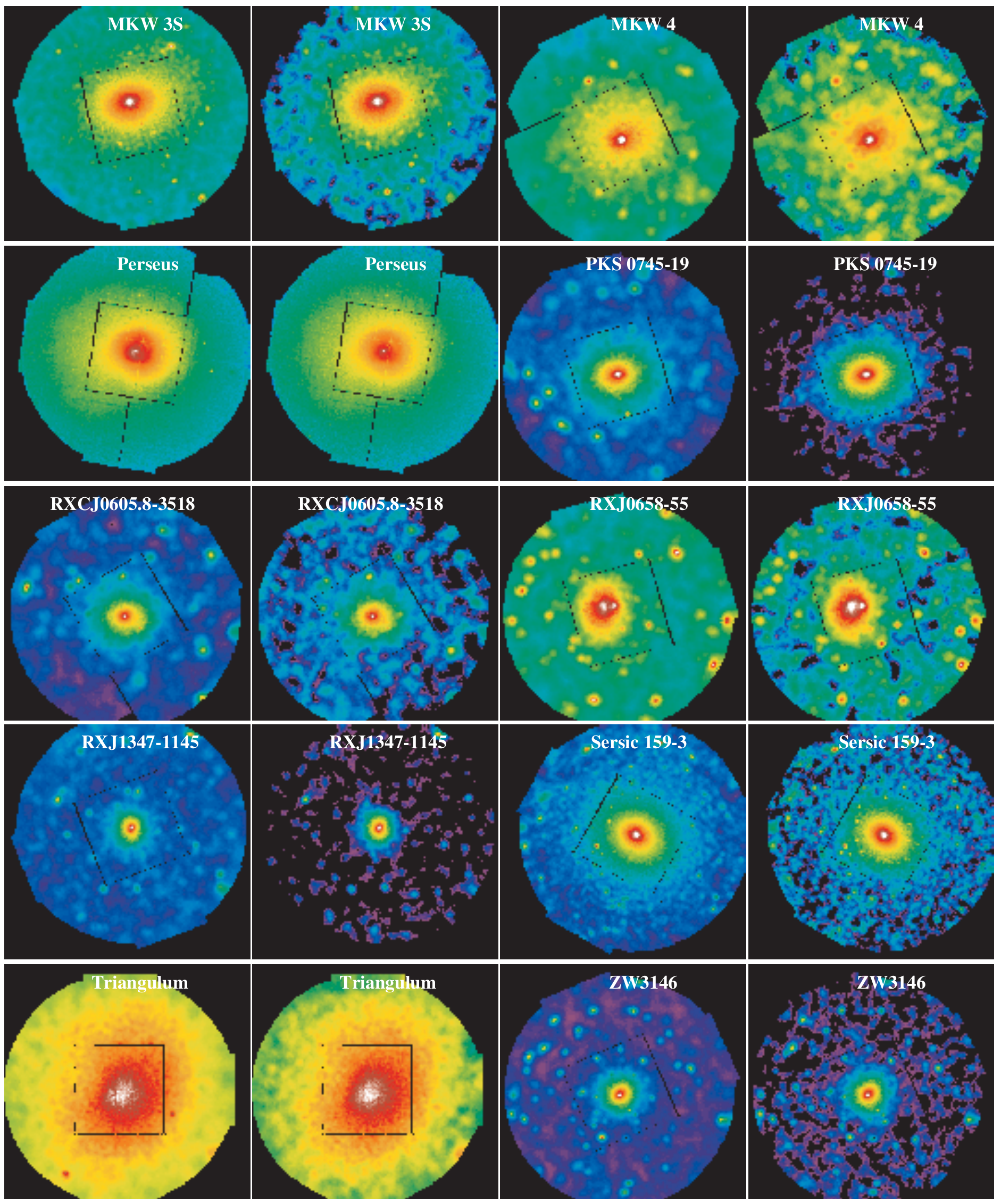}
\caption{Soft (left) and hard (right) band images of the clusters.
\label{fig:im-07}}
\end{figure*}

For the accepted observations, the center of the cluster was
determined from an image, bright point sources were manually 
excluded (typically to the level of a few times 
$10^{-13}$~ergs~cm$^{-2}$~s$^{-1}$, but the level varied due 
to the brightness and angular extent of the cluster),
and the data were processed to produce spectra for the ten annuli 
listed in Table~\ref{tbl:annuli} for both MOS detectors.  The 
count images in the $0.2-1.0$~keV band were examined for evidence 
of the individual CCDs operating in an anomalous state (KS07).  
If so, the individual CCD was excluded from the spectral extraction.
The HEASARC X-ray Background Tool was used to create RASS spectra
in, typically, a $1-2$ degree annulus around the cluster.  For a 
few cases (e.g., the Coma and Virgo clusters) the annulus had to be 
increased in size to fully exclude the cluster.  
The X-ray Background Tool also provided the column density of 
Galactic \ion{H}{i} which was fixed in the spectral fits. The analyzed 
clusters are listed in Table~\ref{tbl:clusters}.  
Included in the table are the fitted X-ray redshifts, {\it XMM-Newton} 
observation identification (ObsID), accepted and initial exposures, 
and the surface brightness limits for the image color bar scalings in 
Figs.~\ref{fig:im-01} through \ref{fig:im-07} of the electronic 
(on-line) version of this paper.

\begin{table}
\begin{minipage}[t]{\columnwidth}
\caption{Annuli Definitions}
\label{tbl:annuli}
\centering
\renewcommand{\footnoterule}{}
\begin{tabular}{ccc}
\hline\hline
Annulus & Inner & Outer \\
 & Radius & Radius \\
\hline
1 & $0'$    & $0.5'$ \\
3 & $0.5'$  & $1'$ \\
3 & $1'$    & $2'$ \\
4 & $2'$    & $3'$ \\
5 & $3'$    & $4'$ \\
6 & $4'$    & $5'$ \\
7 & $5'$    & $7'$ \\
8 & $7'$    & $9'$ \\
9 & $9'$    & $12'$ \\
10 & $12'$  & $14'$ \\
\hline
\end{tabular}
\end{minipage}
Inner and outer radii in arc minutes of the annuli 
used in the analysis of the clusters presented here.
\end{table}

\begin{figure}[h]
\includegraphics[angle=0,width=3.0in]{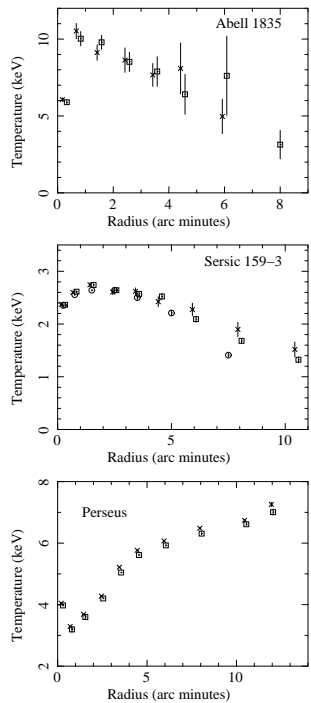}
\caption{Comparison of temperature radial profile results for the two 
observations of Abell~1835 (upper panel), S\'ersic~159-3 (middle 
panel), and Perseus (lower panel).  The radii have been slightly offset in the 
plot for clarity.  For the \object{S\'ersic 159-3} plot the CIE results of 
\citet{dpea06}.  In all panels the box and cross 
symbols represent the results of this paper while in the middle panel the circle
symbols represent the \citet{dpea06} results.
\label{fig:comp}}
\end{figure}

In order to test the reliability of our analysis methods we use second
observations of the clusters \object{Abell 1835}, \object{S\'ersic 159-3} 
and \object{Perseus} for comparisons.  (Note that the second observation of 
\object{S\'ersic 159-3} is under the alternate name \object{AS 1101} and the 
second observation of Perseus is under the alternate name Abell~426).  
Fig.~\ref{fig:comp} shows the fitted temperatures which are in very good 
agreement.  Along with our \object{S\'ersic 159-3} results we have plotted the 
CIE (which are more equivalent to our spectral fitting) results from 
\citet{dpea06}. These data are also in reasonable agreement except at higher 
radii where background subtraction is more problematic and at radii at $0.5-2'$ 
where the cross-talk effect is strongest.  The fitted temperatures for the 
Perseus cluster do so a slight but significant systematic difference with one
observation having consistently higher temperatures by $\sim0.15$~keV. 
However, as the Perseus cluster is very bright, it is very unlikely that this 
systematic difference was caused by errors in the background modeling.

\begin{figure}[h]
\includegraphics[angle=90,width=3.0in]{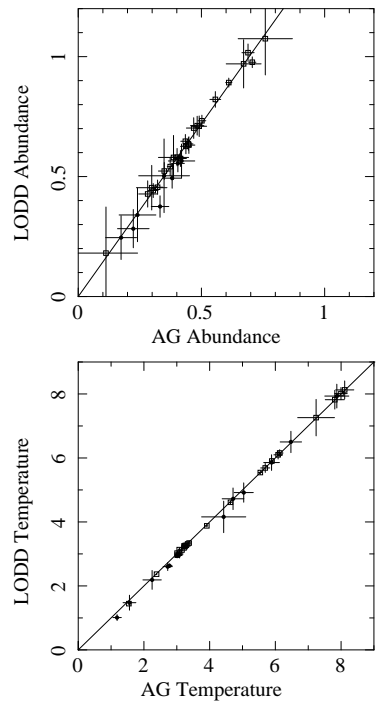}
\caption{The two plots display the comparison results from using \citet{ag89} 
and \citet{lod03} abundances for the fitted values of the abundance (upper panel) 
and temperature (lower panel).  In the upper plot the line is the best-fit scale 
factor of 1.44 while in the lower plot the line shows the one-to-one relationship.
In both plots the filled circles indicate data from the outer annuli.
\label{fig:abund-all}}
\end{figure}

We have tested the robustness of our results to variations in the assumed 
emission abundance model.  As noted above, for the cluster catalog analysis 
we use \citet{ag89} abundances allowing only a single scale factor.  We refit 
the data for four clusters (\object{Abell 665}, \object{Abell 1060}, 
\object{Abell 1795}, and \object{2A 0335+096}) using \citet{lod03} abundances 
with the results shown in Fig.~\ref{fig:abund-all}.  The fits were of similar 
quality and the only 
significant difference were the values of the fitted abundances, which were 
consistent with a simple scaling by a factor of 1.44 with the \citet{lod03} 
abundances greater than those of \citet{ag89}.  The fitted temperatures 
using the two abundance models were all consistent. 

\begin{figure}[h]
\includegraphics[angle=0,width=3.0in]{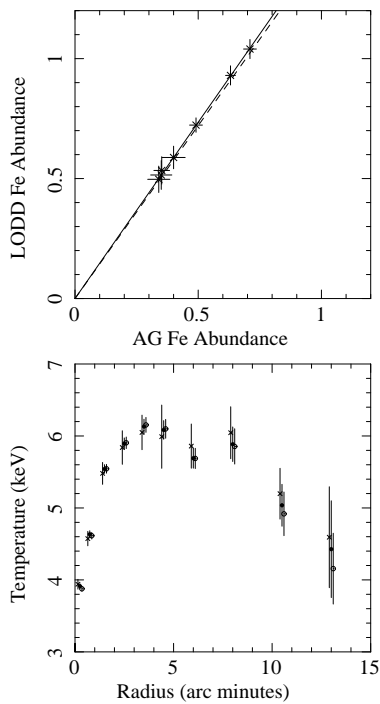}
\caption{Comparison of the fitted values for the annuli iron abundances and 
emission temperatures using \citet{ag89} and \citet{lod03} model abundances 
while allowing all abundances to vary.  The upper panel shows the correlation 
between the fitted values for the iron abundance for the two abundance models.
The solid line is the best-fit correlation of 1.47 while the dashed line shows
the 1.44 correlation of the single abundance normalization.  The lower panel
shows the fitted values for the temperatures where (open circle) abundances 
starting with \citet{ag89} values were allowed to vary independently, (filled 
circle) \citet{ag89} values were allowed to vary only with a single scale factor, 
and (filled triangle) \citet{lod03} values were allowed to vary only with a 
single scale factor.
\label{fig:vapec}}
\end{figure}

We also have used the \object{Abell 1795} data to examine the effect of 
allowing the abundances to vary independently using the VAPEC model of Xspec 
(Fig.~\ref{fig:vapec}).  To examine
the variation in the abundances we considered iron, which showed a simple 
scaling of 1.47, which is consistent with the factor of 1.44 determined for 
the average scale factor between the two abundance models.  The fitted values 
for the temperatures were in very good agreement.  For the \citet{ag89} 
model the fitted Fe abundance when all elements were allowed to vary was 
$\sim10$\% higher than the fitted average value for the abundance. 

\begin{figure*}
\centering\includegraphics[angle=-0,width=5.5in]{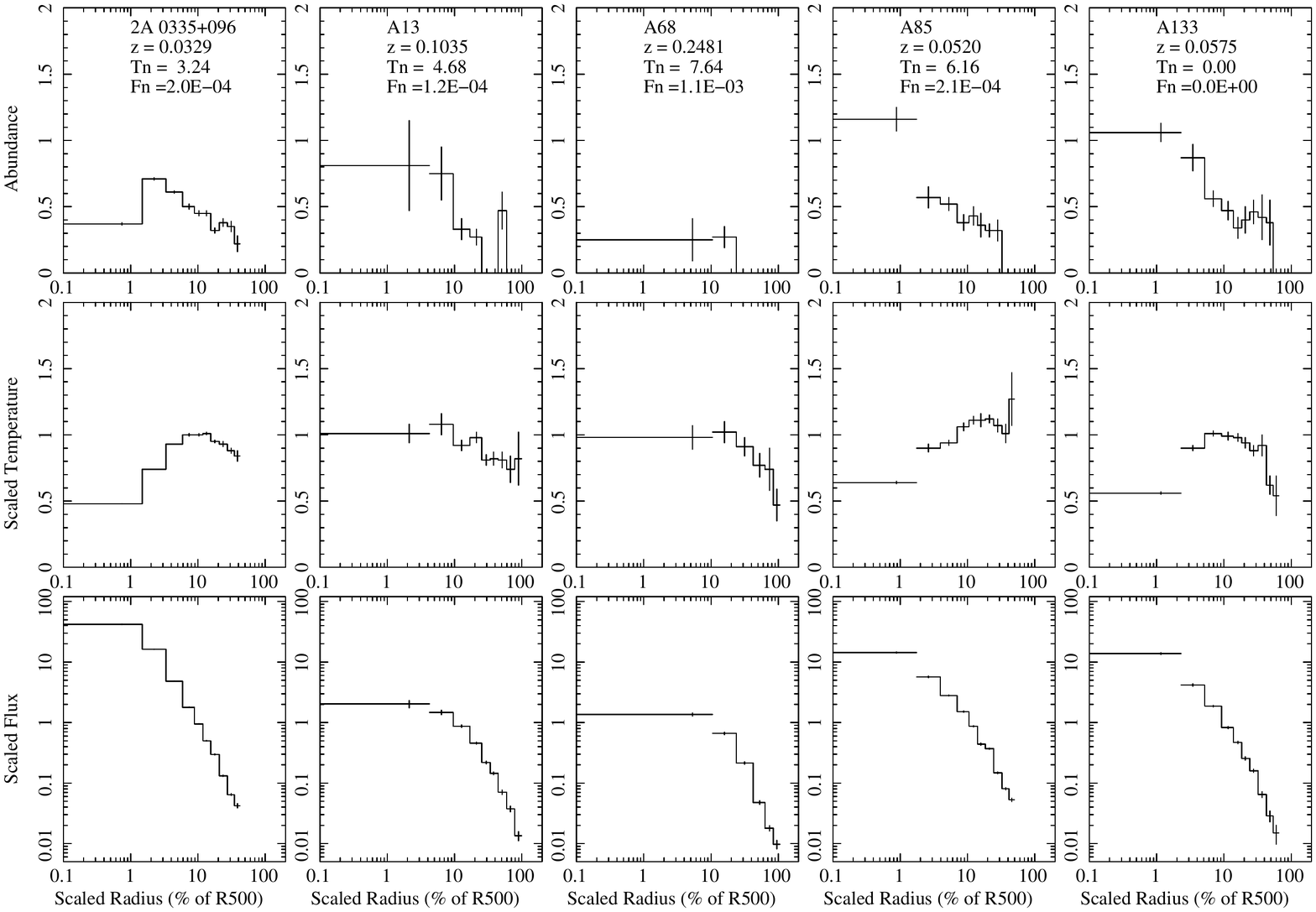}
\caption{Cluster temperature, abundance, and flux radial profiles.  
The name of the cluster, fitted redshift, and values for the temperature 
(${\rm T_N}$) and flux (${\rm T_N}$) used for the normalization of the 
data are provided in the abundance panel. 
\label{fig:prof-01}}
\end{figure*}

\begin{figure*}
\centering\includegraphics[angle=-0,width=5.5in]{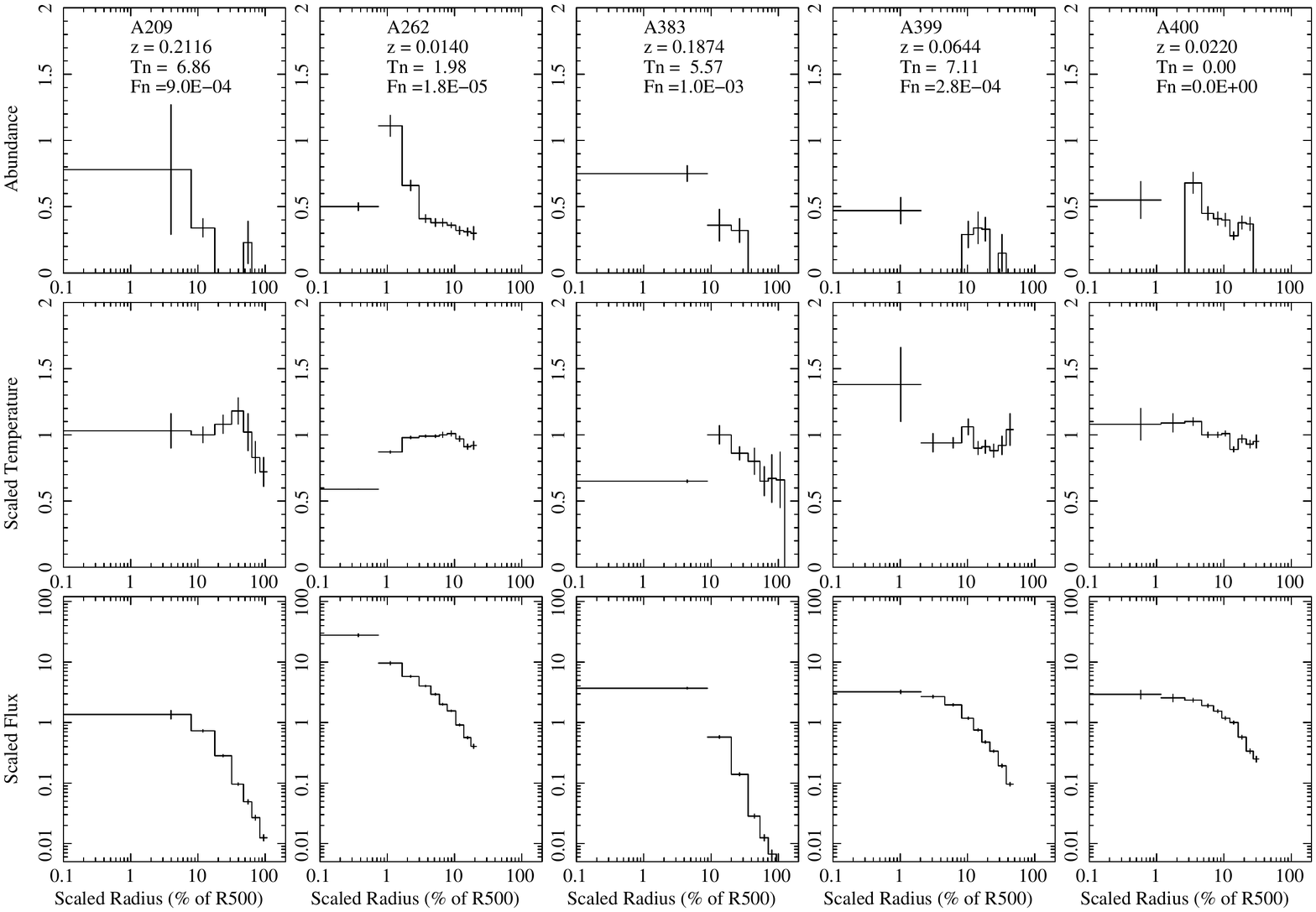}
\caption{Cluster temperature, abundance, and flux radial profiles.  
The name of the cluster, fitted redshift, and values for the temperature 
(${\rm T_N}$) and flux (${\rm T_N}$) used for the normalization of the 
data are provided in the abundance panel. 
\label{fig:prof-02}}
\end{figure*}

\begin{figure*}
\centering\includegraphics[angle=-0,width=5.5in]{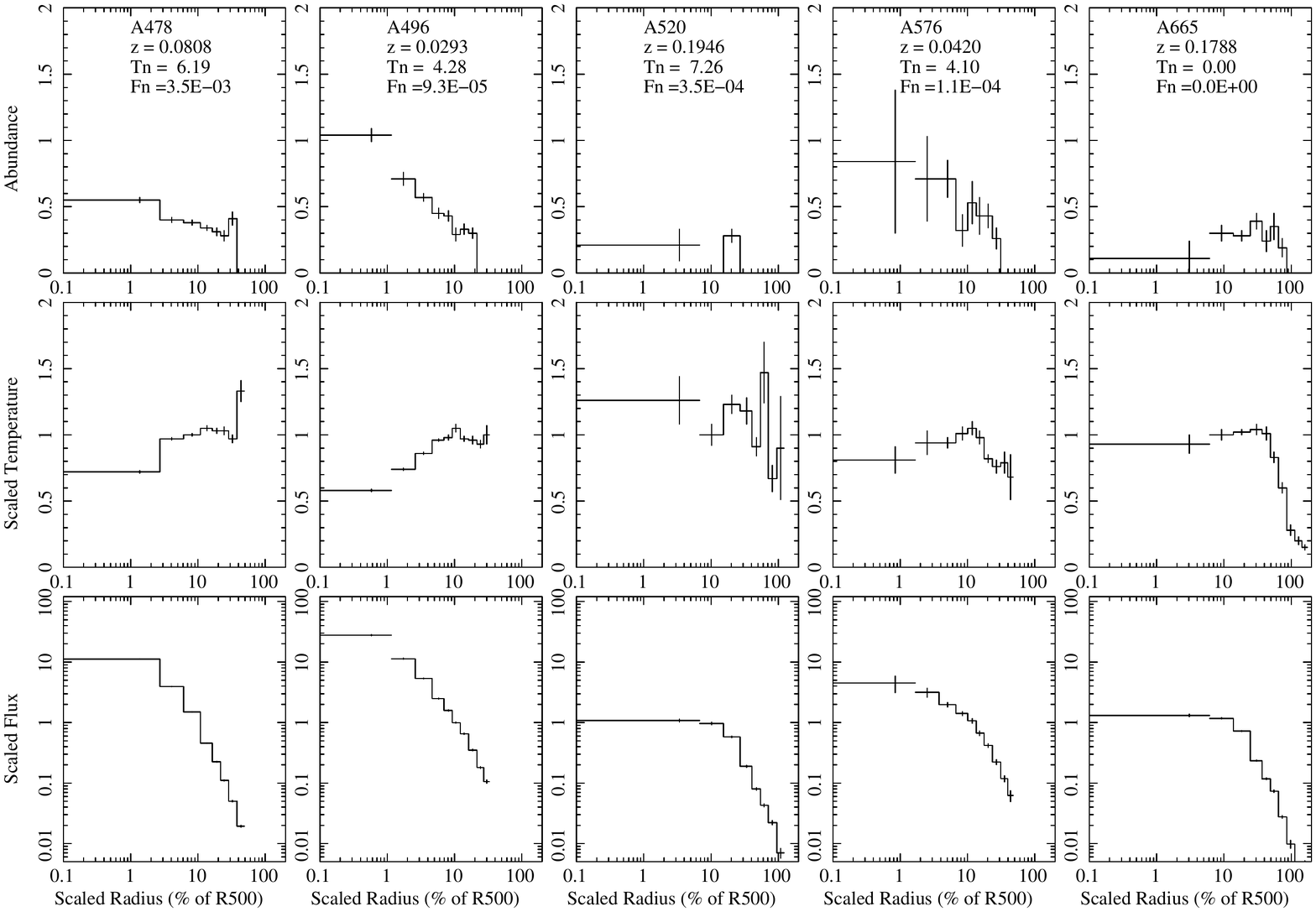}
\caption{Cluster temperature, abundance, and flux radial profiles.  
The name of the cluster, fitted redshift, and values for the temperature 
(${\rm T_N}$) and flux (${\rm T_N}$) used for the normalization of the 
data are provided in the abundance panel. 
\label{fig:prof-03}}
\end{figure*}

\begin{figure*}
\centering\includegraphics[angle=-0,width=5.5in]{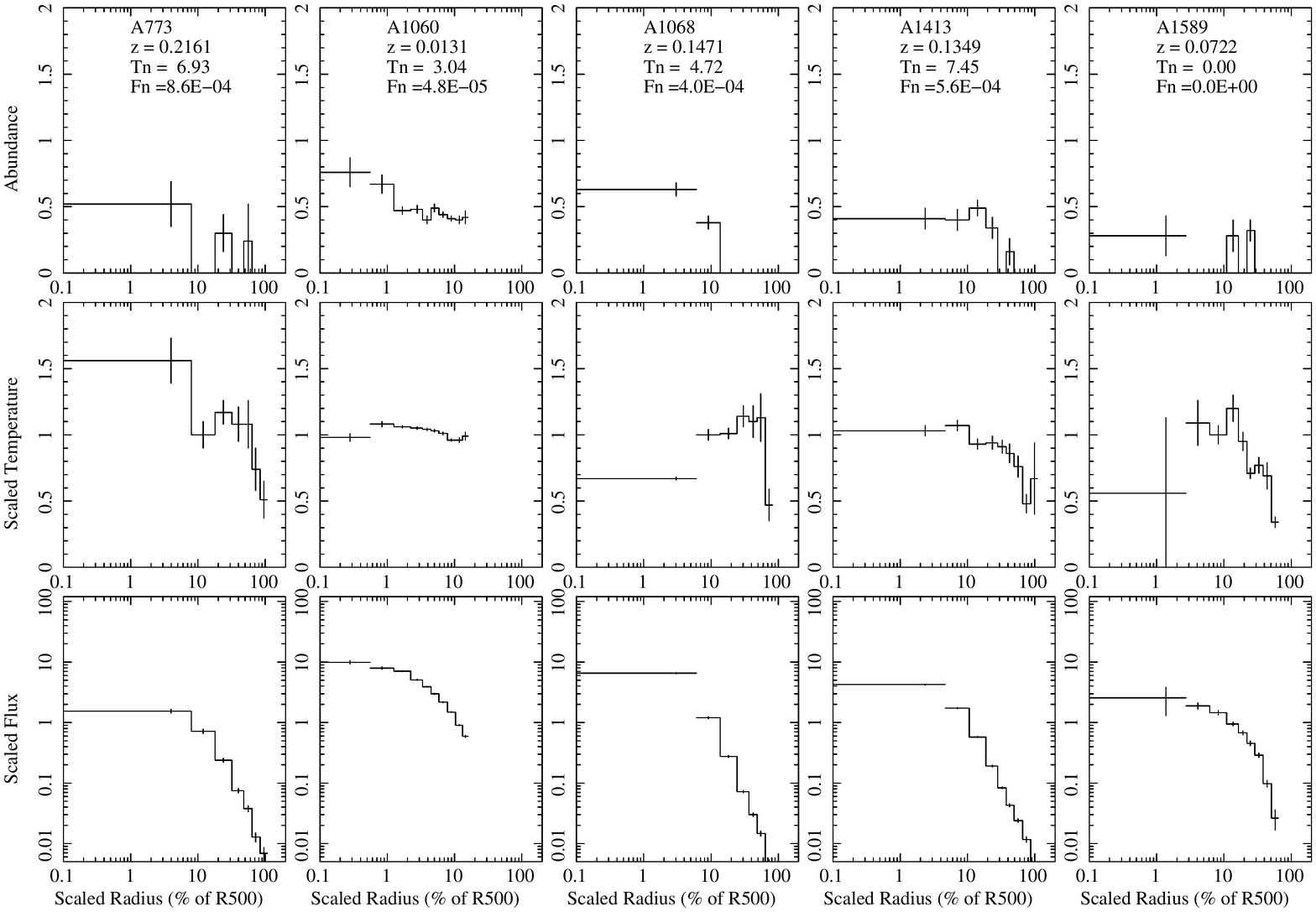}
\caption{Cluster temperature, abundance, and flux radial profiles.  
The name of the cluster, fitted redshift, and values for the temperature 
(${\rm T_N}$) and flux (${\rm T_N}$) used for the normalization of the 
data are provided in the abundance panel. 
\label{fig:prof-04}}
\end{figure*}

\begin{figure*}
\centering\includegraphics[angle=-0,width=5.5in]{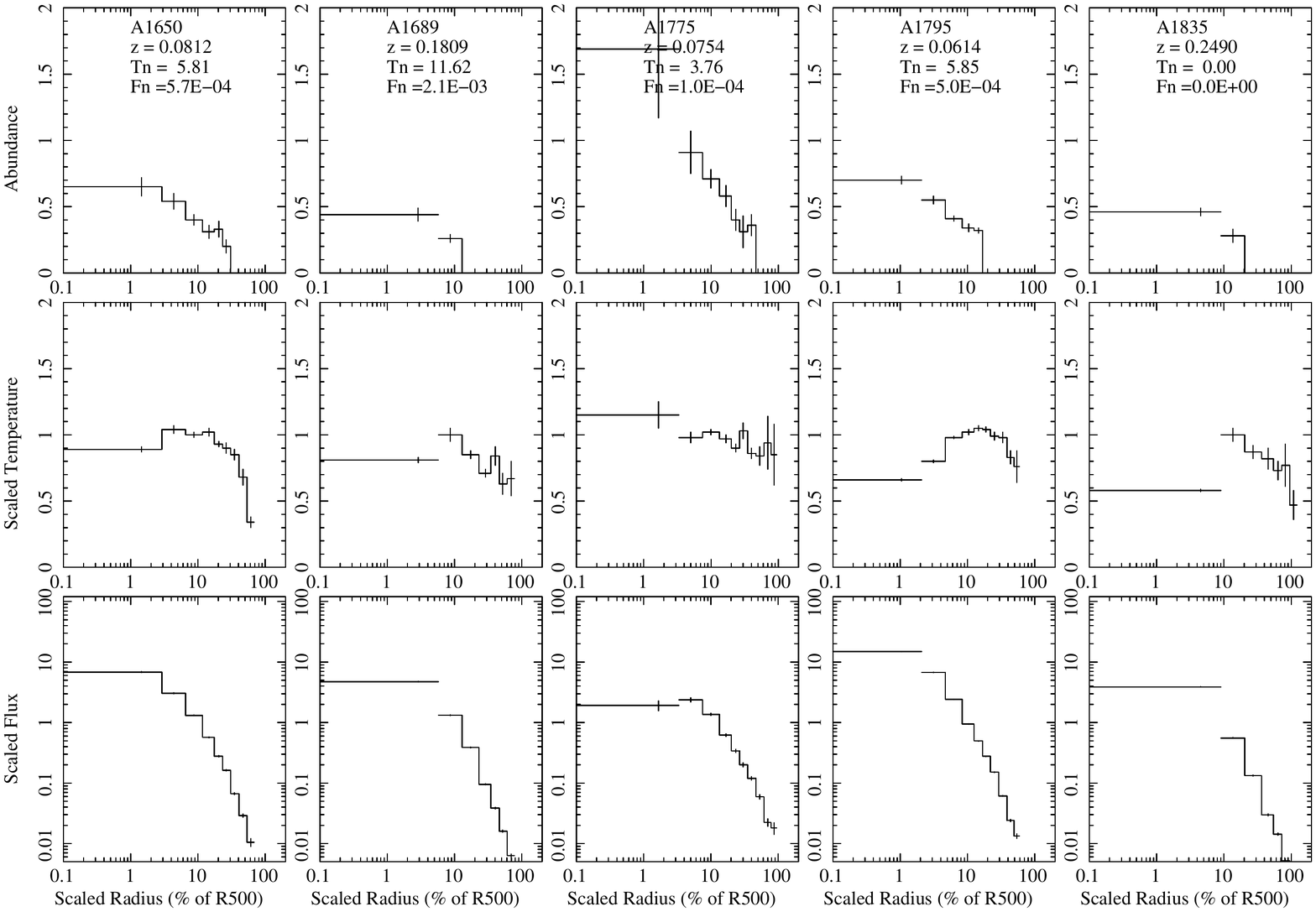}
\caption{Cluster temperature, abundance, and flux radial profiles.  
The name of the cluster, fitted redshift, and values for the temperature 
(${\rm T_N}$) and flux (${\rm T_N}$) used for the normalization of the 
data are provided in the abundance panel. 
\label{fig:prof-05}}
\end{figure*}

\begin{figure*}
\centering\includegraphics[angle=-0,width=5.5in]{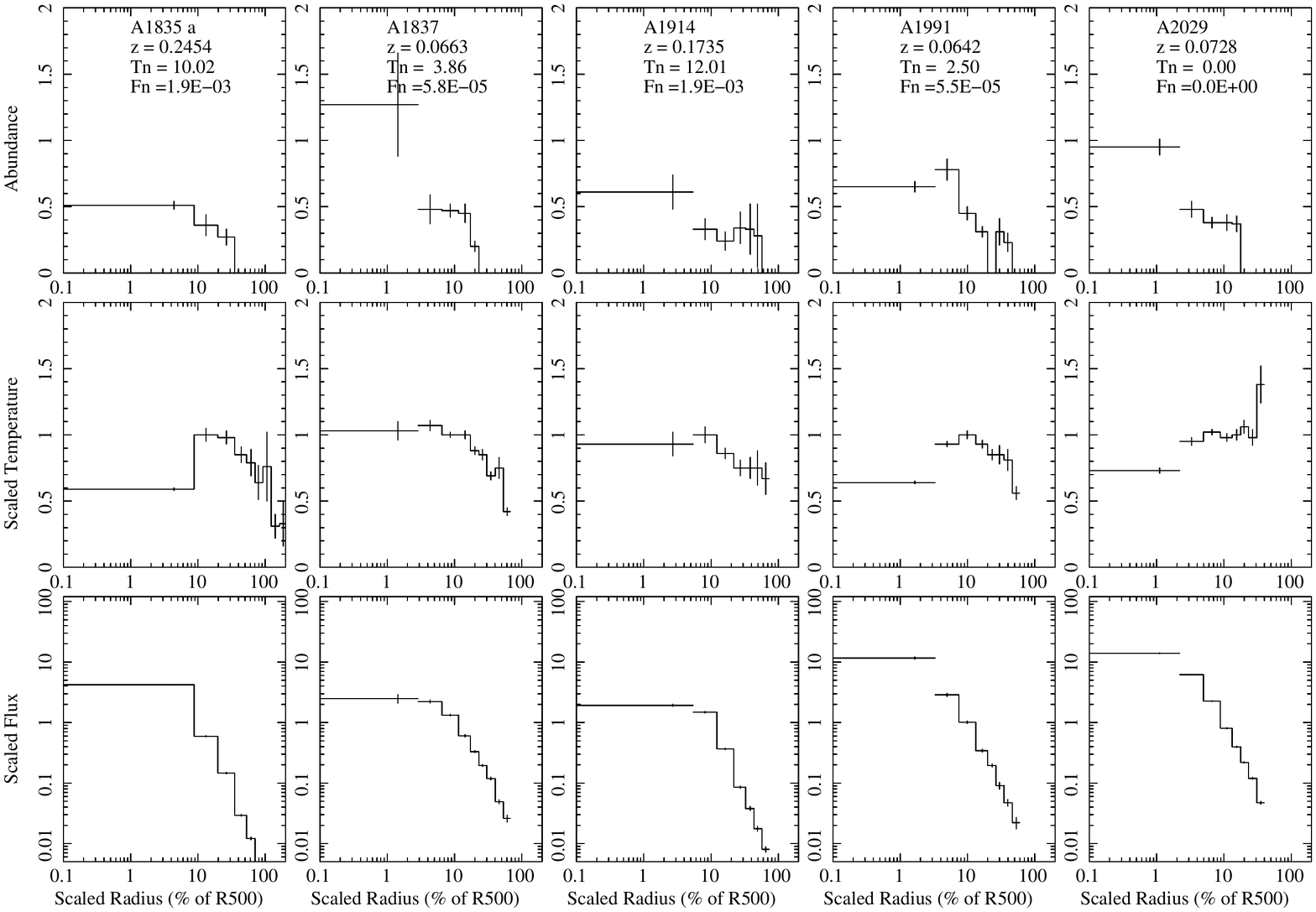}
\caption{Cluster temperature, abundance, and flux radial profiles.  
The name of the cluster, fitted redshift, and values for the temperature 
(${\rm T_N}$) and flux (${\rm T_N}$) used for the normalization of the 
data are provided in the abundance panel. 
\label{fig:prof-06}}
\end{figure*}

\begin{figure*}
\centering\includegraphics[angle=-0,width=5.5in]{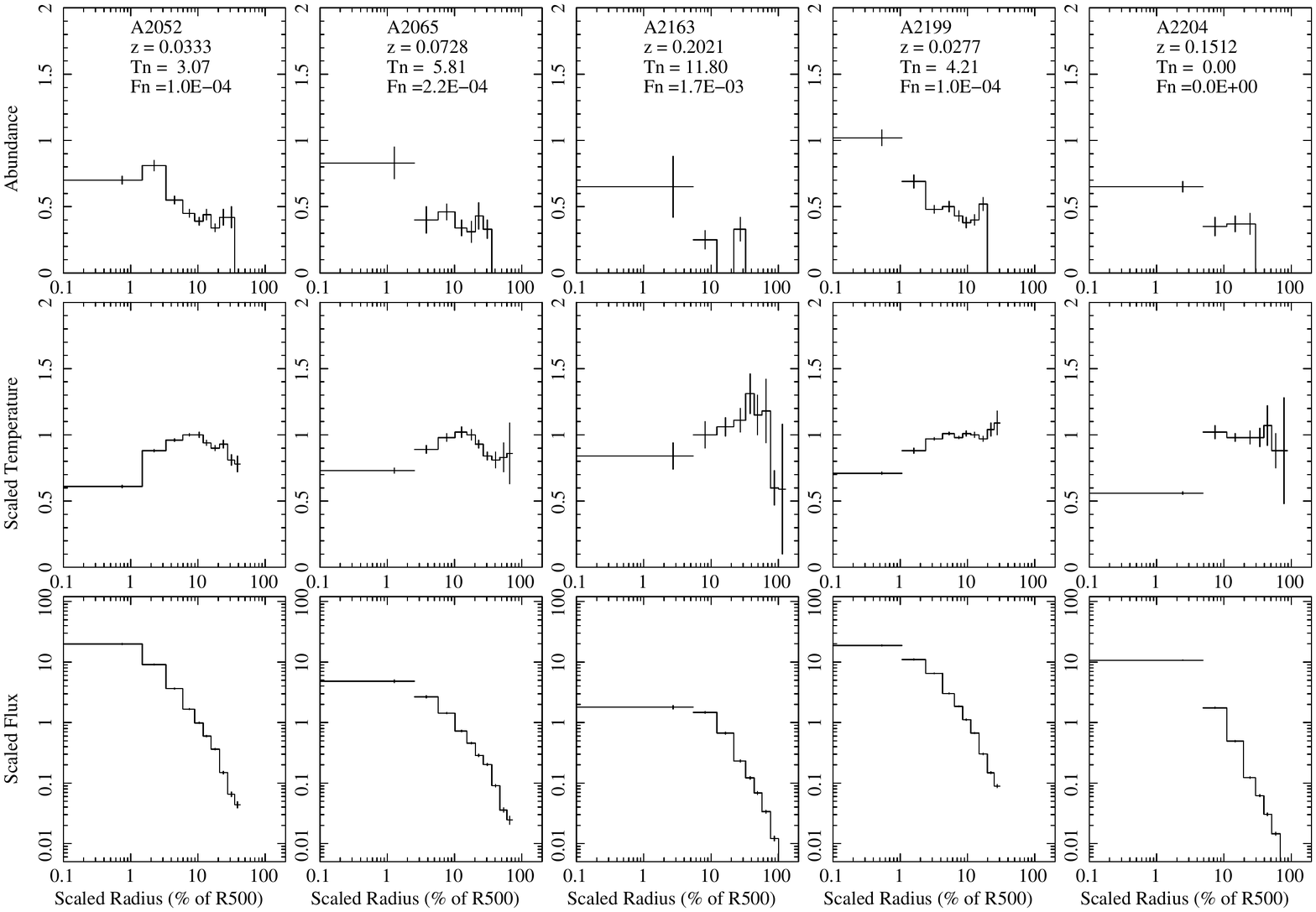}
\caption{Cluster temperature, abundance, and flux radial profiles.  
The name of the cluster, fitted redshift, and values for the temperature 
(${\rm T_N}$) and flux (${\rm T_N}$) used for the normalization of the 
data are provided in the abundance panel. 
\label{fig:prof-07}}
\end{figure*}

\begin{figure*}
\centering\includegraphics[angle=-0,width=5.5in]{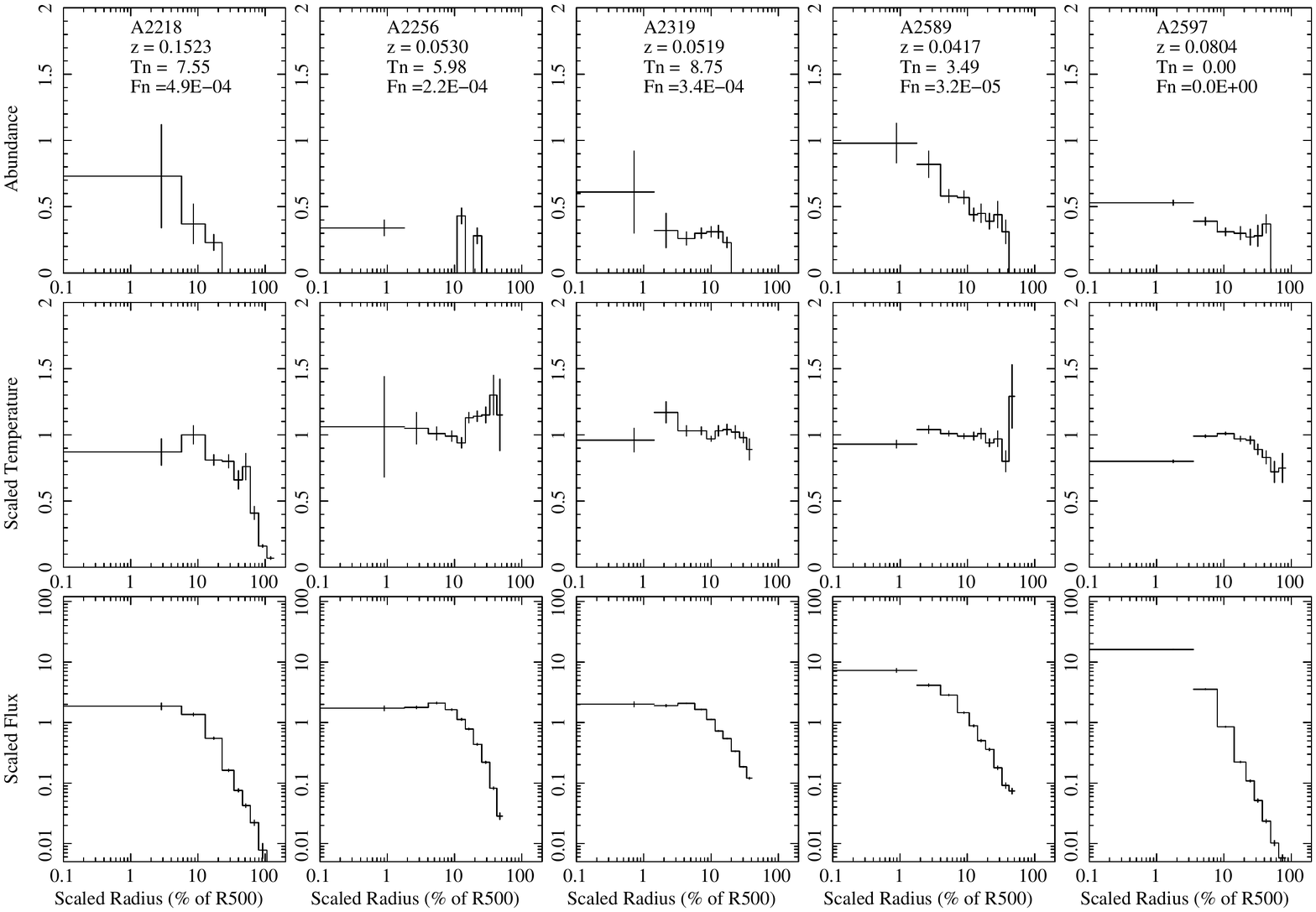}
\caption{Cluster temperature, abundance, and flux radial profiles.  
The name of the cluster, fitted redshift, and values for the temperature 
(${\rm T_N}$) and flux (${\rm T_N}$) used for the normalization of the 
data are provided in the abundance panel. 
\label{fig:prof-08}}
\end{figure*}

\begin{figure*}
\centering\includegraphics[angle=-0,width=5.5in]{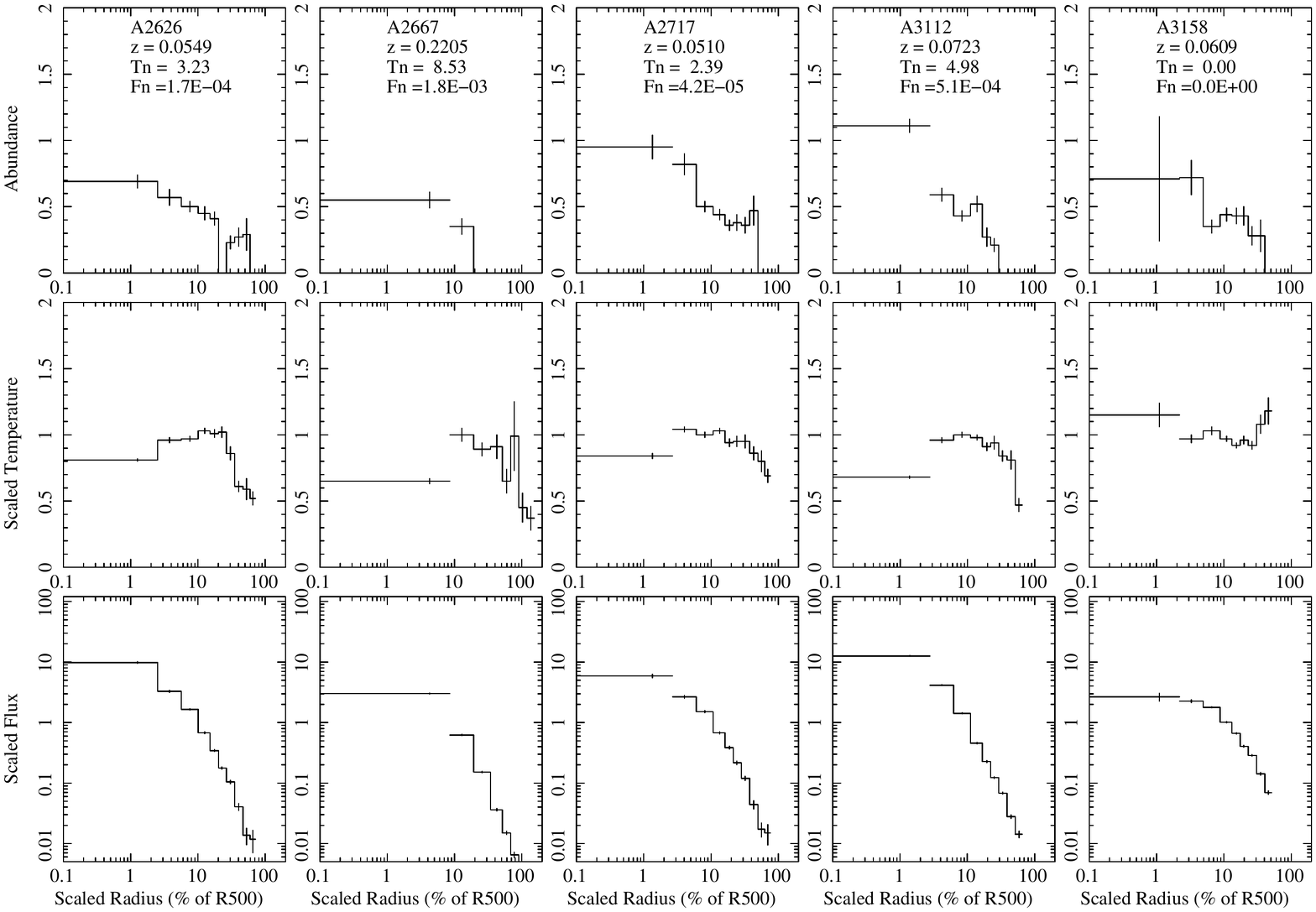}
\caption{Cluster temperature, abundance, and flux radial profiles.  
The name of the cluster, fitted redshift, and values for the temperature 
(${\rm T_N}$) and flux (${\rm T_N}$) used for the normalization of the 
data are provided in the abundance panel. 
\label{fig:prof-09}}
\end{figure*}

\begin{figure*}
\centering\includegraphics[angle=-0,width=5.5in]{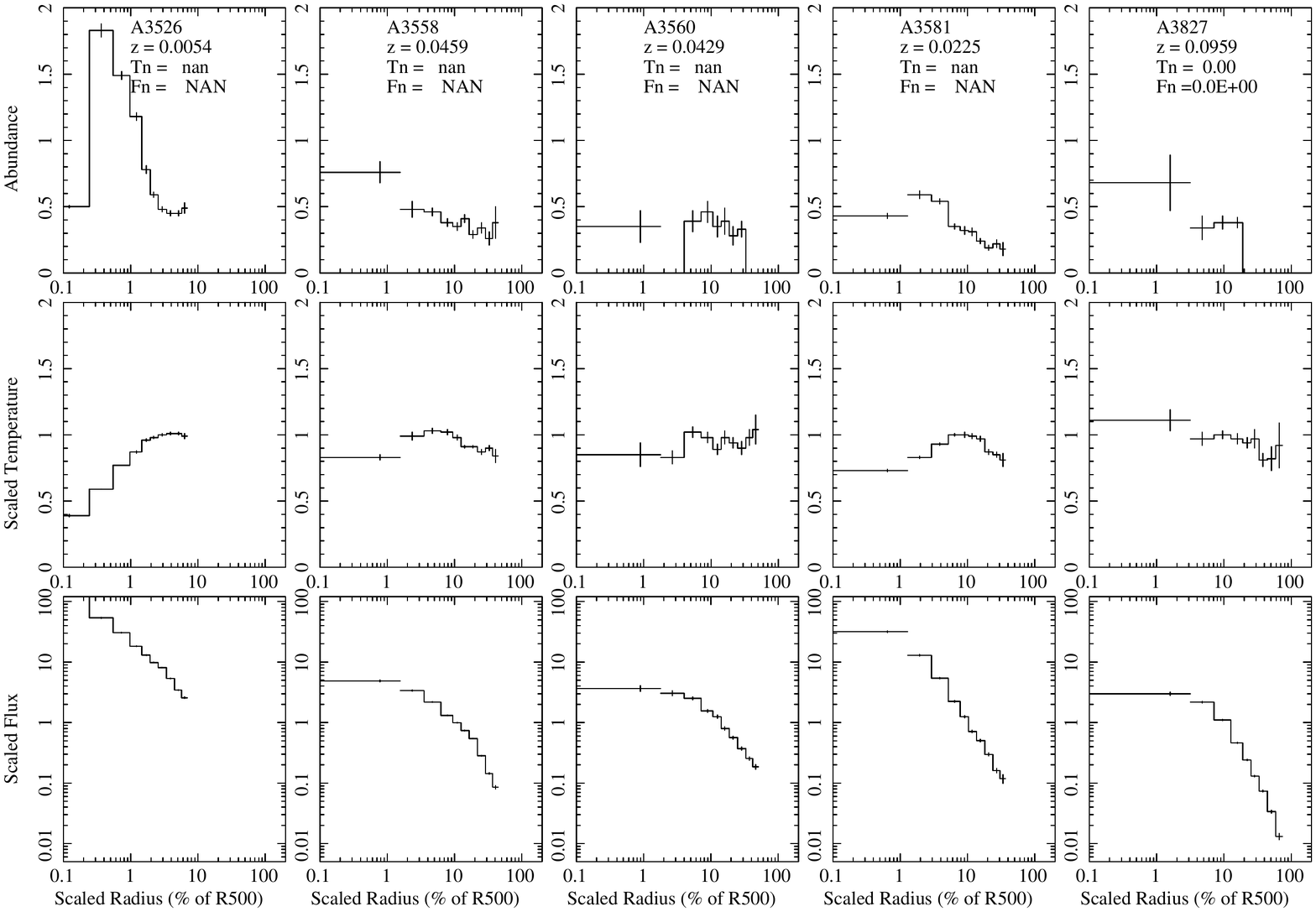}
\caption{Cluster temperature, abundance, and flux radial profiles.  
The name of the cluster, fitted redshift, and values for the temperature 
(${\rm T_N}$) and flux (${\rm T_N}$) used for the normalization of the 
data are provided in the abundance panel. 
\label{fig:prof-10}}
\end{figure*}

\begin{figure*}
\centering\includegraphics[angle=-0,width=5.5in]{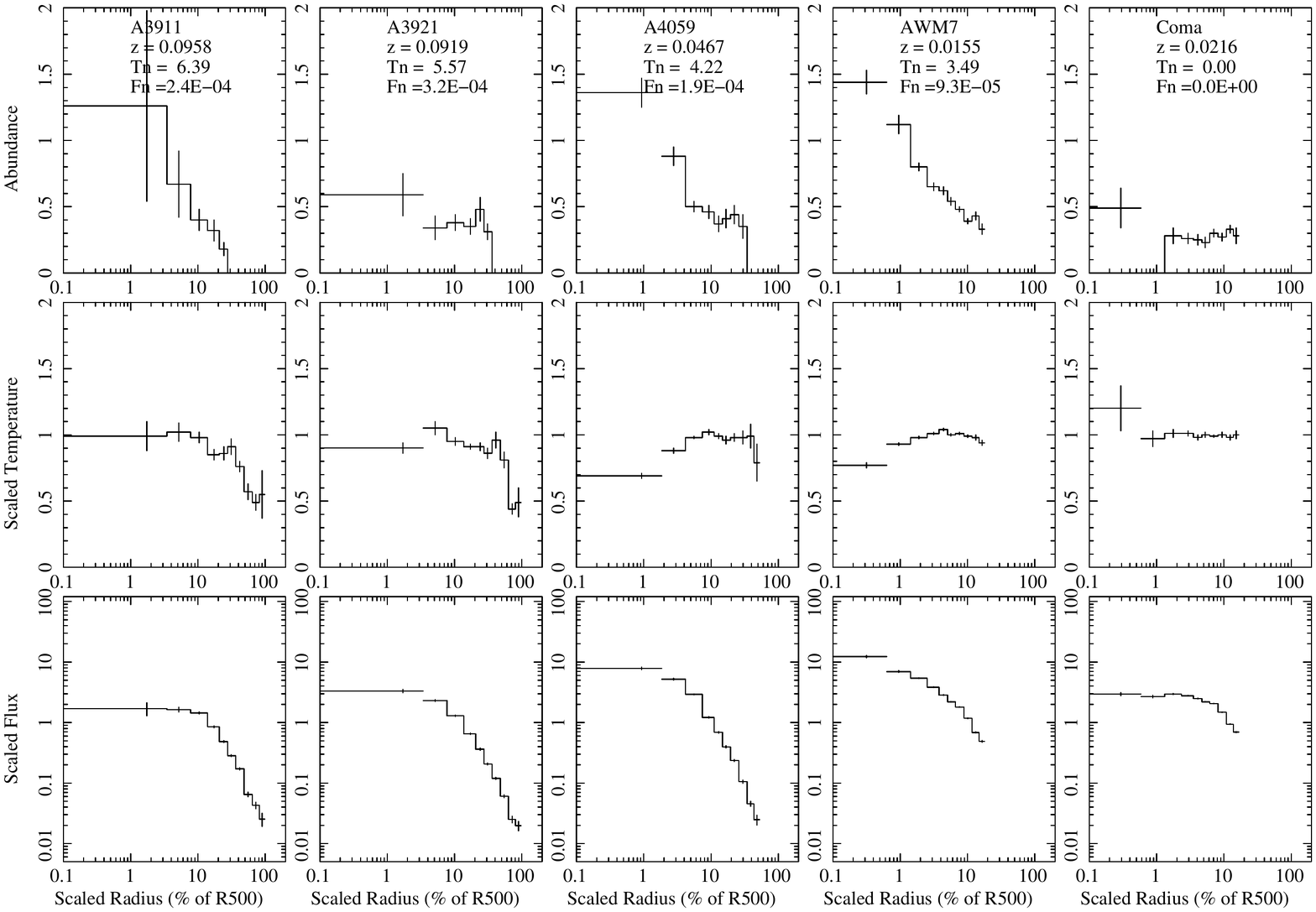}
\caption{Cluster temperature, abundance, and flux radial profiles.  
The name of the cluster, fitted redshift, and values for the temperature 
(${\rm T_N}$) and flux (${\rm T_N}$) used for the normalization of the 
data are provided in the abundance panel. 
\label{fig:prof-11}}
\end{figure*}

\begin{figure*}
\centering\includegraphics[angle=-0,width=5.5in]{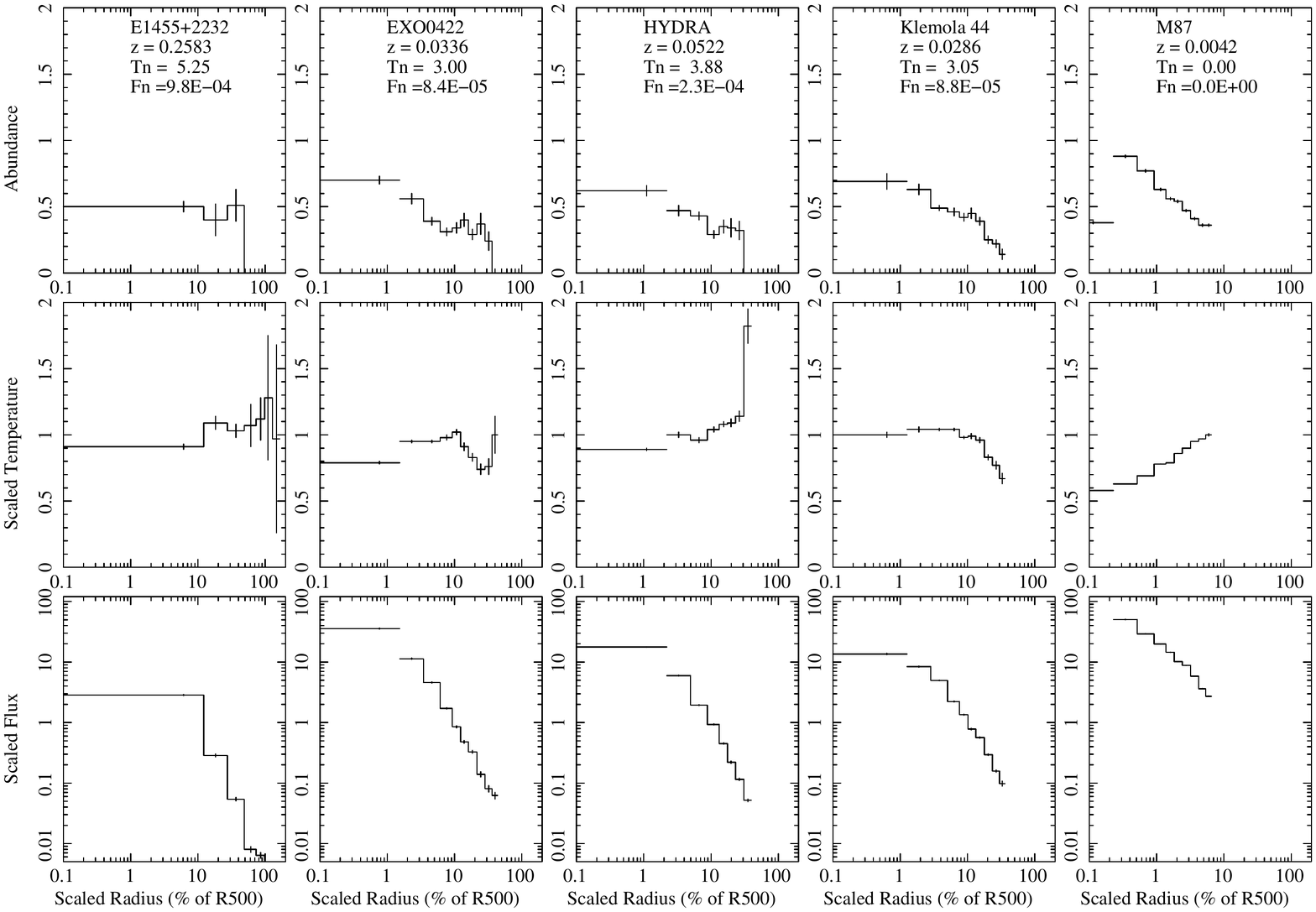}
\caption{Cluster temperature, abundance, and flux radial profiles.  
The name of the cluster, fitted redshift, and values for the temperature 
(${\rm T_N}$) and flux (${\rm T_N}$) used for the normalization of the 
data are provided in the abundance panel. 
\label{fig:prof-12}}
\end{figure*}

\begin{figure*}
\centering\includegraphics[angle=-0,width=5.5in]{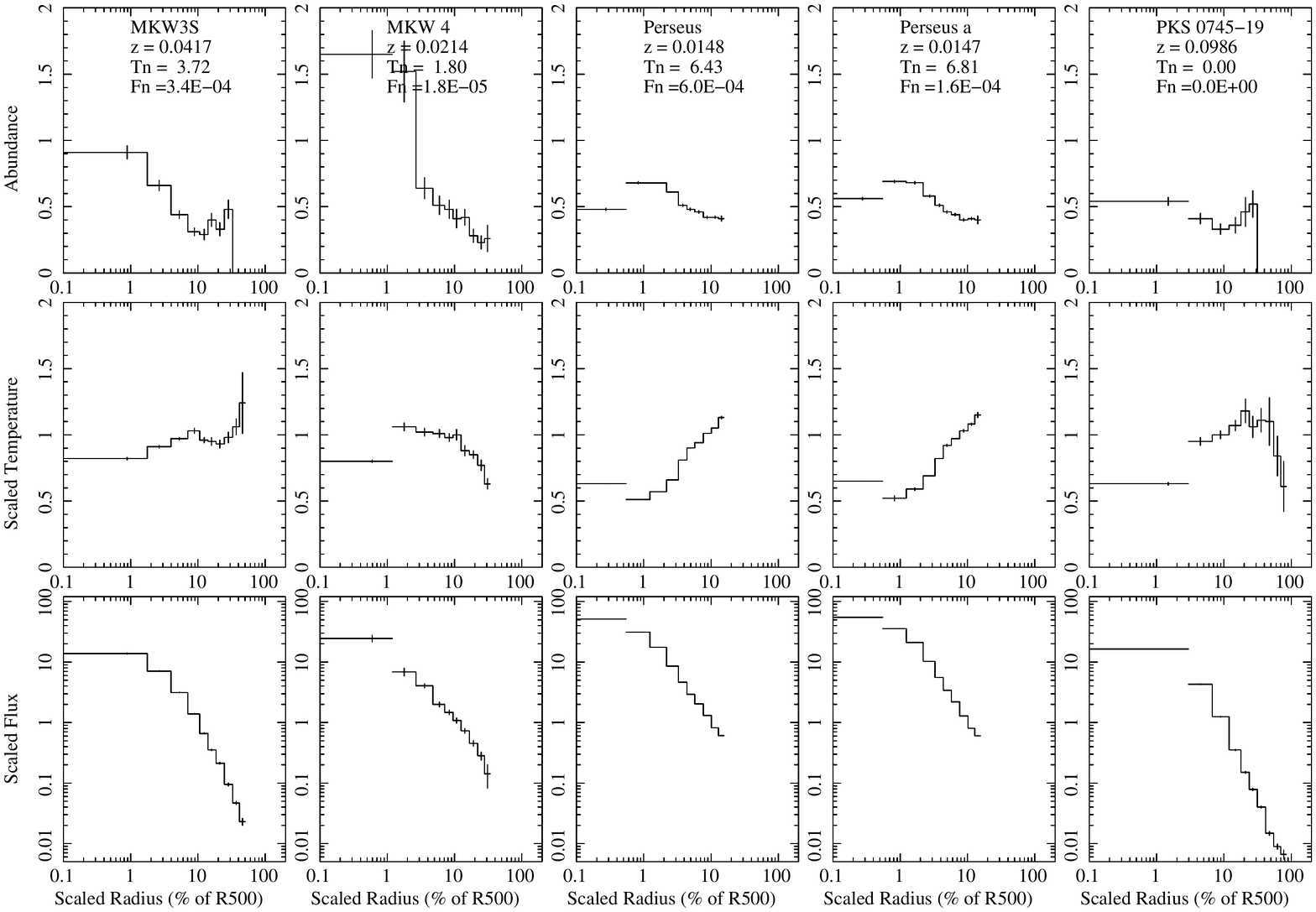}
\caption{Cluster temperature, abundance, and flux radial profiles.  
The name of the cluster, fitted redshift, and values for the temperature 
(${\rm T_N}$) and flux (${\rm T_N}$) used for the normalization of the 
data are provided in the abundance panel. 
\label{fig:prof-13}}
\end{figure*}

\begin{figure*}
\centering\includegraphics[angle=-0,width=5.5in]{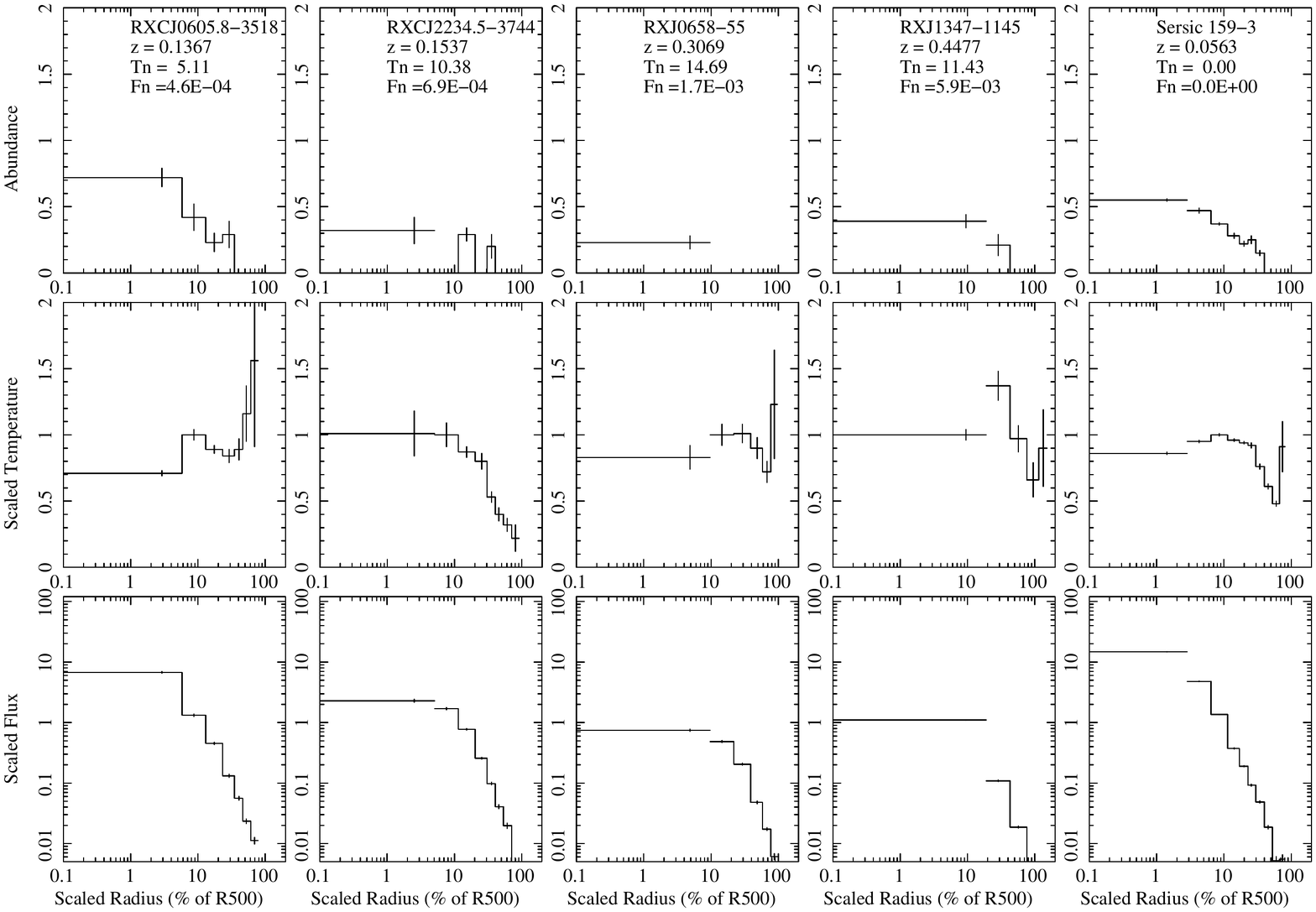}
\caption{Cluster temperature, abundance, and flux radial profiles.  
The name of the cluster, fitted redshift, and values for the temperature 
(${\rm T_N}$) and flux (${\rm T_N}$) used for the normalization of the 
data are provided in the abundance panel. 
\label{fig:prof-14}}
\end{figure*}

\begin{figure*}
\centering\includegraphics[angle=-0,width=5.5in]{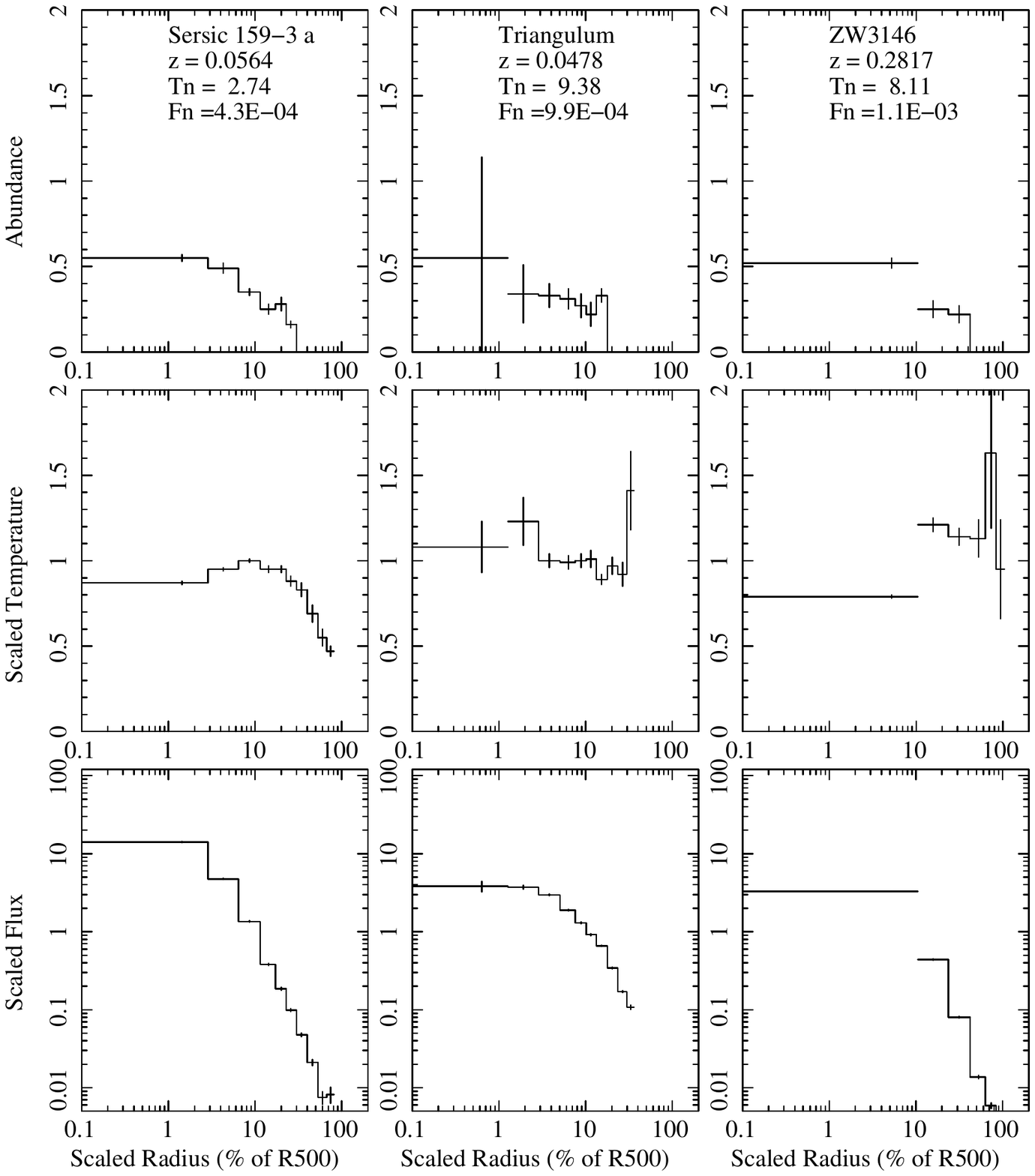}
\caption{Cluster temperature, abundance, and flux radial profiles.  
The name of the cluster, fitted redshift, and values for the temperature 
(${\rm T_N}$) and flux (${\rm T_N}$) used for the normalization of the 
data are provided in the abundance panel. 
\label{fig:prof-15}}
\end{figure*}

The temperature, abundance, and flux radial profiles for the 70 clusters listed 
in Table~\ref{tbl:clusters} are shown in Figs.~\ref{fig:prof-01} through 
\ref{fig:prof-15} in the electronic version of this paper.  The radii of the 
annuli have been scaled to the $R_{500}$ value of the cluster as derived from 
the equation $R_{500} = 2.6\times((1.0 + z)^{-3\over2}\times(T/10.0)^{1\over2}$ 
\citep{emn96} where $z$ is the fitted value for the cluster redshift 
and $T$ is the average fitted value for the cluster temperature in the 
$1'-4'$ annulus.  Both the temperature and flux have been 
normalized to the values in the range 5\% -- 30\% of $R_{500}$.  

We also include soft ($0.35-1.25$~keV) and hard 
($2.0-8.0$~keV) band images of the clusters in the electronic version 
(Figs.~\ref{fig:im-01} through \ref{fig:im-07}).  The images 
combine the MOS1 and MOS2 data and are background subtracted 
(QPB and SP), exposure corrected, and adaptively smoothed.  
Table~\ref{tbl:clusters} provides the upper scaling limits for 
the color coding (purple and blue indicate low intensity while 
red and white indicate high intensity).  The images were produced 
by {\it ds9} where the minimum value of the dynamic range was set 
to zero and the image was logarithmically scaled.  
Units are counts~s$^{-1}$~deg$^{-2}$ where the typical level of the 
cosmic background is $\sim1$ in these units.  The intensities are  
average values of the MOS1 and MOS2 data rather than the sum.  
Table~\ref{tbl:clusterdetails}, also in the electronic version of this 
paper, lists the radial profile data, 
temperature, abundance, and flux, for the clusters.

\section{Results and Conclusions}
\label{sec:conclusions}

In this paper we have outlined a robust and reliable method for 
analyzing extended X-ray sources observed with the {\it XMM-Newton}
EPIC MOS detectors.  The method combines screening of the data for
periods of background enhancements (most notably the soft proton
contamination), detailed modeling of the particle background 
spectrum, and the determination of other background components in
the spectral fitting process (residual SP contamination, fluorescent
particle background lines, and the cosmic background).  

We have demonstrated our method with the bulk processing of the 
observation of 70 clusters of galaxies.  Comparison of the results 
for two separate observations of \object{Abell 1835}, 
\object{S\'ersic 159-3}, and  \object{Perseus}
show good agreement between their fitted temperatures.  However, 
comparison of our results with the {\it Chandra} results of 
\citet{vea05} for the overlapping subset of 
clusters shows a significant discrepancy for higher temperature 
clusters.  The sense of this discrepancy is that the higher the fitted 
temperature, the greater the likelihood that {\it Chandra} will find 
a higher temperature than {\it XMM-Newton}.  The differences can be
over 1~keV at $7-8$~keV.  This effect can increase the apparent 
temperature gradient in the outer annuli of clusters in {\it Chandra} 
data.

\begin{figure}
\centering{\includegraphics[angle=0,width=8.5cm]{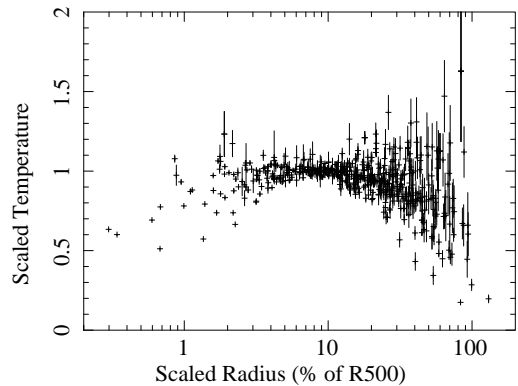}}
\caption{Scaled temperature radial profiles for all of the analyzed 
clusters.  
\label{fig:temp}}
\end{figure}

\begin{figure}
\centering{\includegraphics[angle=0,width=8.5cm]{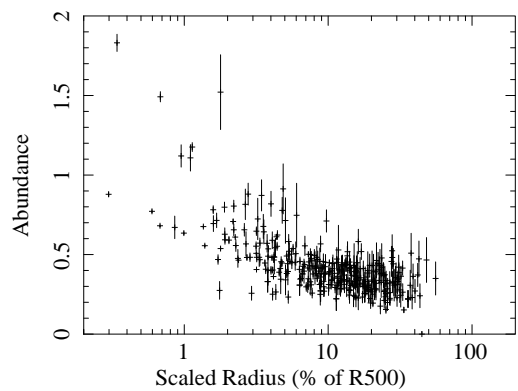}}
\caption{Abundance radial profiles for all of the analyzed 
clusters.  
\label{fig:abund}}
\end{figure}

\begin{figure}
\centering{\includegraphics[angle=0,width=8.5cm]{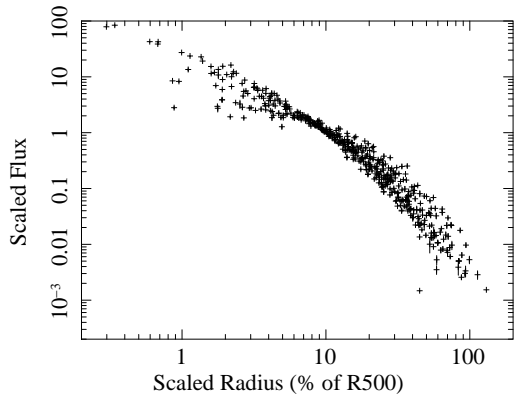}}
\caption{Scaled flux radial profiles for all of the analyzed 
clusters.  
\label{fig:flux}}
\end{figure}

While the detailed scientific analysis and discussion of these results
are deferred to Paper~II, a few aspects are clear from plots of the
entire data set.  For the combined plots, the radii of the annuli 
have been scaled to the $R_{500}$ value in the same manner as the 
individual plots (Sect.~\ref{sec:catalog}).  Figs.~\ref{fig:temp}, 
\ref{fig:abund}, and \ref{fig:flux} show the cumulative plots for the 
temperature, abundance, and flux, respectively.  Again, both the 
temperature and flux have also been normalized to the values in the 
range 5\% -- 30\% of $R_{500}$.  In addition, only points where the 
fitted values are three times the fitted uncertainty are plotted.

Inspection of Fig.~\ref{fig:temp} shows, as seen before 
\citep[e.g.][]{pea07,app05,vea06}, a wide variety 
of temperature profiles inside 5\% of R(500). Most of these can be 
characterized by a temperature drop in the center as has long been 
seen in cooling-flow clusters.  However our single phase analysis 
may produce results slightly different than more detailed analysis. 
Over the range from $0.05-0.2 R_{500}$ the clusters are isothermal 
to better than 5\%. Beyond $\sim0.2 R_{500}$ a significant fraction 
of the clusters (Paper~II) show temperature drops, 
but they are not all self-similar. However a significant fraction of 
the clusters are relatively isothermal out to the largest radii 
measurable.

As noted by \citet{app05}, many of the clusters show a self-similar 
surface brightness profile (Fig.~\ref{fig:flux}).  Inside of 
$\sim0.03 R_{500}$ there is significant scatter in the profile. 
With respect to the overall abundance, 
as was noted for {\it ASCA} spectra of clusters by \citet{fad01} 
and later for many {\it XMM-Newton} and {\it Chandra} spectra \citep{mea07} 
there is, in a significant fraction of the clusters,
an abundance increase in the center. However outside of the 
central $\sim0.05 R_{500}$ there is little evidence for an abundance gradient 
and all the clusters are very close to the average value of $A=0.3$ on 
the \citet{ag89} abundance scale (Fig.~\ref{fig:abund}).  Detailed 
analysis of these results will appear in Paper~II.

\begin{acknowledgements}
We would like to thank Alexey Vihklinin for providing the {\it Chandra}
cluster profiles and Richard Saxton for providing the arfgen modification 
for calculating the ``cross-talk'' between annuli.
We would also like to thank the {\it XMM-Newton} Background Working Group 
for their comments and encouragement.
The data for this work were provided through the HEASARC.
This work was made possible by NASA {\it XMM-Newton} Guest Observer 
grants and the support of the NASA {\it XMM-Newton} Guest Observer Facility 
at the Goddard Space Flight Center.
\end{acknowledgements}

\clearpage
\onecolumn


\begin{longtable}{ccccccccc}
\caption{Clusters}
\label{tbl:clusters}\\
\hline\hline
Cluster & Redshift & ObsID & Filtered & Original & 0.35-1.25 keV & 2.0-8.0 keV 
& $\chi^2_\nu$ & $\nu$ \\ 
 & &  & Exposure & Exposure & Band Scaling & Band Scaling \\ 
 & &  & (s) & (s) & (cnts s$^{-2}$ deg$^{-2}$) & (cnts s$^{-2}$ deg$^{-2}$) \\
\hline
\endfirsthead
\caption{continued.}\\
\hline\hline
Cluster & Redshift & ObsID & Filtered & Original & 0.35-1.25 keV & 2.0-8.0 keV 
& $\chi^2_\nu$ & $\nu$ \\ 
 & &  & Exposure & Exposure & Band Scaling & Band Scaling \\ 
 & &  & (s) & (s) & (cnts s$^{-2}$ deg$^{-2}$) & (cnts s$^{-2}$ deg$^{-2}$) \\
\hline
\hline
\endhead
\hline
\endfoot
\object{2A 0335+096} & 0.0329 & 0147800201 & 74890.7 & 95962.0 & 4500 & 1700 & 1.48 & 5002 \\
\object{A13} & 0.1035  & 0200270101 & 33379.5 & 33870.7 & 200 & 120 & 1.03 & 1063 \\
\object{A68} & 0.2481  & 0084230201 & 23818.7 & 29567.6 & 350 & 200 & 1.25 & 704 \\
\object{A85} & 0.0520  & 0065140101 & 12012.2 & 12524.4 & 5000 & 1800 & 2.02 & 1836 \\
\object{A133} & 0.0575 & 0144310101 & 19042.1 & 33670.5 & 2800 & 700 & 1.17 & 1306 \\
\object{A209} & 0.2116 & 0084230301 & 16847.6 & 21796.1 & 400 & 290 & 0.99 & 643 \\
\object{A262} & 0.0140 & 0109980101 & 22256.9 & 23897.0 & 1900 & 200 & 1.26 & 2067 \\
\object{A383} & 0.1874 & 0084230501 & 25444.9 & 33379.3 & 3000 & 1000 & 1.08 & 942 \\
\object{A399} & 0.0644 & 0112260101 & 10807.6 & 14297.6 & 200 & 200 & 1.03 & 809 \\
\object{A400} & 0.0220 & 0404010101 & 29906.9 & 38620.9 & 200 & 130 & 1.10 & 1431 \\
\object{A478} & 0.0808 & 0109880101 & 49696.1 & 56249.2 & 3500 & 4000 & 1.27 & 4913 \\
\object{A496} & 0.0293 & 0135120201 & 15845.2 & 29448.0 & 3500 & 1400 & 1.19 & 2625 \\
\object{A520} & 0.1946 & 0201510101 & 27227.4 & 46371.6 & 200 & 150 & 1.15 & 1140 \\
\object{A576} & 0.0420 & 0205070301 & 8752.0 & 21671.3 & 300 & 140 & 1.04 & 528 \\
\object{A665} & 0.1788 & 0109890501 & 49696.9 & 78487.2 & 500 & 300 & 1.21 & 2117 \\
\object{A773} & 0.2161 & 0084230601 & 12332.1 & 15082.8 & 350 & 300 & 1.28 & 464 \\
\object{A1060} & 0.0131 & 0206230101 & 32724.8 & 63773.6 & 400 & 200 & 1.24 & 3165 \\
\object{A1068} & 0.1471 & 0147630101 & 19188.9 & 29669.0 & 1000 & 400 & 1.03 & 852 \\
\object{A1413} & 0.1349 & 0112230501 & 23397.2 & 25922.4 & 1000 & 500 & 1.09 & 1311 \\
\object{A1589} & 0.0722 & 0149900301 & 15121.6 & 17170.7 & 150 & 50 & 1.19 & 550 \\
\object{A1650} & 0.0812 & 0093200101 & 34006.0 & 42534.0 & 1200 & 800 & 1.10 & 2470 \\
\object{A1689} & 0.1809 & 0093030101 & 34530.0 & 39169.6 & 2700 & 1900 & 1.17 & 2031 \\
\object{A1775} & 0.0754 & 0108460101 & 22065.8 & 32003.9 & 500 & 200 & 1.18 & 1042 \\
\object{A1795} & 0.0614 & 0097820101 & 35144.6 & 50011.7 & 5000 & 2000 & 1.25 & 3958 \\
\object{A1835} & 0.2490 & 0098010101 & 24849.0 & 40635.5 & 4000 & 2000 & 1.12 & 1301 \\
\object{A1835} a\footnote{Second observation.} & 0.2454 & 147330201 & 27895.1 
      & 83817.3 & $-$\footnote{Dashes indicate that the cluster was 
      not plotted.} & $-$ & 1.17 & 1465 \\
\object{A1837} & 0.0663 & 0109910101 & 46233.9 & 49031.5 & 500 & 200 & 1.14 & 1529 \\
\object{A1914} & 0.1735 & 0112230201 & 19219.6 & 25571.4 & 1000 & 700 & 1.15 & 1334 \\
\object{A1991} & 0.0642 & 0145020101 & 22620.9 & 41790.5 & 2000 & 500 & 1.33 & 1184 \\
\object{A2029} & 0.0728 & 0111270201 & 11088.8 & 17846.7 & 4000 & 3000 & 1.18 & 2195 \\
\object{A2052} & 0.0333 & 0109920101 & 28743.7 & 30397.0 & 4000 & 1000 & 1.37 & 2759 \\
\object{A2065} & 0.0728 & 0202080201 & 19205.3 & 33870.7 & 800 & 500 & 1.09 & 1444 \\
\object{A2163} & 0.2021 & 0112230601 & 10177.1 & 15766.7 & 400 & 550 & 1.16 & 760 \\
\object{A2199} & 0.0277 & 0008030201 & 14190.8 & 20051.5 & 2500 & 1000 & 1.24 & 2745 \\
\object{A2204} & 0.1512 & 0112230301 & 18367.1 & 22097.7 & 4000 & 3000 & 1.30 & 1687 \\
\object{A2218} & 0.1723 & 0112980101 & 17673.1 & 18169.1 & 300 & 200 & 1.30 & 709 \\
\object{A2256} & 0.0530 & 0141380201 & 10233.8 & 18369.3 & 300 & 200 & 1.24 & 987 \\
\object{A2319} & 0.0519 & 0302150101 & 15145.8 & 16668.9 & 650 & 650 & 2.51 & 2672 \\
\object{A2589} & 0.0417 & 0204180101 & 22934.0 & 46670.6 & 400 & 200 & 1.10 & 1574 \\
\object{A2597} & 0.0804 & 0147330101 & 46726.9 & 104451.1 & 5000 & 2000 & 1.13 & 2388 \\
\object{A2626} & 0.0549 & 0148310101 & 38306.4 & 41119.6 & 1000 & 400 & 1.11 & 1746 \\
\object{A2667} & 0.2205 & 0148990101 & 17682.8 & 30914.4 & 3500 & 1500 & 1.21 & 863 \\
\object{A2717} & 0.0510 & 0145020201 & 47414.8 & 54010.5 & 500 & 100 & 1.20 & 1668 \\
\object{A3112} & 0.0723 & 0105660101 & 22271.5 & 23247.0 & 6000 & 2500 & 1.24 & 1919 \\
\object{A3158} & 0.0609 & 0300210201 & 19076.9 & 22149.7 & 400 & 300 & 1.37 & 1755 \\
\object{A3526} & 0.0054 & 0046340101 & 43699.3 & 47182.9 & 5000 & 1000 & 2.32 & 5031 \\
\object{A3558} & 0.0459 & 0107260101 & 40643.4 & 44026.4 & 600 & 400 & 1.23 & 3601 \\
\object{A3560} & 0.0429 & 0205450201 & 27009.7 & 45271.8 & 150 & 70 & 1.10 & 1261 \\
\object{A3581} & 0.0225 & 0205990101 & 33930.6 & 43670.2 & 2400 & 500 & 1.40 & 2206 \\
\object{A3827} & 0.0959 & 0149670101 & 21024.9 & 24970.8 & 500 & 400 & 1.23 & 1437 \\
\object{A3888} & 0.1537 & 0201903101 & 23250.1 & 30469.8 & 500 & 400 & 1.08 & 1564 \\
\object{A3911} & 0.0958 & 0149670301 & 22883.3 & 27269.2 & 200 & 150 & 1.02 & 1236 \\
\object{A3921} & 0.0919 & 0112240101 & 28488.8 & 30763.6 & 400 & 300 & 1.10 & 1467 \\
\object{A4059} & 0.0467 & 0109950201 & 22581.2 & 24398.8 & 1200 & 400 & 1.18 & 1952 \\
\object{AWM 7}        & 0.0155 & 0135950301 & 29418.7 & 31621.6 & 1000 & 500 & 1.35 & 3793 \\
\object{Coma}         & 0.0218 & 0124711401 & 16195.8 & 23598.0 & 500 & 300 & 1.15 & 3692 \\
\object{E1455+2232}   & 0.2583 & 0108670201 & 33785.9 & 39993.7 & 1600 & 1800 & 1.25 & 1100 \\
\object{EXO0422}      & 0.0336 & 0300210401 & 38373.1 & 41070.2 & 3000 & 1500 & 1.26 & 2402 \\
\object{Hydra}        & 0.0521 & 0109980301 & 17944.5 & 31546.3 & 4000 & 2000 & 1.29 & 2180 \\
\object{Klemola 44}   & 0.0286 & 0204460101 & 29668.9 & 29669.0 & 1200 & 500 & 1.28 & 2757 \\
\object{M87}          & 0.0042 & 0114120101 & 35931.1 & 39551.8 & 15000 & 3000 & 3.71 & 5401 \\
\object{MKW 3S}       & 0.0417 & 0109930101 & 33244.3 & 51038.7 & 1900 & 800 & 1.24 & 2699 \\
\object{MKW 4}        & 0.0214 & 0093060101 & 13585.2 & 15368.8 & 1000 & 200 & 1.26 & 673 \\
\object{Perseus}      & 0.0148 & 0305780101 & 101982.0 & 124869.2 & 15000 & 15000 & 3.63 & 10095 \\
\object{Perseus} a\footnote{Second observation, under the name Abell 426 in the archive.}     
                      & 0.0147 & 0085110101 & 47272.9 & 53646.5 & $-$ & $-$ & 2.48 & 8982 \\
\object{PKS 0745-19}  & 0.0986 & 0105870101 & 18043.2 & 26946.8 & 3000 & 5000 & 1.24 & 1872 \\
\object{RXCJ0605.8-3518} & 0.1367 & 201901001 & 17798.1 & 26668.6 & 3000 & 1500 & 1.30 & 904 \\
\object{RXJ0658-55}   & 0.3069 & 0112980201 & 21464.2 & 42770.6 & 600 & 400 & 1.22 & 961 \\
\object{RXJ1347-1145} & 0.4477 & 0112960101 & 30122.7 & 38121.7 & 4000 & 3000 & 1.29 & 1290 \\
\object{S\'ersic 159-3} & 0.0563 & 0147800101 & 81339.4 & 122209.4 & 3500 & 1000 & 1.51 & 3488 \\
\object{S\'ersic 159-3} a\footnote{Second observation, under the name AS 1101 in the archive.}      
                      & 0.0564 & 0123900101 & 30643.1 & 60996.9 & $-$ & $-$ & 1.14 & 1982 \\
\object{Triangulum}   & 0.0478 & 0093620101 & 9168.9 & 14497.8 & 500 & 600 & 1.02 & 1890 \\
\object{ZW3146}       & 0.2817 & 0108670101 & 51450.6 & 53597.1 & 4000 & 2000 & 1.36 & 1802 \\
\end{longtable}

\Online

\clearpage

\clearpage

\clearpage

\clearpage

\clearpage

\clearpage

\clearpage

\clearpage

\clearpage

\clearpage

\clearpage

\clearpage

\clearpage

\clearpage

\clearpage

\clearpage

\clearpage

\clearpage

\clearpage

\clearpage

\clearpage

\clearpage

\clearpage
\onecolumn

\begin{longtable}{ccccccccc}
\caption{Cluster Details}
\label{tbl:clusterdetails}\\
\hline\hline
Cluster & Annulus & $T$ & $\sigma_{T}$ & $A$ & $\sigma_{A}$ & $F$ & $\sigma_{F}$ \\ 
 & & (keV) & (keV) & & & ergs cm$^{-2}$ s$^{-1}$ arcmin$^{-2}$ & ergs cm$^{-2}$ s$^{-1}$ arcmin$^{-2}$ \\
\hline
\endfirsthead
\caption{continued.}\\
\hline\hline
Cluster & Annulus & $T$ & $\sigma_{T}$ & $A$ & $\sigma_{A}$ & $F$ & $\sigma_{F}$ \\ 
 & & (keV) & (keV) & & & ergs cm$^{-2}$ s$^{-1}$ arcmin$^{-2}$ & ergs cm$^{-2}$ s$^{-1}$ arcmin$^{-2}$ \\
\hline
\endhead
\hline
\endfoot
2A 0335+096 & 1 & 1.542 & 0.008 & 0.371 & 0.011 & 1.876E-11 & 0.034E-11 \\
- & 2 & 2.383 & 0.011 & 0.708 & 0.015 & 8.398E-12 & 0.079E-12 \\
- & 3 & 3.011 & 0.014 & 0.612 & 0.013 & 2.607E-12 & 0.024E-12 \\
- & 4 & 3.225 & 0.020 & 0.502 & 0.016 & 9.532E-13 & 0.121E-13 \\
- & 5 & 3.233 & 0.025 & 0.449 & 0.018 & 4.965E-13 & 0.076E-13 \\
- & 6 & 3.257 & 0.033 & 0.447 & 0.024 & 2.619E-13 & 0.053E-13 \\
- & 7 & 3.072 & 0.041 & 0.321 & 0.021 & 1.557E-13 & 0.033E-13 \\
- & 8 & 3.019 & 0.053 & 0.381 & 0.031 & 6.525E-14 & 0.199E-14 \\
- & 9 & 2.826 & 0.078 & 0.346 & 0.037 & 3.062E-14 & 0.127E-14 \\
- & 10 & 2.721 & 0.133 & 0.221 & 0.063 & 1.871E-14 & 0.153E-14 \\
Abell 13  & 1 &  5.184 & 0.348 & 0.807 & 0.336 & 5.463e-13 & 0.764e-13 \\
- & 2 & 5.541 & 0.391 & 0.747 & 0.201 & 3.992e-13 & 0.348e-13 \\
- & 3 & 4.682 & 0.182 & 0.326 & 0.078 & 1.980e-13 & 0.092e-13 \\
- & 4 & 4.985 & 0.219 & 0.266 & 0.063 & 1.043e-13 & 0.044e-13 \\
- & 5 & 4.152 & 0.225 &   -\footnote{Dashes in the abundance column 
signifies that the abundance for the given annulus has been 
linked to the abundance above.}   &   -   & 4.627e-14 & 0.248e-14 \\
- & 6 & 4.174 & 0.273 &   -   &   -   & 3.081e-14 & 0.177e-14 \\
- & 7 & 4.150 & 0.300 & 0.466 & 0.141 & 1.595e-14 & 0.142e-14 \\
- & 8 & 3.758 & 0.493 &   -   &   -   & 8.115e-15 & 0.899e-15 \\
- & 9 & 4.191 & 1.012 &   -   &   -   & 3.035e-15 & 0.517e-15 \\
Abell 68 & 1 & 7.487 & 0.726 & 0.254 & 0.162 & 1.351E-12 & 0.087E-12 \\
- & 2 & 7.803 & 0.635 & 0.274 & 0.081 & 6.687E-13 & 0.319E-13 \\
- & 3 & 6.966 & 0.541 & - & - & 2.102E-13 & 0.103E-13 \\
- & 4 & 5.897 & 0.711 & - & - & 4.401E-14 & 0.315E-14 \\
- & 5 & 5.621 & 1.241 & - & - & 1.627E-14 & 0.170E-14 \\
- & 6 & 3.606 & 0.893 & - & - & 7.168E-15 & 1.210E-15 \\
Abell 85 & 1 & 3.618 & 0.084 & 1.161 & 0.089 & 1.043E-11 & 0.030E-11 \\
- & 2 & 5.089 & 0.145 & 0.567 & 0.077 & 3.921E-12 & 0.120E-12 \\
- & 3 & 5.321 & 0.103 & 0.521 & 0.053 & 1.946E-12 & 0.046E-12 \\
- & 4 & 5.972 & 0.185 & 0.379 & 0.062 & 1.043E-12 & 0.029E-12 \\
- & 5 & 6.269 & 0.195 & 0.428 & 0.069 & 6.139E-13 & 0.185E-13 \\
- & 6 & 6.246 & 0.273 & 0.356 & 0.095 & 3.046E-13 & 0.126E-13 \\
- & 7 & 6.305 & 0.157 & 0.319 & 0.047 & 2.548E-13 & 0.058E-13 \\
- & 8 & 6.047 & 0.292 & 0.322 & 0.080 & 1.005E-13 & 0.039E-13 \\
- & 9 & 5.679 & 0.404 &   -   &   -   & 5.382E-14 & 0.252E-14 \\
- & 10 & 7.177 & 1.115 & - & - & 3.768E-14 & 2.496E-15 \\
Abell 133 & 1 & 2.376 & 0.044 & 1.056 & 0.070 & 4.090E-12 & 0.169E-12 \\
- & 2 & 3.848 & 0.100 & 0.872 & 0.098 & 1.338E-12 & 0.066E-12 \\
- & 3 & 4.300 & 0.087 & 0.555 & 0.061 & 5.686E-13 & 0.196E-13 \\
- & 4 & 4.222 & 0.107 & 0.472 & 0.072 & 2.436E-13 & 0.102E-13 \\
- & 5 & 4.159 & 0.137 & 0.338 & 0.081 & 1.322E-13 & 0.066E-13 \\
- & 6 & 3.997 & 0.176 & 0.397 & 0.105 & 7.203E-14 & 0.471E-14 \\
- & 7 & 3.738 & 0.180 & 0.457 & 0.095 & 4.473E-14 & 0.286E-14 \\
- & 8 & 3.940 & 0.337 & 0.423 & 0.171 & 1.829E-14 & 0.209E-14 \\
- & 9 & 2.658 & 0.296 & 0.381 & 0.172 & 6.742E-15 & 1.381E-15 \\
- & 10 & 2.314 & 0.640 & - & - & 3.300E-15 & 1.160E-15 \\
Abell 209 & 1 & 6.946 & 0.855 & 0.778 & 0.486 & 1.608E-12 & 0.260E-12 \\
- & 2 & 6.766 & 0.422 & 0.336 & 0.072 & 7.642E-13 & 0.384E-13 \\
- & 3 & 7.303 & 0.470 & - & - & 3.035E-13 & 0.147E-13 \\
- & 4 & 7.993 & 0.668 & - & - & 1.053E-13 & 0.056E-13 \\
- & 5 & 6.892 & 0.915 & 0.233 & 0.164 & 5.042E-14 & 0.421E-14 \\
- & 6 & 5.603 & 0.821 & - & - & 2.559E-14 & 0.253E-14 \\
- & 7 & 4.867 & 0.776 & - & - & 1.127E-14 & 0.135E-14 \\
Abell 262 & 1 & 1.270 & 0.010 & 0.495 & 0.030 & 2.636E-12 & 0.143E-12 \\
- & 2 & 1.889 & 0.032 & 1.108 & 0.083 & 1.182E-12 & 0.080E-12 \\
- & 3 & 2.121 & 0.022 & 0.656 & 0.037 & 5.865E-13 & 0.220E-13 \\
- & 4 & 2.140 & 0.025 & 0.408 & 0.028 & 3.609E-13 & 0.126E-13 \\
- & 5 & 2.141 & 0.027 & 0.381 & 0.027 & 2.596E-13 & 0.092E-13 \\
- & 6 & 2.161 & 0.034 & 0.380 & 0.030 & 1.783E-13 & 0.070E-13 \\
- & 7 & 2.198 & 0.040 & 0.359 & 0.024 & 1.380E-13 & 0.046E-13 \\
- & 8 & 2.102 & 0.034 & 0.322 & 0.027 & 7.796E-14 & 0.311E-14 \\
- & 9 & 1.982 & 0.041 & 0.306 & 0.027 & 4.649E-14 & 0.220E-14 \\
- & 10 & 1.983 & 0.072 & 0.299 & 0.046 & 3.336E-14 & 0.270E-14 \\
Abell 383 & 1 & 3.643 & 0.078 & 0.746 & 0.062 & 3.913E-12 & 0.132E-12 \\
- & 2 & 5.573 & 0.374 & 0.363 & 0.116 & 6.474E-13 & 0.352E-13 \\
- & 3 & 4.812 & 0.251 & 0.325 & 0.095 & 1.463E-13 & 0.084E-13 \\
- & 4 & 4.449 & 0.543 & - & - & 2.879E-14 & 0.236E-14 \\
- & 5 & 3.615 & 0.628 & - & - & 1.149E-14 & 0.135E-14 \\
- & 6 & 3.717 & 0.976 & - & - & 6.276E-15 & 0.983E-15 \\
- & 7 & 3.678 & 1.145 & - & - & 2.919E-15 & 0.645E-15 \\
- & 8 & 0.000 & 0.000 & - & - & 2.322E-15 & 0.578E-15 \\
Abell 399 & 1 & 9.840 & 1.980 & 0.465 & 0.097 & 1.277E-12 & 0.088E-12 \\
- & 2 & 6.665 & 0.525 & - & - & 9.888E-13 & 0.527E-13 \\
- & 3 & 6.704 & 0.279 & - & - & 7.264E-13 & 0.333E-13 \\
- & 4 & 7.509 & 0.451 & 0.287 & 0.103 & 4.289E-13 & 0.199E-13 \\
- & 5 & 6.425 & 0.332 & 0.343 & 0.123 & 2.669E-13 & 0.145E-13 \\
- & 6 & 6.445 & 0.375 & 0.330 & 0.093 & 1.680E-13 & 0.082E-13 \\
- & 7 & 6.258 & 0.333 & - & - & 1.174E-13 & 0.055E-13 \\
- & 8 & 6.520 & 0.464 & 0.154 & 0.135 & 6.528E-14 & 0.413E-14 \\
- & 9 & 7.380 & 0.854 & - & - & 3.345E-14 & 0.230E-14 \\
Abell 400 & 1 & 2.303 & 0.259 & 0.551 & 0.137 & 2.112e-13 & 0.353e-13 \\
- & 2 & 2.326 & 0.153 & - & - & 1.851e-13 & 0.267e-13 \\
- & 3 & 2.340 & 0.071 & 0.677 & 0.076 & 1.792e-13 & 0.134e-13 \\ 
- & 4 & 2.126 & 0.044 & 0.452 & 0.051 & 1.273e-13 & 0.082e-13 \\
- & 5 & 2.122 & 0.046 & 0.414 & 0.047 & 1.012e-13 & 0.063e-13 \\
- & 6 & 2.142 & 0.050 & 0.398 & 0.050 & 7.716e-14 & 0.600e-14 \\
- & 7 & 1.906 & 0.051 & 0.277 & 0.030 & 5.831e-14 & 0.338e-14 \\
- & 8 & 2.067 & 0.060 & 0.384 & 0.052 & 3.667e-14 & 0.274e-14 \\
- & 9 & 1.989 & 0.073 & 0.373 & 0.053 & 2.116e-14 & 0.193e-14 \\
- & 10 & 2.027 & 0.110 & - & - & 1.575e-14 & 1.770e-15 \\
Abell 478 & 1 & 4.499 & 0.049 & 0.551 & 0.019 & 2.252E-11 & 0.016E-11 \\
- & 2 & 6.086 & 0.085 & 0.403 & 0.020 & 6.245E-12 & 0.055E-12 \\
- & 3 & 6.290 & 0.069 & 0.383 & 0.016 & 2.400E-12 & 0.022E-12 \\
- & 4 & 6.610 & 0.108 & 0.335 & 0.023 & 7.404E-13 & 0.108E-13 \\
- & 5 & 6.497 & 0.138 & 0.307 & 0.028 & 3.604E-13 & 0.067E-13 \\
- & 6 & 6.508 & 0.201 & 0.276 & 0.040 & 1.766E-13 & 0.047E-13 \\
- & 7 & 6.124 & 0.217 & 0.415 & 0.045 & 8.236E-14 & 0.238E-14 \\
- & 8 & 8.367 & 0.533 & - & - & 3.657E-14 & 0.125E-14 \\
Abell 496 & 1 & 2.479 & 0.032 & 1.038 & 0.055 & 9.749E-12 & 0.240E-12 \\
- & 2 & 3.176 & 0.048 & 0.715 & 0.046 & 3.833E-12 & 0.100E-12 \\
- & 3 & 3.682 & 0.050 & 0.572 & 0.032 & 1.821E-12 & 0.036E-12 \\
- & 4 & 4.135 & 0.063 & 0.449 & 0.038 & 8.506E-13 & 0.195E-13 \\
- & 5 & 4.233 & 0.070 & 0.428 & 0.042 & 5.433E-13 & 0.138E-13 \\
- & 6 & 4.536 & 0.122 & 0.290 & 0.047 & 3.372E-13 & 0.094E-13 \\
- & 7 & 4.185 & 0.080 & 0.333 & 0.042 & 2.164E-13 & 0.058E-13 \\
- & 8 & 4.130 & 0.111 & 0.300 & 0.041 & 1.146E-13 & 0.036E-13 \\
- & 9 & 4.020 & 0.144 & - & - & 5.824E-14 & 0.231E-14 \\
- & 10 & 4.289 & 0.305 & - & - & 3.519E-14 & 0.225E-14 \\
Abell 520 & 1 & 9.176 & 1.306 & 0.210 & 0.121 & 7.735E-13 & 0.470E-13 \\
- & 2 & 7.264 & 0.593 & - & - & 6.521E-13 & 0.352E-13 \\
- & 3 & 8.960 & 0.509 & 0.285 & 0.054 & 4.198E-13 & 0.150E-13 \\
- & 4 & 8.576 & 0.744 & - & - & 1.357E-13 & 0.056E-13 \\
- & 5 & 6.629 & 0.516 & - & - & 5.353E-14 & 0.272E-14 \\
- & 6 & 10.682 & 1.638 & - & - & 3.200E-14 & 0.199E-15 \\
- & 7 & 4.873 & 0.709 & - & - & 1.326E-14 & 0.108E-14 \\
- & 8 & 6.524 & 2.797 & - & - & 4.687E-15 & 0.856E-15 \\
Abell 576 & 1 & 3.270 & 0.411 & 0.838 & 0.536 & 9.522E-13 & 2.921E-13 \\
- & 2 & 3.782 & 0.364 & 0.712 & 0.320 & 6.781E-13 & 1.163E-13 \\
- & 3 & 3.804 & 0.165 & 0.714 & 0.143 & 4.255E-13 & 0.363E-13 \\
- & 4 & 4.068 & 0.192 & 0.323 & 0.122 & 2.755E-13 & 0.218E-13 \\
- & 5 & 4.262 & 0.188 & 0.529 & 0.155 & 2.270E-13 & 0.198E-13 \\
- & 6 & 3.962 & 0.212 & 0.430 & 0.139 & 1.335E-13 & 0.120E-13 \\
- & 7 & 3.322 & 0.137 & 0.432 & 0.091 & 7.780E-14 & 0.577E-14 \\
- & 8 & 3.054 & 0.216 & 0.261 & 0.081 & 3.749E-14 & 0.352E-14 \\
- & 9 & 3.212 & 0.304 & - & - & 2.039E-14 & 0.241E-14 \\
- & 10 & 2.759 & 0.681 & - & - & 9.953E-15 & 2.048E-15 \\
Abell 665 & 1 & 7.240 & 0.561 & 0.112 & 0.128 & 1.017E-12 & 0.051E-12 \\
- & 2 & 7.810 & 0.291 & 0.298 & 0.062 & 9.722E-13 & 0.265E-13 \\
- & 3 & 7.998 & 0.191 & 0.282 & 0.036 & 5.992E-13 & 0.111E-13 \\
- & 4 & 8.114 & 0.277 & 0.388 & 0.061 & 2.001E-13 & 0.055E-13 \\
- & 5 & 7.881 & 0.367 & 0.241 & 0.075 & 9.623E-14 & 0.320E-14 \\
- & 6 & 6.475 & 0.327 & 0.350 & 0.104 & 5.830E-14 & 0.275E-14 \\
- & 7 & 4.689 & 0.331 & 0.187 & 0.066 & 1.847E-14 & 0.103E-14 \\
- & 8 & 2.222 & 0.281 & - & - & 4.497E-15 & 0.702E-15 \\
- & 9 & 1.540 & 0.198 & - & - & 1.300E-15 & 0.439E-15 \\
- & 10 & 1.160 & 0.135 & - & - & 1.309E-15 & 0.593E-15 \\
Abell 773 & 1 & 10.816 & 1.210 & 0.525 & 0.173 & 1.888E-12 & 0.139E-12 \\
- & 2 & 6.933 & 0.725 & - & - & 8.006E-13 & 0.067E-13 \\
- & 3 & 8.095 & 0.651 & 0.299 & 0.144 & 2.624E-13 & 0.202E-13 \\
- & 4 & 7.505 & 0.907 & - & - & 8.074E-14 & 0.687E-14 \\
- & 5 & 7.463 & 1.243 & 0.238 & 0.277 & 3.993E-14 & 0.487E-14 \\
- & 6 & 5.162 & 1.127 & - & - & 1.194E-14 & 0.209E-14 \\
- & 7 & 3.538 & 0.959 & - & - & 5.400E-15 & 1.328E-15 \\
Abell 1060 & 1 & 3.081 & 0.108 & 0.759 & 0.110 & 1.194E-12 & 0.084E-12 \\
- & 2 & 3.374 & 0.071 & 0.671 & 0.071 & 9.606E-13 & 0.429E-13 \\
- & 3 & 3.330 & 0.040 & 0.469 & 0.030 & 7.914E-13 & 0.177E-13 \\
- & 4 & 3.291 & 0.037 & 0.483 & 0.027 & 5.704E-13 & 0.121E-13 \\
- & 5 & 3.250 & 0.039 & 0.401 & 0.025 & 4.243E-13 & 0.090E-14 \\
- & 6 & 3.220 & 0.042 & 0.491 & 0.030 & 3.336E-13 & 0.080E-14 \\
- & 7 & 3.158 & 0.035 & 0.437 & 0.022 & 2.361E-13 & 0.045E-14 \\
- & 8 & 3.012 & 0.043 & 0.415 & 0.024 & 1.573E-13 & 0.035E-14 \\
- & 9 & 3.011 & 0.051 & 0.404 & 0.026 & 9.503E-14 & 0.234E-14 \\
- & 10 & 3.087 & 0.101 & 0.419 & 0.054 & 6.314E-14 & 0.300E-14 \\
Abell 1068 & 1 & 3.123 & 0.055 & 0.631 & 0.048 & 5.377E-12 & 0.148E-12 \\
- & 2 & 4.691 & 0.197 & 0.384 & 0.053 & 1.078E-12 & 0.038E-12 \\
- & 3 & 4.743 & 0.201 & - & - & 2.483E-13 & 0.094E-13 \\
- & 4 & 5.366 & 0.352 & - & - & 6.827E-14 & 0.335E-14 \\
- & 5 & 5.163 & 0.543 & - & - & 2.801E-14 & 0.183E-14 \\
- & 6 & 5.287 & 0.867 & - & - & 1.384E-14 & 0.129E-14 \\
- & 7 & 2.220 & 0.562 & - & - & 2.919E-15 & 0.557E-15 \\
Abell 1413 & 1 & 7.682 & 0.332 & 0.409 & 0.079 & 4.639E-12 & 0.012E-12 \\
- & 2 & 7.995 & 0.327 & 0.396 & 0.076 & 1.899E-12 & 0.056E-12 \\
- & 3 & 6.898 & 0.269 & 0.492 & 0.061 & 6.280E-13 & 0.176E-13 \\
- & 4 & 7.013 & 0.403 & 0.338 & 0.075 & 2.006E-13 & 0.071E-13 \\
- & 5 & 6.806 & 0.366 & - & - & 8.671E-14 & 0.345E-14 \\
- & 6 & 6.423 & 0.530 & 0.163 & 0.104 & 4.190E-14 & 0.244E-14 \\
- & 7 & 5.650 & 0.615 & - & - & 2.249E-14 & 0.146E-14 \\
- & 8 & 3.559 & 0.528 & - & - & 8.913E-15 & 1.024E-15 \\
- & 9 & 5.015 & 2.030 & - & - & 2.495E-15 & 0.755E-15 \\
Abell 1589 & 1 & 2.634 & 2.668 & 0.276 & 0.146 & 2.234E-13 & 1.104E-13 \\
- & 2 & 5.080 & 0.780 & - & - & 2.203E-13 & 0.259E-13 \\
- & 3 & 4.676 & 0.308 & - & - & 1.645E-13 & 0.147E-13 \\
- & 4 & 5.614 & 0.463 & 0.281 & 0.115 & 1.150E-13 & 0.080E-13 \\
- & 5 & 4.443 & 0.339 & - & - & 7.515E-14 & 0.602E-14 \\
- & 6 & 3.326 & 0.203 & 0.317 & 0.079 & 4.498E-14 & 0.391E-14 \\
- & 7 & 3.583 & 0.291 & - & - & 2.948E-14 & 0.245E-14 \\
- & 8 & 3.232 & 0.445 & - & - & 9.579E-15 & 1.204E-15 \\
- & 9 & 1.588 & 0.204 & - & - & 1.953E-15 & 0.719E-15 \\
Abell 1650 & 1 & 5.145 & 0.114 & 0.654 & 0.069 & 4.832E-12 & 0.122E-12 \\
- & 2 & 6.064 & 0.158 & 0.543 & 0.059 & 2.196E-12 & 0.053E-12 \\
- & 3 & 5.812 & 0.126 & 0.396 & 0.040 & 9.004E-13 & 0.173E-13 \\
- & 4 & 5.951 & 0.163 & 0.306 & 0.049 & 3.821E-13 & 0.090E-13 \\
- & 5 & 5.401 & 0.141 & 0.332 & 0.061 & 1.828E-13 & 0.056E-13 \\
- & 6 & 5.235 & 0.206 & 0.199 & 0.051 & 1.017E-13 & 0.033E-13 \\
- & 7 & 4.967 & 0.244 & - & - & 4.088E-14 & 0.149E-14 \\
- & 8 & 3.938 & 0.337 & - & - & 1.606E-14 & 0.100E-14 \\
- & 9 & 1.991 & 0.219 & - & - & 4.182E-15 & 0.642E-15 \\
Abell 1689 & 1 & 9.435 & 0.280 & 0.437 & 0.048 & 9.087E-12 & 0.117E-12 \\
- & 2 & 11.619 & 0.593 & 0.258 & 0.031 & 2.505E-12 & 0.047E-12 \\
- & 3 & 9.927 & 0.309 & - & - & 7.202E-13 & 0.122E-13 \\
- & 4 & 8.234 & 0.360 & - & - & 1.702E-13 & 0.039E-13 \\
- & 5 & 9.743 & 0.780 & - & - & 7.142E-14 & 0.216E-14 \\
- & 6 & 7.357 & 0.913 & - & - & 2.785E-14 & 0.127E-14 \\
- & 7 & 7.804 & 1.481 & - & - & 1.113E-14 & 0.078E-14 \\
Abell 1775 & 1 & 4.341 & 0.382 & 1.687 & 0.517 & 7.223E-13 & 1.356E-13 \\
- & 2 & 3.716 & 0.155 & 0.913 & 0.157 & 7.036E-13 & 0.563E-13 \\
- & 3 & 3.862 & 0.089 & 0.711 & 0.070 & 3.867E-13 & 0.165E-13 \\
- & 4 & 3.660 & 0.116 & 0.582 & 0.077 & 1.647E-13 & 0.085E-13 \\
- & 5 & 3.413 & 0.116 & 0.404 & 0.083 & 8.308E-14 & 0.522E-14 \\
- & 6 & 3.896 & 0.221 & 0.315 & 0.120 & 5.016E-14 & 0.403E-14 \\
- & 7 & 3.276 & 0.166 & 0.363 & 0.078 & 2.834E-14 & 0.196E-14 \\
- & 8 & 3.201 & 0.268 & - & - & 1.386E-14 & 0.119E-14 \\
- & 9 & 3.546 & 0.744 & - & - & 5.478E-15 & 0.793E-15 \\
- & 10 & 3.217 & 0.880 & - & - & 4.229E-15 & 0.947E-15 \\
Abell 1795 & 1 & 3.881 & 0.037 & 0.702 & 0.027 & 1.328E-11 & 0.011E-11 \\
- & 2 & 4.707 & 0.054 & 0.553 & 0.027 & 6.189E-12 & 0.064E-12 \\
- & 3 & 5.742 & 0.070 & 0.407 & 0.022 & 2.267E-12 & 0.022E-12 \\
- & 4 & 5.958 & 0.092 & 0.338 & 0.028 & 8.825E-13 & 0.115E-13 \\
- & 5 & 6.124 & 0.112 & 0.316 & 0.021 & 4.640E-13 & 0.056E-13 \\
- & 6 & 6.060 & 0.141 & - & - & 2.574E-13 & 0.036E-13 \\
- & 7 & 5.805 & 0.150 & - & - & 1.394E-13 & 0.020E-13 \\
- & 8 & 5.744 & 0.248 & - & - & 5.622E-14 & 0.121E-14 \\
- & 9 & 4.880 & 0.292 & - & - & 2.091E-14 & 0.086E-14 \\
- & 10 & 4.426 & 0.693 & - & - & 1.115E-14 & 0.099E-14 \\
Abell 1835 & 1 & 6.062 & 0.121 & 0.457 & 0.034 & 1.099E-11 & 0.012E-11 \\
- & 2 & 10.524 & 0.495 & 0.284 & 0.046 & 1.743E-12 & 0.383E-12 \\
- & 3 & 9.123 & 0.497 & - & - & 4.054E-13 & 0.101E-13 \\
- & 4 & 8.632 & 0.793 & - & - & 8.954E-14 & 0.390E-14 \\
- & 5 & 7.667 & 0.741 & - & - & 4.165E-14 & 0.177E-14 \\
- & 6 & 8.092 & 1.650 & - & - & 1.544E-14 & 0.126E-14 \\
- & 7 & 4.969 & 1.125 & - & - & 6.206E-15 & 0.763E-15 \\
Abell 1835 a\footnote{Second observation.} & 1 & 5.896 & 0.114 
& 0.513 & 0.034 & 1.078E-11 & 0.011E-11 \\
- & 2 & 10.024 & 0.466 & 0.364 & 0.082 & 1.682E-12 & 0.049E-12 \\
- & 3 & 9.790 & 0.456 & 0.266 & 0.062 & 4.036E-13 & 0.110E-13 \\
- & 4 & 8.517 & 0.630 & - & - & 7.844E-14 & 0.313E-14 \\
- & 5 & 7.891 & 0.958 & - & - & 3.176E-14 & 0.170E-14 \\
- & 6 & 6.411 & 1.301 & - & - & 1.120E-14 & 0.110E-14 \\
- & 7 & 7.612 & 2.568 & - & - & 6.764E-15 & 0.828E-15 \\
- & 8 & 3.138 & 0.926 & - & - & 2.740E-15 & 0.613E-15 \\
- & 9 & 3.325 & 1.748 & - & - & 1.290E-15 & 0.554E-15 \\
Abell 1837 & 1 & 3.959 & 0.256 & 1.268 & 0.386 & 5.620E-13 & 0.921E-13 \\
- & 2 & 4.110 & 0.143 & 0.484 & 0.106 & 4.097E-13 & 0.253E-13 \\
- & 3 & 3.855 & 0.088 & 0.469 & 0.053 & 2.367E-13 & 0.089E-13 \\
- & 4 & 3.860 & 0.112 & 0.453 & 0.068 & 1.071E-13 & 0.050E-13 \\
- & 5 & 3.393 & 0.122 & 0.196 & 0.035 & 5.081E-14 & 0.218E-14 \\
- & 6 & 3.286 & 0.160 & - & - & 2.941E-14 & 0.129E-14 \\
- & 7 & 2.678 & 0.122 & - & - & 1.616E-14 & 0.079E-14 \\
- & 8 & 2.893 & 0.320 & - & - & 6.978E-15 & 0.554E-15 \\
- & 9 & 1.621 & 0.099 & - & - & 2.805E-15 & 0.382E-15 \\
Abell 1914 & 1 & 11.219 & 1.021 & 0.614 & 0.126 & 4.565E-12 & 0.188E-12 \\  
- & 2 & 12.007 & 0.728 & 0.329 & 0.083 & 3.317E-12 & 0.114E-12 \\
- & 3 & 10.327 & 0.422 & 0.242 & 0.070 & 7.948E-13 & 0.248E-13 \\
- & 4 & 8.967 & 0.712 & 0.336 & 0.117 & 1.845E-13 & 0.090E-13 \\
- & 5 & 9.057 & 1.015 & 0.326 & 0.194 & 8.221E-14 & 0.627E-14 \\
- & 6 & 8.996 & 1.556 & 0.280 & 0.238 & 3.770E-14 & 0.364E-14 \\
- & 7 & 8.053 & 1.404 & - & - & 1.673E-14 & 0.175E-14 \\
Abell 1991 & 1 & 1.785 & 0.033 & 0.654 & 0.043 & 3.296E-12 & 0.143E-12 \\
- & 2 & 2.611 & 0.057 & 0.778 & 0.082 & 9.264E-13 & 0.568E-13 \\
- & 3 & 2.795 & 0.075 & 0.451 & 0.050 & 2.954E-13 & 0.139E-13 \\
- & 4 & 2.603 & 0.087 & 0.310 & 0.044 & 9.172E-14 & 0.505E-14 \\
- & 5 & 2.388 & 0.106 & - & - & 5.005E-14 & 0.321E-14 \\
- & 6 & 2.377 & 0.183 & 0.308 & 0.100 & 2.319E-14 & 0.301E-14 \\
- & 7 & 2.262 & 0.235 & 0.230 & 0.072 & 1.140E-14 & 0.146E-14 \\
- & 8 & 1.573 & 0.126 & - & - & 4.611E-15 & 1.003E-15 \\
Abell 2029 & 1 & 5.585 & 0.134 & 0.951 & 0.065 & 1.900E-11 & 0.027E-11 \\
- & 2 & 7.273 & 0.224 & 0.483 & 0.058 & 8.060E-12 & 0.134E-12 \\
- & 3 & 7.827 & 0.189 & 0.377 & 0.044 & 2.905E-12 & 0.047E-12 \\
- & 4 & 7.558 & 0.250 & 0.382 & 0.061 & 1.029E-12 & 0.025E-12 \\
- & 5 & 7.695 & 0.322 & 0.367 & 0.064 & 5.066E-13 & 0.149E-13 \\
- & 6 & 8.187 & 0.403 & - & - & 2.837E-13 & 0.084E-13 \\
- & 7 & 7.531 & 0.429 & - & - & 1.515E-13 & 0.064E-13 \\
- & 8 & 10.599 & 1.042 & - & - & 6.345E-14 & 0.342E-14 \\
Abell 2052 & 1 & 1.968 & 0.021 & 0.701 & 0.031 & 5.517E-12 & 0.127E-12 \\
- & 2 & 2.839 & 0.038 & 0.805 & 0.037 & 2.871E-12 & 0.067E-12 \\
- & 3 & 3.105 & 0.033 & 0.552 & 0.026 & 1.085E-12 & 0.021E-12 \\
- & 4 & 3.254 & 0.041 & 0.454 & 0.029 & 4.902E-13 & 0.113E-13 \\
- & 5 & 3.230 & 0.049 & 0.393 & 0.032 & 2.827E-13 & 0.075E-13 \\
- & 6 & 3.053 & 0.058 & 0.442 & 0.038 & 1.698E-13 & 0.055E-13 \\
- & 7 & 2.913 & 0.061 & 0.341 & 0.031 & 9.754E-14 & 0.300E-14 \\
- & 8 & 3.027 & 0.103 & 0.420 & 0.057 & 4.174E-14 & 0.217E-14 \\
- & 9 & 2.633 & 0.119 & 0.421 & 0.076 & 1.733E-14 & 0.151E-14 \\
- & 10 & 2.517 & 0.197 & - & - & 1.146E-14 & 0.136E-14 \\
Abell 2065 & 1 & 4.236 & 0.123 & 0.829 & 0.124 & 2.491E-12 & 1.383E-13 \\
- & 2 & 5.182 & 0.190 & 0.402 & 0.096 & 1.306E-12 & 0.058E-12 \\
- & 3 & 5.704 & 0.176 & 0.461 & 0.059 & 7.381E-13 & 0.214E-13 \\
- & 4 & 5.906 & 0.210 & 0.343 & 0.065 & 3.649E-13 & 0.118E-13 \\
- & 5 & 5.823 & 0.242 & 0.309 & 0.080 & 2.276E-13 & 0.086E-13 \\
- & 6 & 5.374 & 0.188 & 0.430 & 0.103 & 1.441E-13 & 0.070E-13 \\
- & 7 & 4.880 & 0.192 & 0.331 & 0.066 & 9.532E-14 & 0.377E-14 \\
- & 8 & 4.719 & 0.339 & - & - & 4.190E-14 & 0.215E-14 \\
- & 9 & 4.824 & 0.655 & - & - & 1.682E-14 & 0.145E-14 \\
- & 10 & 4.967 & 1.320 & - & - & 1.160E-14 & 0.179E-14 \\
Abell 2163 & 1 & 9.871 & 1.153 & 0.648 & 0.227 & 3.807E-12 & 0.255E-12 \\
- & 2 & 11.801 & 1.177 & 0.250 & 0.068 & 2.939E-12 & 0.112E-12 \\
- & 3 & 12.472 & 0.835 & - & - & 1.337E-12 & 0.049E-12 \\
- & 4 & 13.101 & 1.117 & 0.331 & 0.095 & 4.721E-13 & 0.230E-13 \\
- & 5 & 15.446 & 1.803 & - & - & 2.505E-13 & 0.122E-13 \\
- & 6 & 13.591 & 1.745 & - & - & 1.397E-13 & 0.083E-13 \\
- & 7 & 13.884 & 2.783 & - & - & 6.871E-14 & 0.420E-14 \\
- & 8 & 7.125 & 1.593 & - & - & 2.234E-14 & 0.205E-14 \\
- & 9 & 6.906 & 5.839 & - & - & 3.493E-15 & 1.253E-15 \\
Abell 2199 & 1 & 3.011 & 0.052 & 1.023 & 0.063 & 8.126E-12 & 0.211E-12 \\
- & 2 & 3.721 & 0.066 & 0.695 & 0.050 & 4.617E-12 & 0.107E-12 \\
- & 3 & 4.129 & 0.046 & 0.483 & 0.029 & 2.649E-12 & 0.041E-12 \\
- & 4 & 4.278 & 0.053 & 0.498 & 0.035 & 1.256E-12 & 0.024E-12 \\
- & 5 & 4.161 & 0.061 & 0.430 & 0.037 & 7.471E-13 & 0.161E-13 \\
- & 6 & 4.272 & 0.074 & 0.383 & 0.044 & 4.469E-13 & 0.114E-13 \\
- & 7 & 4.251 & 0.071 & 0.396 & 0.042 & 2.695E-13 & 0.066E-13 \\
- & 8 & 4.117 & 0.099 & 0.519 & 0.050 & 1.245E-13 & 0.039E-13 \\
- & 9 & 4.422 & 0.192 & - & - & 6.276E-14 & 0.229E-14 \\
- & 10 & 4.611 & 0.390 & - & - & 3.822E-14 & 0.230E-14 \\
Abell 2204 & 1 & 4.807 & 0.071 & 0.655 & 0.035 & 1.879E-11 & 0.020E-11 \\
- & 2 & 8.706 & 0.433 & 0.348 & 0.068 & 3.387E-12 & 0.080E-12 \\
- & 3 & 8.335 & 0.269 & 0.370 & 0.061 & 9.448E-13 & 0.238E-13 \\
- & 4 & 8.384 & 0.463 & 0.372 & 0.078 & 2.374E-13 & 0.085E-13 \\
- & 5 & 8.323 & 0.575 & - & - & 1.190E-13 & 0.048E-13 \\
- & 6 & 9.085 & 1.276 & - & - & 5.940E-14 & 0.298E-14 \\
- & 7 & 7.499 & 1.109 & - & - & 2.721E-14 & 0.166E-14 \\
- & 8 & 7.486 & 3.432 & - & - & 6.861E-15 & 1.134E-15 \\
Abell 2218 & 1 & 7.295 & 0.806 & 0.730 & 0.386 & 1.393E-12 & 0.180E-12 \\
- & 2 & 8.342 & 0.599 & 0.368 & 0.148 & 9.645E-13 & 0.579E-13 \\
- & 3 & 6.759 & 0.304 & 0.230 & 0.059 & 3.572E-13 & 0.157E-13 \\
- & 4 & 6.675 & 0.451 & - & - & 1.048E-13 & 0.054E-13 \\
- & 5 & 5.484 & 0.610 & - & - & 4.587E-14 & 0.287E-14 \\
- & 6 & 6.357 & 0.857 & - & - & 2.687E-14 & 0.199E-14 \\
- & 7 & 3.387 & 0.392 & - & - & 1.077E-14 & 0.113E-14 \\
- & 8 & 1.323 & 0.095 & - & - & 2.543E-15 & 0.720E-15 \\
- & 9 & 0.609 & 0.098 & - & - & 7.458E-16 & 4.554E-16 \\
Abell 2256 & 1 & 6.241 & 2.232 & 0.342 & 0.063 & 8.576E-13 & 0.800E-13 \\
- & 2 & 6.173 & 0.722 & - & - & 8.819E-13 & 0.474E-13 \\
- & 3 & 5.937 & 0.302 & - & - & 1.031E-12 & 0.039E-12 \\
- & 4 & 5.813 & 0.246 & - & - & 7.972E-13 & 0.289E-13 \\
- & 5 & 5.497 & 0.236 & 0.434 & 0.061 & 5.563E-13 & 0.231E-13 \\
- & 6 & 6.633 & 0.263 & - & - & 4.054E-13 & 0.148E-13 \\
- & 7 & 6.704 & 0.240 & 0.278 & 0.063 & 2.179E-13 & 0.079E-13 \\
- & 8 & 6.767 & 0.345 & - & - & 1.103E-13 & 0.044E-13 \\
- & 9 & 7.647 & 0.889 & - & - & 4.244E-14 & 0.223E-14 \\
- & 10 & 6.727 & 1.572 & - & - & 1.428E-14 & 0.181E-14 \\
Abell 2319 & 1 & 8.310 & 0.811 & 0.611 & 0.315 & 2.251E-12 & 0.221E-12 \\
- & 2 & 10.162 & 0.716 & 0.316 & 0.132 & 2.052E-12 & 0.089E-12 \\
- & 3 & 8.944 & 0.308 & 0.265 & 0.048 & 2.182E-12 & 0.039E-12 \\
- & 4 & 8.894 & 0.272 & 0.303 & 0.041 & 1.738E-12 & 0.028E-12 \\
- & 5 & 8.427 & 0.195 & 0.312 & 0.044 & 1.181E-12 & 0.021E-12 \\
- & 6 & 8.919 & 0.333 & 0.313 & 0.053 & 7.701E-13 & 0.163E-13 \\
- & 7 & 9.006 & 0.313 & 0.233 & 0.037 & 5.671E-13 & 0.099E-13 \\
- & 8 & 8.871 & 0.443 & - & - & 3.489E-13 & 0.072E-13 \\
- & 9 & 8.448 & 0.362 & - & - & 1.907E-13 & 0.045E-13 \\
- & 10 & 7.671 & 0.703 & - & - & 1.211E-13 & 0.047E-13 \\
Abell 2589 & 1 & 3.349 & 0.107 & 0.975 & 0.150 & 1.749E-12 & 0.134E-12 \\
- & 2 & 3.730 & 0.110 & 0.816 & 0.096 & 9.821E-13 & 0.515E-13 \\
- & 3 & 3.622 & 0.071 & 0.582 & 0.045 & 6.220E-13 & 0.189E-13 \\
- & 4 & 3.575 & 0.081 & 0.567 & 0.050 & 3.139E-13 & 0.108E-13 \\
- & 5 & 3.579 & 0.099 & 0.437 & 0.055 & 1.826E-13 & 0.072E-13 \\
- & 6 & 3.625 & 0.129 & 0.449 & 0.070 & 1.050E-13 & 0.051E-13 \\
- & 7 & 3.378 & 0.098 & 0.392 & 0.061 & 7.137E-14 & 0.337E-14 \\
- & 8 & 3.473 & 0.217 & 0.442 & 0.097 & 3.668E-14 & 0.260E-14 \\
- & 9 & 2.863 & 0.276 & 0.307 & 0.095 & 1.631E-14 & 0.174E-14 \\
- & 10 & 4.660 & 0.880 & - & - & 1.628E-14 & 0.174E-14 \\
Abell 2597 & 1 & 3.054 & 0.027 & 0.530 & 0.020 & 1.113E-11 & 0.011E-11 \\
- & 2 & 3.800 & 0.055 & 0.394 & 0.030 & 2.562E-12 & 0.045E-12 \\
- & 3 & 3.869 & 0.056 & 0.310 & 0.028 & 6.003E-13 & 0.117E-13 \\
- & 4 & 3.722 & 0.092 & 0.299 & 0.045 & 1.543E-13 & 0.050E-13 \\
- & 5 & 3.665 & 0.120 & 0.268 & 0.057 & 7.358E-14 & 0.307E-14 \\
- & 6 & 3.399 & 0.142 & 0.278 & 0.079 & 3.380E-14 & 0.215E-14 \\
- & 7 & 3.170 & 0.178 & 0.372 & 0.073 & 1.552E-14 & 0.104E-14 \\
- & 8 & 2.751 & 0.311 & - & - & 6.340E-15 & 0.593E-15 \\
- & 9 & 2.877 & 0.438 & - & - & 3.688E-15 & 0.427E-15 \\
Abell 2626 & 1 & 2.610 & 0.042 & 0.693 & 0.054 & 2.492E-12 & 0.097E-12 \\
- & 2 & 3.084 & 0.066 & 0.571 & 0.056 & 8.441E-13 & 0.354E-13 \\
- & 3 & 3.134 & 0.049 & 0.501 & 0.036 & 4.176E-13 & 0.126E-13 \\
- & 4 & 3.323 & 0.064 & 0.448 & 0.047 & 1.722E-13 & 0.064E-13 \\
- & 5 & 3.251 & 0.086 & 0.413 & 0.047 & 8.561E-14 & 0.352E-14 \\
- & 6 & 3.288 & 0.119 & - & - & 4.422E-14 & 0.198E-14 \\
- & 7 & 2.762 & 0.149 & 0.229 & 0.055 & 2.233E-14 & 0.148E-14 \\
- & 8 & 1.973 & 0.133 & 0.271 & 0.073 & 7.593E-15 & 1.039E-15 \\
- & 9 & 1.901 & 0.261 & 0.289 & 0.124 & 2.561E-15 & 0.784E-15 \\
- & 10 & 1.685 & 0.158 & - & - & 2.120E-15 & 0.859E-15 \\
Abell 2667 & 1 & 5.530 & 0.196 & 0.554 & 0.061 & 5.890E-12 & 0.142E-12 \\
- & 2 & 8.529 & 0.455 & 0.352 & 0.061 & 1.321E-12 & 0.041E-12 \\
- & 3 & 7.616 & 0.442 & - & - & 3.116E-13 & 0.108E-13 \\
- & 4 & 7.788 & 0.810 & - & - & 7.479E-14 & 0.366E-14 \\
- & 5 & 5.558 & 0.810 & - & - & 2.791E-14 & 0.190E-14 \\
- & 6 & 8.407 & 2.186 & - & - & 1.370E-14 & 0.143E-14 \\
- & 7 & 3.807 & 0.973 & - & - & 5.288E-15 & 0.786E-15 \\
- & 8 & 3.145 & 0.777 & - & - & 3.624E-15 & 0.628E-15 \\
Abell 2717 & 1 & 2.040 & 0.039 & 0.953 & 0.085 & 8.896E-13 & 0.607E-13 \\
- & 2 & 2.538 & 0.057 & 0.819 & 0.079 & 3.955E-13 & 0.240E-13 \\
- & 3 & 2.430 & 0.044 & 0.495 & 0.037 & 1.952E-13 & 0.079E-13 \\
- & 4 & 2.497 & 0.059 & 0.437 & 0.044 & 8.545E-14 & 0.407E-14 \\
- & 5 & 2.291 & 0.078 & 0.359 & 0.043 & 4.538E-14 & 0.250E-14 \\
- & 6 & 2.303 & 0.107 & 0.384 & 0.058 & 2.579E-14 & 0.184E-14 \\
- & 7 & 2.310 & 0.128 & 0.364 & 0.063 & 1.419E-14 & 0.114E-14 \\
- & 8 & 2.085 & 0.119 & 0.473 & 0.112 & 5.339E-15 & 0.850E-15 \\
- & 9 & 1.954 & 0.199 & - & - & 2.055E-15 & 0.542E-15 \\
- & 10 & 1.682 & 0.118 & - & - & 1.729E-15 & 0.642E-15 \\
Abell 3112 & 1 & 3.379 & 0.036 & 1.113 & 0.053 & 9.632E-12 & 0.184E-12 \\
- & 2 & 4.797 & 0.100 & 0.590 & 0.053 & 3.081E-12 & 0.071E-13 \\
- & 3 & 4.979 & 0.091 & 0.427 & 0.040 & 1.020E-12 & 0.021E-13 \\
- & 4 & 4.876 & 0.123 & 0.518 & 0.063 & 3.337E-13 & 0.108E-13 \\
- & 5 & 4.552 & 0.165 & 0.270 & 0.071 & 1.499E-13 & 0.060E-13 \\
- & 6 & 4.697 & 0.226 & 0.212 & 0.050 & 8.087E-14 & 0.300E-14 \\
- & 7 & 4.184 & 0.180 & - & - & 4.278E-14 & 0.184E-14 \\
- & 8 & 4.048 & 0.328 & - & - & 1.723E-14 & 0.115E-14 \\
- & 9 & 2.354 & 0.246 & - & - & 6.746E-15 & 0.834E-15 \\
Abell 3158 & 1 & 6.253 & 0.470 & 0.710 & 0.472 & 1.515E-12 & 0.230E-12 \\
- & 2 & 5.305 & 0.183 & 0.724 & 0.130 & 1.239E-12 & 0.066E-12 \\
- & 3 & 5.627 & 0.160 & 0.350 & 0.050 & 9.060E-13 & 0.221E-13 \\
- & 4 & 5.293 & 0.114 & 0.444 & 0.054 & 5.160E-13 & 0.140E-13 \\
- & 5 & 5.018 & 0.125 & 0.433 & 0.060 & 3.296E-13 & 0.104E-13 \\
- & 6 & 5.230 & 0.151 & 0.432 & 0.071 & 2.049E-13 & 0.073E-13 \\
- & 7 & 5.014 & 0.163 & 0.285 & 0.066 & 1.362E-13 & 0.049E-13 \\
- & 8 & 5.875 & 0.404 & 0.279 & 0.118 & 7.133E-14 & 0.406E-14 \\
- & 9 & 6.460 & 0.525 & - & - & 3.595E-14 & 0.244E-14 \\
Abell 3526 & 1 & 1.299 & 0.037 & 0.503 & 0.012 & 1.013E-11 & 0.014E-11 \\
- & 2 & 1.983 & 0.010 & 1.831 & 0.054 & 4.244E-12 & 0.120E-12 \\
- & 3 & 2.558 & 0.012 & 1.492 & 0.032 & 2.147E-12 & 0.037E-12 \\
- & 4 & 2.907 & 0.020 & 1.176 & 0.028 & 1.194E-12 & 0.020E-12 \\
- & 5 & 3.207 & 0.023 & 0.782 & 0.025 & 7.746E-13 & 0.126E-13 \\
- & 6 & 3.268 & 0.027 & 0.591 & 0.023 & 5.541E-13 & 0.093E-13 \\
- & 7 & 3.324 & 0.024 & 0.482 & 0.018 & 4.419E-13 & 0.061E-13 \\
- & 8 & 3.369 & 0.030 & 0.447 & 0.021 & 2.902E-13 & 0.047E-13 \\
- & 9 & 3.372 & 0.033 & 0.447 & 0.023 & 1.864E-13 & 0.033E-13 \\
- & 10 & 3.300 & 0.056 & 0.487 & 0.039 & 1.404E-13 & 0.043E-13 \\
Abell 3558 & 1 & 4.677 & 0.136 & 0.761 & 0.082 & 2.637E-12 & 0.094E-12 \\
- & 2 & 5.552 & 0.168 & 0.476 & 0.056 & 1.790E-12 & 0.045E-12 \\
- & 3 & 5.757 & 0.099 & 0.455 & 0.032 & 1.161E-12 & 0.018E-12 \\
- & 4 & 5.724 & 0.106 & 0.378 & 0.032 & 6.826E-13 & 0.108E-13 \\
- & 5 & 5.492 & 0.105 & 0.353 & 0.031 & 5.075E-13 & 0.081E-13 \\
- & 6 & 5.114 & 0.078 & 0.412 & 0.034 & 3.728E-13 & 0.068E-13 \\
- & 7 & 5.122 & 0.071 & 0.294 & 0.028 & 2.667E-13 & 0.042E-13 \\
- & 8 & 4.878 & 0.097 & 0.337 & 0.040 & 1.373E-13 & 0.031E-13 \\
- & 9 & 5.043 & 0.133 & 0.264 & 0.053 & 6.947E-14 & 0.210E-14 \\
- & 10 & 4.710 & 0.264 & 0.376 & 0.119 & 4.147E-14 & 0.278E-14 \\
Abell 3560 & 1 & 3.331 & 0.359 & 0.346 & 0.115 & 2.771E-13 & 0.306E-13 \\
- & 2 & 3.253 & 0.190 & - & - & 2.286E-13 & 0.224E-13 \\
- & 3 & 3.997 & 0.139 & 0.393 & 0.083 & 2.083E-13 & 0.118E-13 \\
- & 4 & 3.841 & 0.141 & 0.458 & 0.083 & 1.293E-13 & 0.075E-13 \\
- & 5 & 3.479 & 0.164 & 0.351 & 0.080 & 9.652E-14 & 0.617E-14 \\
- & 6 & 3.827 & 0.182 & 0.392 & 0.100 & 6.497E-14 & 0.454E-14 \\
- & 7 & 3.693 & 0.163 & 0.285 & 0.069 & 4.344E-14 & 0.244E-14 \\
- & 8 & 3.528 & 0.207 & 0.329 & 0.065 & 2.858E-14 & 0.176E-14 \\
- & 9 & 3.856 & 0.246 & - & - & 2.028E-14 & 0.138E-14 \\
- & 10 & 4.093 & 0.442 & - & - & 1.519E-14 & 0.141E-14 \\
Abell 3581 & 1 & 1.396 & 0.017 & 0.432 & 0.022 & 3.243E-12 & 0.113E-12 \\
- & 2 & 1.585 & 0.013 & 0.593 & 0.028 & 1.484E-12 & 0.056E-12 \\
- & 3 & 1.764 & 0.020 & 0.536 & 0.022 & 5.915E-13 & 0.185E-13 \\
- & 4 & 1.895 & 0.028 & 0.346 & 0.022 & 2.212E-13 & 0.080E-13 \\
- & 5 & 1.909 & 0.035 & 0.320 & 0.025 & 1.221E-13 & 0.052E-13 \\
- & 6 & 1.886 & 0.046 & 0.307 & 0.030 & 6.876E-14 & 0.366E-14 \\
- & 7 & 1.843 & 0.046 & 0.238 & 0.023 & 4.600E-14 & 0.222E-14 \\
- & 8 & 1.648 & 0.034 & 0.195 & 0.023 & 2.526E-14 & 0.154E-14 \\
- & 9 & 1.623 & 0.042 & 0.225 & 0.031 & 1.388E-14 & 0.121E-14 \\
- & 10 & 1.535 & 0.092 & 0.176 & 0.048 & 9.583E-15 & 1.583E-15 \\
Abell 3827 & 1 & 7.938 & 0.565 & 0.682 & 0.206 & 2.633E-12 & 0.169E-12 \\
- & 2 & 6.944 & 0.373 & 0.345 & 0.091 & 1.718E-12 & 0.063E-12 \\
- & 3 & 7.144 & 0.247 & 0.378 & 0.054 & 8.827E-13 & 0.218E-13 \\
- & 4 & 6.920 & 0.304 & 0.385 & 0.044 & 3.679E-13 & 0.092E-13 \\
- & 5 & 6.683 & 0.257 & - & - & 1.902E-13 & 0.053E-13 \\
- & 6 & 6.926 & 0.498 & - & - & 1.038E-13 & 0.033E-13 \\
- & 7 & 5.804 & 0.382 & - & - & 5.573E-14 & 0.203E-14 \\
- & 8 & 5.867 & 0.650 & - & - & 2.549E-14 & 0.146E-14 \\
- & 9 & 6.561 & 1.230 & - & - & 1.033E-14 & 0.129E-14 \\
Abell 3911 & 1 & 6.454 & 0.707 & 1.265 & 0.716 & 5.754E-13 & 1.407E-13 \\
- & 2 & 6.636 & 0.467 & 0.666 & 0.245 & 4.875E-13 & 0.452E-13 \\
- & 3 & 6.389 & 0.233 & 0.401 & 0.080 & 3.971E-13 & 0.150E-13 \\
- & 4 & 5.538 & 0.246 & 0.316 & 0.077 & 2.185E-13 & 0.087E-13 \\
- & 5 & 5.591 & 0.301 & 0.176 & 0.049 & 1.203E-13 & 0.042E-13 \\
- & 6 & 5.953 & 0.405 & - & - & 7.168E-14 & 0.284E-14 \\
- & 7 & 4.981 & 0.290 & - & - & 4.071E-14 & 0.185E-14 \\
- & 8 & 3.697 & 0.420 & - & - & 1.351E-14 & 0.120E-14 \\
- & 9 & 3.215 & 0.379 & - & - & 8.326E-15 & 1.054E-15 \\
- & 10 & 3.611 & 1.182 & - & - & 5.228E-15 & 1.288E-15 \\
Abell 3921 & 1 & 5.234 & 0.245 & 0.588 & 0.160 & 1.322E-12 & 0.091E-12 \\
- & 2 & 6.095 & 0.287 & 0.341 & 0.095 & 9.003E-13 & 0.382E-13 \\
- & 3 & 5.570 & 0.183 & 0.377 & 0.057 & 4.962E-13 & 0.145E-13 \\
- & 4 & 5.302 & 0.144 & 0.354 & 0.059 & 2.444E-13 & 0.078E-13 \\
- & 5 & 5.312 & 0.170 & 0.479 & 0.090 & 1.409E-13 & 0.062E-13 \\
- & 6 & 5.036 & 0.232 & 0.313 & 0.059 & 7.505E-14 & 0.303E-14 \\
- & 7 & 5.577 & 0.337 & - & - & 4.503E-14 & 0.178E-14 \\
- & 8 & 4.711 & 0.371 & - & - & 2.138E-14 & 0.122E-14 \\
- & 9 & 2.551 & 0.231 & - & - & 6.745E-15 & 0.851E-15 \\
- & 10 & 2.854 & 0.638 & - & - & 5.575E-15 & 0.989E-15 \\
Abell 4059 & 1 & 2.868 & 0.071 & 1.356 & 0.108 & 2.976E-12 & 0.155E-12 \\
- & 2 & 3.688 & 0.077 & 0.880 & 0.069 & 1.852E-12 & 0.066E-12 \\
- & 3 & 4.087 & 0.060 & 0.503 & 0.038 & 9.636E-13 & 0.220E-13 \\
- & 4 & 4.285 & 0.078 & 0.456 & 0.049 & 4.045E-13 & 0.115E-13 \\
- & 5 & 4.154 & 0.096 & 0.371 & 0.056 & 2.194E-13 & 0.075E-13 \\
- & 6 & 4.038 & 0.118 & 0.411 & 0.071 & 1.268E-13 & 0.055E-13 \\
- & 7 & 4.093 & 0.121 & 0.443 & 0.071 & 7.628E-14 & 0.330E-14 \\
- & 8 & 4.118 & 0.211 & 0.355 & 0.093 & 3.329E-14 & 0.219E-14 \\
- & 9 & 4.143 & 0.363 & - & - & 1.443E-14 & 0.138E-14 \\
- & 10 & 3.326 & 0.576 & - & - & 7.179E-15 & 1.351E-15 \\
AWM7 & 1 & 2.780 & 0.056 & 1.440 & 0.094 & 3.722E-12 & 0.169E-12 \\
- & 2 & 3.351 & 0.046 & 1.120 & 0.070 & 2.023E-12 & 0.072E-12 \\
- & 3 & 3.526 & 0.040 & 0.798 & 0.032 & 1.451E-12 & 0.027E-12 \\
- & 4 & 3.616 & 0.040 & 0.647 & 0.027 & 9.924E-13 & 0.175E-13 \\
- & 5 & 3.728 & 0.041 & 0.623 & 0.028 & 7.379E-13 & 0.134E-13 \\
- & 6 & 3.609 & 0.044 & 0.538 & 0.027 & 5.481E-13 & 0.102E-13 \\
- & 7 & 3.631 & 0.039 & 0.476 & 0.021 & 4.421E-13 & 0.068E-13 \\
- & 8 & 3.547 & 0.049 & 0.391 & 0.024 & 2.772E-13 & 0.050E-13 \\
- & 9 & 3.528 & 0.057 & 0.434 & 0.027 & 1.639E-13 & 0.034E-13 \\
- & 10 & 3.391 & 0.079 & 0.334 & 0.044 & 1.108E-13 & 0.041E-13 \\
Coma & 1 & 9.933 & 1.389 & 0.490 & 0.154 & 1.500E-12 & 0.091E-12 \\
- & 2 & 8.079 & 0.528 & - & - & 1.326E-12 & 0.071E-12 \\
- & 3 & 8.396 & 0.222 & 0.275 & 0.056 & 1.400E-12 & 0.029E-12 \\
- & 4 & 8.395 & 0.184 & 0.257 & 0.044 & 1.310E-12 & 0.022E-12 \\
- & 5 & 8.156 & 0.168 & 0.251 & 0.038 & 1.165E-12 & 0.018E-12 \\
- & 6 & 8.324 & 0.166 & 0.233 & 0.039 & 1.032E-12 & 0.016E-12 \\
- & 7 & 8.231 & 0.121 & 0.297 & 0.028 & 9.749E-13 & 0.111E-13 \\
- & 8 & 8.339 & 0.132 & 0.270 & 0.032 & 6.986E-13 & 0.088E-13 \\
- & 9 & 8.140 & 0.132 & 0.333 & 0.033 & 4.489E-13 & 0.065E-13 \\
- & 10 & 8.320 & 0.222 & 0.278 & 0.063 & 3.279E-13 & 0.078E-13 \\
E1455+2232 & 1 & 4.409 & 0.095 & 0.503 & 0.040 & 4.869E-12 & 0.098E-12 \\
- & 2 & 5.247 & 0.244 & 0.401 & 0.117 & 5.104E-13 & 0.297E-13 \\
- & 3 & 4.954 & 0.263 & 0.509 & 0.125 & 9.724E-14 & 0.664E-14 \\
- & 4 & 5.165 & 0.796 & - & - & 1.468E-14 & 0.168E-14 \\
- & 5 & 5.408 & 0.772 & - & - & 1.197E-14 & 0.129E-14 \\
- & 6 & 6.167 & 2.284 & - & - & 5.321E-15 & 0.939E-15 \\
- & 7 & 4.679 & 3.423 & - & - & 1.590E-15 & 0.604E-15 \\
EXO0422 & 1 & 2.429 & 0.027 & 0.705 & 0.033 & 6.726E-12 & 0.144E-12 \\
- & 2 & 2.938 & 0.045 & 0.557 & 0.037 & 2.158E-12 & 0.060E-12 \\
- & 3 & 2.953 & 0.039 & 0.389 & 0.025 & 8.194E-13 & 0.192E-13 \\
- & 4 & 3.047 & 0.055 & 0.310 & 0.032 & 3.005E-13 & 0.094E-13 \\
- & 5 & 3.140 & 0.072 & 0.342 & 0.044 & 1.531E-13 & 0.062E-13 \\
- & 6 & 2.825 & 0.097 & 0.402 & 0.055 & 8.436E-14 & 0.457E-14 \\
- & 7 & 2.564 & 0.078 & 0.290 & 0.044 & 5.283E-14 & 0.295E-14 \\
- & 8 & 2.296 & 0.134 & 0.367 & 0.075 & 2.223E-14 & 0.225E-14 \\
- & 9 & 2.366 & 0.189 & 0.244 & 0.074 & 1.221E-14 & 0.152E-14 \\
- & 10 & 3.089 & 0.434 & - & - & 1.069E-14 & 0.135E-14 \\
Hydra & 1 & 3.251 & 0.040 & 0.620 & 0.035 & 1.179E-11 & 0.017E-11 \\
- & 2 & 3.653 & 0.064 & 0.468 & 0.037 & 3.962E-12 & 0.081E-12 \\
- & 3 & 3.503 & 0.057 & 0.426 & 0.030 & 1.247E-12 & 0.027E-12 \\
- & 4 & 3.810 & 0.070 & 0.293 & 0.034 & 5.933E-13 & 0.143E-13 \\
- & 5 & 3.943 & 0.088 & 0.346 & 0.046 & 2.957E-13 & 0.091E-13 \\
- & 6 & 3.970 & 0.127 & 0.340 & 0.070 & 1.451E-13 & 0.066E-13 \\
- & 7 & 4.170 & 0.142 & 0.322 & 0.071 & 7.686E-14 & 0.355E-14 \\
- & 8 & 6.665 & 0.484 & - & - & 4.059E-14 & 0.214E-14 \\
Klemola 44 & 1 & 3.141 & 0.067 & 0.687 & 0.063 & 3.555E-12 & 0.125E-12 \\
- & 2 & 3.238 & 0.047 & 0.628 & 0.040 & 2.171E-12 & 0.055E-12 \\
- & 3 & 3.240 & 0.033 & 0.490 & 0.024 & 1.225E-12 & 0.022E-12 \\
- & 4 & 3.241 & 0.039 & 0.457 & 0.027 & 5.435E-13 & 0.116E-13 \\
- & 5 & 3.060 & 0.045 & 0.417 & 0.028 & 3.156E-13 & 0.076E-13 \\
- & 6 & 3.082 & 0.056 & 0.451 & 0.036 & 1.857E-13 & 0.055E-13 \\
- & 7 & 3.002 & 0.055 & 0.393 & 0.030 & 1.301E-13 & 0.035E-13 \\
- & 8 & 2.610 & 0.062 & 0.251 & 0.031 & 6.013E-14 & 0.224E-14 \\
- & 9 & 2.402 & 0.086 & 0.215 & 0.031 & 3.037E-14 & 0.150E-14 \\
- & 10 & 2.093 & 0.128 & 0.140 & 0.044 & 1.693E-14 & 0.166E-14 \\
M87 & 1 & 1.517 & 0.006 & 0.379 & 0.007 & 2.240E-11 & 0.015E-11 \\
- & 2 & 1.650 & 0.004 & 0.879 & 0.015 & 7.215E-12 & 0.080E-12 \\
- & 3 & 1.802 & 0.006 & 0.772 & 0.010 & 3.905E-12 & 0.033E-12 \\
- & 4 & 2.033 & 0.006 & 0.635 & 0.009 & 2.494E-12 & 0.020E-12 \\
- & 5 & 2.065 & 0.007 & 0.556 & 0.008 & 1.756E-12 & 0.015E-12 \\
- & 6 & 2.231 & 0.011 & 0.538 & 0.009 & 1.242E-12 & 0.011E-12 \\
- & 7 & 2.347 & 0.009 & 0.465 & 0.007 & 1.060E-12 & 0.008E-12 \\
- & 8 & 2.470 & 0.011 & 0.406 & 0.008 & 6.960E-13 & 0.058E-13 \\
- & 9 & 2.526 & 0.013 & 0.362 & 0.008 & 4.294E-13 & 0.039E-13 \\
- & 10 & 2.604 & 0.021 & 0.362 & 0.014 & 3.265E-13 & 0.050E-13 \\
MKW 3S & 1 & 3.043 & 0.044 & 0.910 & 0.048 & 4.954E-12 & 0.120E-12 \\
- & 2 & 3.397 & 0.043 & 0.656 & 0.040 & 2.432E-12 & 0.057E-12 \\
- & 3 & 3.619 & 0.047 & 0.443 & 0.026 & 1.032E-12 & 0.019E-12 \\
- & 4 & 3.812 & 0.060 & 0.312 & 0.030 & 4.453E-13 & 0.095E-13 \\
- & 5 & 3.549 & 0.076 & 0.287 & 0.036 & 2.038E-13 & 0.056E-13 \\
- & 6 & 3.530 & 0.099 & 0.401 & 0.052 & 1.131E-13 & 0.042E-13 \\
- & 7 & 3.473 & 0.101 & 0.327 & 0.046 & 6.595E-14 & 0.239E-14 \\
- & 8 & 3.625 & 0.157 & 0.482 & 0.070 & 3.148E-14 & 0.168E-14 \\
- & 9 & 3.955 & 0.225 & - & - & 1.610E-14 & 0.105E-14 \\
- & 10 & 4.597 & 0.866 & - & - & 8.398E-15 & 1.077E-15 \\
MKW 4 & 1 & 1.573 & 0.023 & 1.651 & 0.184 & 2.340E-12 & 0.302E-12 \\
- & 2 & 2.096 & 0.056 & 1.521 & 0.235 & 5.578E-13 & 0.820E-13 \\
- & 3 & 2.008 & 0.052 & 0.643 & 0.076 & 2.307E-13 & 0.192E-13 \\
- & 4 & 2.001 & 0.064 & 0.506 & 0.074 & 1.056E-13 & 0.096E-13 \\
- & 5 & 1.944 & 0.067 & 0.483 & 0.070 & 7.648E-14 & 0.703E-14 \\
- & 6 & 1.984 & 0.078 & 0.412 & 0.070 & 5.420E-14 & 0.525E-14 \\
- & 7 & 1.741 & 0.078 & 0.419 & 0.056 & 3.543E-14 & 0.327E-14 \\
- & 8 & 1.679 & 0.053 & 0.278 & 0.052 & 1.979E-14 & 0.219E-14 \\
- & 9 & 1.523 & 0.079 & 0.235 & 0.052 & 1.160E-14 & 0.177E-14 \\
- & 10 & 1.244 & 0.078 & 0.259 & 0.103 & 5.885E-15 & 2.469E-15 \\
Perseus & 1 & 4.043 & 0.014 & 0.477 & 0.008 & 5.565E-11 & 0.149E-11 \\
- & 2 & 3.292 & 0.007 & 0.681 & 0.007 & 3.332E-11 & 0.008E-11 \\
- & 3 & 3.686 & 0.006 & 0.676 & 0.005 & 1.964E-11 & 0.003E-11 \\
- & 4 & 4.276 & 0.007 & 0.610 & 0.005 & 9.889E-12 & 0.020E-12 \\
- & 5 & 5.215 & 0.014 & 0.508 & 0.006 & 5.551E-12 & 0.014E-12 \\
- & 6 & 5.766 & 0.023 & 0.482 & 0.008 & 3.584E-12 & 0.011E-12 \\
- & 7 & 6.071 & 0.022 & 0.456 & 0.007 & 2.514E-12 & 0.008E-12 \\
- & 8 & 6.484 & 0.028 & 0.416 & 0.009 & 1.631E-12 & 0.006E-12 \\
- & 9 & 6.741 & 0.029 & 0.424 & 0.010 & 1.030E-12 & 0.004E-12 \\
- & 10 & 7.258 & 0.072 & 0.405 & 0.018 & 7.729E-13 & 0.056E-13 \\
Perseus a\footnote{Second observation, under the name Abell 426 in the 
archive.} & 1 & 3.973 & 0.021 & 0.560 & 0.014 & 4.996E-11 & 0.022E-11 \\
- & 2 & 3.195 & 0.099 & 0.689 & 0.009 & 3.128E-11 & 0.011E-11 \\
- & 3 & 3.598 & 0.088 & 0.675 & 0.006 & 1.912E-11 & 0.004E-11 \\
- & 4 & 4.199 & 0.011 & 0.584 & 0.007 & 9.600E-12 & 0.027E-12 \\
- & 5 & 5.044 & 0.019 & 0.506 & 0.009 & 5.425E-12 & 0.020E-12 \\
- & 6 & 5.612 & 0.033 & 0.460 & 0.011 & 3.417E-12 & 0.015E-12 \\
- & 7 & 5.926 & 0.030 & 0.441 & 0.010 & 2.217E-12 & 0.009E-12 \\
- & 8 & 6.309 & 0.038 & 0.401 & 0.012 & 1.298E-12 & 0.007E-12 \\
- & 9 & 6.612 & 0.040 & 0.410 & 0.014 & 8.264E-13 & 0.050E-13 \\
- & 10 & 7.006 & 0.099 & 0.397 & 0.025 & 6.279E-13 & 0.065E-13 \\
PKS 0745-19 & 1 & 4.875 & 0.080 & 0.538 & 0.033 & 2.748E-11 & 0.030E-11 \\
- & 2 & 7.354 & 0.226 & 0.413 & 0.044 & 7.893E-12 & 0.121E-12 \\
- & 3 & 7.733 & 0.210 & 0.332 & 0.039 & 2.275E-12 & 0.041E-12 \\
- & 4 & 8.278 & 0.326 & 0.362 & 0.065 & 6.574E-13 & 0.194E-13 \\
- & 5 & 9.087 & 0.659 & 0.463 & 0.111 & 2.923E-13 & 0.128E-13 \\
- & 6 & 8.200 & 0.616 & 0.524 & 0.101 & 1.514E-13 & 0.070E-13 \\
- & 7 & 8.579 & 0.687 & - & - & 7.834E-14 & 0.380E-14 \\
- & 8 & 8.506 & 1.375 & - & - & 2.868E-14 & 0.219E-14 \\
- & 9 & 6.472 & 1.167 & - & - & 1.626E-14 & 0.165E-14 \\
- & 10 & 4.732 & 1.481 & - & - & 1.090E-14 & 0.202E-14 \\
RXCJ0605.8-3518 & 1 & 3.866 & 0.089 & 0.717 & 0.069 & 5.409E-12 & 0.171E-12 \\
- & 2 & 5.408 & 0.220 & 0.416 & 0.103 & 1.104E-12 & 0.054E-12 \\
- & 3 & 4.811 & 0.189 & 0.227 & 0.074 & 3.412E-13 & 0.153E-13 \\
- & 4 & 4.537 & 0.286 & 0.295 & 0.103 & 9.945E-14 & 0.645E-14 \\
- & 5 & 4.834 & 0.437 & - & - & 4.333E-14 & 0.313E-14 \\
- & 6 & 6.248 & 1.149 & - & - & 1.980E-14 & 0.180E-14 \\
- & 7 & 8.425 & 3.500 & - & - & 1.031E-14 & 0.132E-14 \\
RXCJ2234.5-3744 & 1 & 11.165 & 1.896 & 0.323 & 0.102 & 2.133E-12 & 0.124E-12 \\
- & 2 & 11.087 & 1.011 & - & - & 1.577E-12 & 0.069E-12 \\
- & 3 & 9.664 & 0.415 & 0.290 & 0.055 & 6.989E-13 & 0.192E-13 \\
- & 4 & 8.816 & 0.651 & - & - & 2.285E-13 & 0.076E-13 \\
- & 5 & 5.890 & 0.477 & 0.198 & 0.094 & 7.595E-14 & 0.405E-14 \\
- & 6 & 4.478 & 0.584 & - & - & 2.822E-14 & 0.216E-14 \\
- & 7 & 3.566 & 0.594 & - & - & 1.229E-14 & 0.128E-14 \\
- & 8 & 2.412 & 1.099 & - & - & 2.150E-15 & 0.697E-15 \\
RXJ0658-55 & 1 & 12.257 & 1.286 & 0.229 & 0.050 & 2.409E-12 & 0.109E-12 \\
- & 2 & 14.686 & 1.205 & - & - & 1.592E-12 & 0.064E-12 \\
- & 3 & 14.799 & 0.975 & - & - & 6.728E-13 & 0.276E-13 \\
- & 4 & 13.160 & 1.248 & - & - & 1.555E-13 & 0.082E-13 \\
- & 5 & 10.503 & 1.156 & - & - & 5.489E-14 & 0.314E-14 \\
- & 6 & 18.091 & 6.064 & - & - & 2.052E-14 & 0.243E-14 \\
RXJ1347-1145 & 1 & 11.426 & 0.446 & 0.390 & 0.045 & 9.748E-12 & 0.123E-12 \\
- & 2 & 15.643 & 1.239 & 0.214 & 0.081 & 9.690E-13 & 0.329E-13 \\
- & 3 & 11.086 & 1.179 & - & - & 1.594E-13 & 0.072E-13 \\
- & 4 & 7.533 & 1.484 & - & - & 1.935E-14 & 0.180E-14 \\
- & 5 & 10.277 & 3.281 & - & - & 9.269E-15 & 1.140E-15 \\
S\'ersic 159-3 & 1 & 2.365 & 0.016 & 0.554 & 0.015 & 7.273E-12 & 0.074E-12 \\
- & 2 & 2.614 & 0.018 & 0.466 & 0.016 & 2.359E-12 & 0.033E-12 \\
- & 3 & 2.741 & 0.024 & 0.366 & 0.014 & 6.529E-13 & 0.092E-13 \\
- & 4 & 2.642 & 0.031 & 0.278 & 0.019 & 1.670E-13 & 0.037E-13 \\
- & 5 & 2.572 & 0.041 & 0.224 & 0.021 & 8.322E-14 & 0.222E-14 \\
- & 6 & 2.523 & 0.061 & 0.255 & 0.030 & 4.061E-14 & 0.153E-14 \\
- & 7 & 2.093 & 0.056 & 0.152 & 0.018 & 1.848E-14 & 0.076E-14 \\
- & 8 & 1.681 & 0.053 & - & - & 6.327E-15 & 0.443E-15 \\
- & 9 & 1.322 & 0.059 & - & - & 1.637E-15 & 0.291E-15 \\
- & 10 & 2.491 & 0.528 & - & - & 2.306E-15 & 0.436E-15 \\
S\'ersic 159-3 a\footnote{Second observation, under the name AS~1101 in the 
archive.} & 1 & 2.374 & 0.026 & 0.554 & 0.024 & 7.165E-12 & 0.112E-12 \\
- & 2 & 2.600 & 0.030 & 0.486 & 0.027 & 2.424E-12 & 0.054E-12 \\
- & 3 & 2.744 & 0.039 & 0.351 & 0.022 & 6.678E-13 & 0.148E-13 \\
- & 4 & 2.606 & 0.051 & 0.246 & 0.030 & 1.739E-13 & 0.061E-13 \\
- & 5 & 2.620 & 0.065 & 0.278 & 0.039 & 8.679E-14 & 0.376E-14 \\
- & 6 & 2.424 & 0.094 & 0.155 & 0.024 & 4.177E-14 & 0.187E-14 \\
- & 7 & 2.276 & 0.121 & - & - & 1.952E-14 & 0.107E-14 \\
- & 8 & 1.898 & 0.136 & - & - & 7.853E-15 & 0.706E-15 \\
- & 9 & 1.519 & 0.140 & - & - & 2.530E-15 & 0.478E-15 \\
- & 10 & 1.287 & 0.085 & - & - & 2.632E-15 & 0.615E-15 \\
Triangulum & 1 & 10.192 & 1.381 & 0.554 & 0.588 & 3.323E-12 & 0.510E-12 \\
- & 2 & 11.615 & 1.334 & 0.344 & 0.172 & 3.126E-12 & 0.183E-12 \\
- & 3 & 9.432 & 0.388 & 0.330 & 0.066 & 2.436E-12 & 0.059E-12 \\
- & 4 & 9.337 & 0.388 & 0.308 & 0.065 & 1.538E-12 & 0.039E-12 \\
- & 5 & 9.405 & 0.410 & 0.272 & 0.067 & 1.053E-12 & 0.028E-12 \\
- & 6 & 9.528 & 0.450 & 0.222 & 0.074 & 7.344E-13 & 0.213E-13 \\
- & 7 & 8.379 & 0.252 & 0.331 & 0.043 & 5.297E-13 & 0.111E-13 \\
- & 8 & 9.156 & 0.495 & - & - & 2.796E-13 & 0.065E-13 \\
- & 9 & 8.701 & 0.633 & - & - & 1.385E-13 & 0.038E-13 \\
- & 10 & 13.256 & 2.205 & - & - & 9.028E-14 & 0.600E-14 \\
ZW3146 & 1 & 5.289 & 0.068 & 0.517 & 0.028 & 7.745E-12 & 0.083E-12 \\
- & 2 & 8.106 & 0.259 & 0.255 & 0.054 & 1.116E-12 & 0.027E-12 \\
- & 3 & 7.668 & 0.312 & 0.221 & 0.053 & 1.985E-13 & 0.056E-13 \\
- & 4 & 7.542 & 0.746 & - & - & 3.378E-14 & 0.162E-14 \\
- & 5 & 10.906 & 2.961 & - & - & 1.589E-14 & 0.140E-14 \\
- & 6 & 6.385 & 1.955 & - & - & 4.422E-15 & 0.642E-15 \\
\end{longtable}

\end{document}